\documentclass[11pt]{amsart}

\usepackage{latexsym}
\usepackage{amssymb}
\usepackage{amsmath}
\usepackage{color}
\usepackage{xcolor}

\usepackage{tikz-cd}

\calclayout

\usepackage{amscd} 

\usepackage{comment}

\usepackage{bbm}
\usepackage{hyperref}

\usepackage{appendix}
\newtheorem{theorem}{Theorem}[section]
\newtheorem{lemma}[theorem]{Lemma}
\newtheorem{proposition}[theorem]{Proposition}
\newtheorem{corollary}[theorem]{Corollary}
\newtheorem{definition}[theorem]{Definition}

\newtheorem{remark}[theorem]{Remark}

\newtheorem*{theorem*}{Theorem}
\newtheorem*{definition*}{Definition}

\newcommand\id{\mathop{\rm id}}

\newcommand\conv{\mathop{\rm conv}}

\newcommand\Tr{\mathop{\rm Tr}}

\newcommand{\A}{\mathcal{A}}

\newcommand{\cl}[1]{\mathcal{#1}}
\newcommand{\bb}[1]{\mathbb{#1}}

\begin{document}

\title{Cantor correlations II. Synchronicity and bisynchronicity}

\author[G. Baziotis]{Georgios Baziotis}
\address{
Research Institute of Mathematics and NextQuantum Innovation Research Center, Department of Physics and Astronomy\\ Seoul National University, Seoul 08826, Korea}
\email{baziotis@snu.ac.kr}

\author[A. Chatzinikolaou]{Alexandros Chatzinikolaou}
\address{Institute of Mathematics of the Polish Academy of Sciences\\ ul. Śniadeckich 8\\
00--656 Warszawa\\ Poland}
\email{achatzinikolaou@impan.pl}

\author[I. G. Todorov]{Ivan G. Todorov}
\address{
School of Mathematical Sciences, University of Delaware, 501 Ewing Hall,
Newark, DE 19716, USA}
\email{todorov@udel.edu}

\author[L. Turowska]{Lyudmila Turowska}
\address{Department of Mathematical Sciences, Chalmers University
of Technology and the University of Gothenburg, Gothenburg SE-412 96, Sweden}
\email{turowska@chalmers.se}

\date{16 November 2026}

\begin{abstract}
We study synchronous and bisynchronous no-signalling correlations, defined over Cantor topological spaces, 
from the perspectives of 
both continuous variable quantum information and operator algebra theories. 
We focus on strategies, satisfying a natural no-signalling condition
across the Cantor space levels, called graded, 
identifying a universal C*-algebra, whose tracial states of 
different types characterise synchronous correlations of 
respective types. The universal C*-algebra for bisynchronous graded 
correlations is shown to coincide with the C*-algebra of a 
projective limit of free wreath products of 
quantum permutation groups,  
introduced by Bhowmick, Goswami and Skalski, and 
a type description, analogous to the one in the synchronous case, 
is provided. Extending results in the finite case, 
we show that bisynchronous graded correlations 
canonically give rise to factorisable maps on type I factors, 
defined via Cantor permutation matrices.
We examine non-local games over Cantor topological spaces, 
with inputs/outputs that are finite words of equal length, 
define corresponding game algebras, and characterise the perfect
strategies of the games via the respective game algebras. 
In the case of the Cantor graph isomorphism game, 
we provide equivalent descriptions of the existence of 
perfect strategies of quantum commuting type, 
in terms of bigalois correspondences, factorisable maps, 
Cantor permutation intertwiners and game algebra non-triviality. 
Our results supply a quantum information interpretation 
of a notion of a quantum isomorphism of locally finite graphs, 
introduced by Rollier and Vaes, canonically arising from Cantor graphs. 
\end{abstract}

\maketitle

\tableofcontents


\section{Introduction}\label{s_intro}

The correlations between the behaviours of two spatially separated and non-communicating, but possibly entangled, parties, 
referred to hereafter simply as \emph{no-signalling correlations}, have been the focus of numerous investigations from both the quantum physical and the mathematical perspectives. They are of primary importance 
in entanglement detection and self-testing of quantum devices, 
and were the main theoretical tool behind the fundamental Bell Theorem \cite{bell}. They were at the heart of the 
separation between the tensor, the liminal tensor, 
and the commuting operator models of quantum mechanics \cite{fri-CEP, jnvwy, jnppsw, oz}, which 
in particular settled
(in the negative) the long-standing Tsirelson Problem 
in quantum physics and Connes Embedding Problem in operator algebra theory. 
They have, in addition, been instrumental in showing the distinctive features of general probabilistic theories and their 
comparison to quantum mechanical and classical theories \cite{plavala}. 

One of the main theoretical gadgets in these investigations are 
\emph{non-local games}, which allow for concrete frameworks, 
where the power and limitations of quantum mechanical models can be 
tested. These are cooperative games, played by a tandem of players against a verifier, whose aim is demonstrating the knowledge of 
an underlying classical reality by allowing to be interrogated 
with the aim of producing answers which match the rules of the game
(see e.g. \cite{harris-paulsen}). 
No-signalling correlations fulfill the role of being 
the \emph{game strategies} of the players \cite{harris-paulsen, lmprsstw}. 
In the lack of a perfect strategy, that is, a strategy that yields a win with probability one, one is interested in the optimal strategies, that is, the strategies that yield a maximum winning probability, the latter parameter being known as the \emph{game value} \cite{harris-paulsen, cltt}. The game value depends, of course, on the strategy pool to which the players have access, and the access to different physical models yields different values (see \cite{cmnsw, hmps, jnvwy, pt}). 
Remarkably, the (normalised) game value can strictly increase
if several rounds are played in parallel. The hosting paradigm for this phenomenon, known as 
\emph{parallel repetition}, has given rise to fruitful 
investigations, focusing on the level of advantage that 
repetition allows for a given game \cite{csuu, jpy, r}. At the 
heart of the phenomenon is the fact that the pool of joint 
no-signalling correlations, allowed by a given quantum mechanical model for $n$ parallel rounds, is strictly larger than the $n$-fold product 
of the correlation sets of each individual round. 

This was the motivation behind the paper \cite{btt} where, 
using the theory of quantum information channels over abstract 
alphabets \cite{kakihara},
a broad framework of \emph{measurable no-signalling correlations},
serving as strategies for games with memory, was 
introduced, and different types of such correlations, 
arising from the various physical models, were defined. 
This study was specialised to the setup that most directly 
relates to parallel repetition in \cite{bctt}, where 
\emph{Cantor correlations}, that is, measurable no-signalling 
correlations over Cantor topological spaces, were investigated
and, following fruitful developments in the 
finite case \cite{lmprsstw, pt}, characterised according to type via 
tensor products of inductive limit operator systems \cite{mt}. 
By analogy to the finite alphabet case, Cantor correlations 
can be viewed as no-signalling strategies for games over 
Cantor topological spaces, where the players receive strings 
of questions and are expected to produce 
strings of answers of length 
equal to that of the question strings. This is a far-reaching scope which includes the parallel repetition paradigm, but goes beyond it to capture games with memory. Each Cantor game can be thought of as the compound game of a sequence of finite 
games with $n$-string questions and answers each; in the limit as 
$n\to\infty$ one obtains the overlying compound game. 

The present work continues this line of investigation, 
focusing on \emph{synchronous} and \emph{bisynchronous} Cantor
games and correlations. 
We note that synchronous non-local games and synchronous 
no-signalling correlations over \emph{finite} alphabets 
have undergone an intensive 
development after their introduction in \cite{psstw}; we 
refer to \cite{bhtt, dpp, jnvwy, vidick} for a sample thereof. 
Bisynchronicity, introduced in \cite{paulsen-rahaman}, has made equally 
vibrant progress through its connections to factorisable maps (see 
\cite{paulsen-rahaman}), graph isomorphism games over classical and 
quantum \cite{bhtt2} finite alphabets and connections to 
compact quantum groups \cite{bigalois}.
In the present paper, we undertake a detailed study of both properties 
for Cantor correlations. 
Our study of (bi)synchronicity in an infinite dimensional setting
reveals fruitful connections between continuous variable quantum 
information theory, operator algebras and compact quantum groups. 

We consider a particular operationally meaningful setting, 
imposing the restriction that, for every $n$, 
independently of the inputs at the $(n+1)$-st round, 
the total probability of the outputs that can be given by the players should reduce to the accumulated probability 
of inputs/outputs up to the $n$-th round (see condition (\ref{eq:graded_ns})). 
Our condition can be thought of as a 
condition of no-signalling from the future to the present, 
over successive rounds of a game with memory; we call the resulting correlations \emph{graded}, as they correspond precisely, in the 
Heisenberg picture, to information channels over Cantor spaces 
that respect each of the terms of the inductive sequences
that give rise to the compound Cantor space. 

More concretely, given  inductive sequences $X = (X_n)_{n\in \bb{N}}$ and 
$A = (A_n)_{n\in \bb{N}}$ of finite sets, the corresponding 
Cantor topological spaces $\Omega_X$ and $\Omega_A$ play the role 
of the input and outputs sets for the players, respectively. 
We identify a C*-algebra $\cl B_{X,A}$ whose tracial states 
are in canonical correspondence with the graded Cantor correlations
over the pair $(X,A)$; this extends to the Cantor setting 
well-known characterisations in the finite case \cite{psstw, dpp}. 
The C*-algebra $\cl B_{X,A}$ is an inductive limit in the C*-algebra category of a sequence $(\cl B_{X_n,A_n})_{n\in \bb{N}}$ of 
finitely generated C*-algebras, in which the $n$-term $\cl B_{X_n,A_n}$
captures the no-signalling conditions up to level $n$. Given a Cantor game over the pair $(X,A)$, we identify a suitable quotient of $\cl B_{X,A}$ that plays the role of the game algebra from the finite case \cite{hmps} and 
show that the perfect strategies of the Cantor game of a specific type correspond to tracial states of the Cantor game algebra with a specified type property; namely, local strategies correspond to characters, quantum commuting -- to arbitrary tracial states, while quantum approximate -- to 
traces factorisable through an ultrapower of the hyperfinite II$_1$ factor. 

We, similarly, identify a universal C*-algebra $\cl B_X$ for bisynchronous graded Cantor correlations over $X$, 
whose traces of specific type correspond canonically to bisynchronous correlation types. 
Interestingly, $\cl B_X$ is the overlying C*-algebra of a 
compact quantum group studied previously in the literature 
as the projective limit of quantum permutation groups \cite{bgs, ss}. Our results thus provide an operational interpretation of 
the latter quantum groups, bridging quantum information theory and 
the theory of compact quantum groups. Given a bisynchronous Cantor game, 
we identify its game algebra and establish corresponding 
perfect strategy tracial characterisations. 

Factorisable maps between von Neumann algebras, first introduced in 
\cite{ade}, have received substantial attention in the literature 
\cite{hm1,hm2,mror,dlem, ctt}. These are completely positive
trace-preserving maps between semi-finite von Neumann algebras 
that admit factorisation through a third von Neumann algebra
via a normal trace-preserving *-homomorphism and an adjoint 
thereof. The study of factorisable maps from the perspective 
of quantum information theory was initiated in \cite{hm1} and 
carried out further in \cite{hm2} and \cite{paulsen-rahaman}. 
As an illustrious 
example, they were used in \cite{hm2} to settle (in the negative) the asymptotic quantum Birkhoff conjecture \cite{svw}. The scope of these studies was, however, somewhat restricted, in that only 
maps between (finite dimensional) 
matrix algebras were considered. Our theory opens up a route 
for such connections in the infinite dimensional case as well. 
Our particular setup
allows us, namely, to demonstrate that factorisable maps (of different types) 
on the algebra $\cl B(H)$ of bounded operators on a separable Hilbert space
can be obtained from Cantor bisynchronous correlations 
(of the respective type). 
The classes of maps thus obtained depend on the particular choice of 
coordinates for the underlying Hilbert space. 
We point out that the ideas presented here
can be extended in order to obtain factorisable maps over 
non-type I von Neumann algebras as well, 
by imposing suitable no-signalling conditions on the underlying 
correlations (of which gradedness is only one of the possible natural choices); these directions are left out of the present work 
as they are out of its immediate scope.

The present paper includes a detailed study of a particular example 
of a bisynchronous Cantor game, namely, the Cantor graph isomorphism game.
We note that the isomorphism game for 
finite graphs has been one of the prime examples demonstrating the 
close connections between no-signalling correlations, non-local games 
and combinatorics \cite{atserias-etal, lmr, mr-planar}; these 
articles all have as a starting point the operational, non-local game, 
definition of graph isomorphisms of quantum types. 
For locally finite graphs, a purely operator-algebraic 
approach to quantum isomorphisms was proposed in \cite{rollier-vaes}; 
it is therefore natural to attempt connecting the latter approach to the former. The tools we described so far and developed herein 
allow us to think of infinite, Cantor graph isomorphisms 
entirely from the perspective of continuous variable 
quantum information theory. We prove that the operational 
definitions of Cantor graph isomorphisms of various types 
have neat operator-algebraic characterisations, and 
apply this in particular to the case of locally finite graphs, 
arising from Cantor graphs, obtaining an 
operational interpretation of the definitions from \cite{rollier-vaes}. 

\smallskip

We next describe the organisation of the paper. 
In Section \ref{s_gradedcor} we recall the types of 
local, quantum, approximately quantum, quantum commuting and general 
no-signalling correlations, first over finite alphabets and then in the 
measurable case. We define the main class of correlations studied thereafter, 
namely, graded Cantor correlations, and provide a lacking characterisation 
of the ones of local type (Proposition \ref{p_indforlo}). 
The Cantor setup involves two families, $X = (X_n)_{n\in \bb{N}}$ and 
$A = (A_n)_{n\in \bb{N}}$, of finite sets, where the cardinality of $X_n$
(resp. $A_n$) divides that of $X_{n+1}$ (resp. $A_{n+1}$) for each $n$ (this property is referred to as
\emph{inductivity}). Graded Cantor correlations are thus 
families $(p_n)_{n\in \bb{N}}$ of no-signalling correlations, 
where $p_n$ has an underlying quadruple $(X_n,X_n,A_n,A_n)$,
satisfying further the aforementioned no-signalling condition 
from the future to the present (see (\ref{eq:graded_ns}) for the 
precise formulas). 
We provide the perspectives in both 
the Heisenberg and the Schrödinger picture and the passage from one to another
for the case of graded correlations; both pictures will be used thereafter according to which one is more suitable for the specific task at hand. 

In Section \ref{s_syn} we introduce synchronous graded Cantor correlations
and define the universal C*-algebra $\cl B_{X,A}$, showing that its 
tracial state spaces of various types correspond precisely to 
graded synchronous Cantor correlations of corresponding type. 
The result (Theorem \ref{th_trBXA}) extends the well-known 
characterisations in the finite case; while the quantum commuting 
case can be obtained using inductive limit properties, 
the approximately quantum case, in which we achieve a characterisation 
via amenable tracial states, exhibits subtleties due to 
the lack of injectivity, as so does the quantum case due to the 
fact that the size of entanglement allowed at level $n$ may 
grow to infinity with $n$. A central tool here is establishing that 
the $n$-level universal algebra $\cl B_{X_n,A_n}$ injects into 
the $n+1$-level algebra $\cl B_{X_{n+1},A_{n+1}}$; this requires an 
alternative direct description of $\cl B_{X_n,A_n}$ as 
an inductively defined amalgamated free product. The construction 
is new in spirit and can be thought of as a version of a semi-wreath 
product. This analogy is clearly seen in the chatacterisation of 
the character space of $\cl B_{X,A}$ (Proposition \ref{p_charactBXA}). 

In Section \ref{s_cqpg}, we restrict to the case where $X = A$ 
and introduce bisynchronous graded Cantor correlations and an accompanying universal C*-algebra $\cl B_X$, showing that its tracial state spaces of different types correspond to bisynchronous graded Cantor correlations 
of respective types (Theorem \ref{th_trBX}). The C*-algebra $\cl B_X$ has three alternative equivalent 
descriptions: as an inductive limit of quotients of 
the $n$-level synchronous algebras $\cl B_{X_n,A_n}$, as 
a quotient of the synchronous universal C*-algebra $\cl B_{X,A}$, 
and as the C*-algebra of a projective limit of quantum permutation 
groups \cite{bgs}. 

The latter description is discussed as part of Section \ref{s_fmsc}, 
namely in Subsection \ref{subs_projlims}; while the algebra $\cl B_X$ has been studied from the perspective of quantum group theory in \cite{bgs}, 
a detailed description via wreath products makes connection
with the factorisable maps on which Section \ref{s_fmsc} is focused
(see Theorem \ref{thm:corepsandcoacts}). 
Bisynchronous graded Cantor correlations give rise to factorisable maps 
via an inductive procedure; however, a direct integral formula 
can also be provided, as demonstrated in Subsection \ref{ss_intformu}. 

Section \ref{s_cantorgames} hosts definitions and results 
about general Cantor games. We confirm that the existence of a 
perfect local strategy for such a game is equivalent to
deterministically winning the game (Proposition \ref{p_perfect_cantor}), 
and proceed by defining the game algebra of a synchronous Cantor game. Following the well-known paradigm in the case of finite games, 
we provide tracial descriptions of the perfect strategies of different types, 
and achieve a similar result for 
a bisynchronous version of the game algebra. 

Section \ref{s_cicg} is dedicated to the study of quantum isomorphisms 
of Cantor graphs. The latter
can be thought of as inductive limits of successively embedded finite graphs
or, alternatively, as locally finite graphs. The main characterisation 
result comprises Theorem \ref{thm:main_equivalence} and 
Corollary \ref{cor:iso_fact}, where we show that 
four perspectives on quantum commuting isomorphisms of Cantor graphs are 
equivalent: a purely algebraic one requiring the non-triviality of the 
game algebra, an operator-theoretic one requiring an intertwining relation 
with a Cantor permutation matrix, a quantum information one
asking for the existence of a perfect quantum commuting  
strategy of the underlying Cantor graph isomorphism game, and 
an operator-algebraic one postulating the existence of a single 
factorisable map that allows for a mutual transformation of the rule functions of the Cantor games into one another. 
Following a similar development in the case of the isomorphism game over finite graphs \cite{bigalois}, we show in 
Theorem \ref{thm:COH-bi-Galois-quotients-kernels-commute}
that the bisynchronous game algebra of the Cantor graph 
isomorphism game yields a bigalois extension between 
the automorphism quantum groups of the individual Cantor graphs. In Section~\ref{sec:loc_fin} we relate the isomorphism game for Cantor graphs to the notion of
quantum isomorphism of locally finite graphs introduced by Rollier and Vaes.
To this end, we associate to every Cantor graph $\frak G$ a locally finite level
graph $\Pi_{\frak G}$. In Theorem~\ref{level_iso} we show that two Cantor graphs $\frak G$ and
$\frak H$ admit a perfect graded quantum commuting (qc) strategy for their
isomorphism game if and only if $\Pi_{\frak G}$ and $\Pi_{\frak H}$ are quantum
isomorphic via a level-preserving quantum isomorphism. The classical
counterpart is Proposition~\ref{p_gradedlocallyloc}: a perfect graded local strategy exists if and
only if $\Pi_{\frak G}$ and $\Pi_{\frak H}$ are isomorphic as level graphs, which in
turn is equivalent to the existence of a graded homeomorphism between $\frak G$
and $\frak H$. Finally, using pairs of finite graphs that are quantum isomorphic
but not isomorphic, we construct Cantor graphs that are not graded homeomorphic
but nevertheless admit a perfect graded qc strategy for their isomorphism game.
This shows that the quantum and classical notions of graded isomorphism for
Cantor graphs are genuinely different.

\smallskip

We finish this section by introducing notation. 
We denote by $\cl U\otimes\cl V$ the algebraic 
tensor product of complex vector spaces $\cl U$ and $\cl V$, unless $\cl U$ and $\cl V$ 
are Hilbert spaces, in which case the notation is used for their Hilbertian tensor products, or C*-algebras, in which case the 
symbol will be reserved for their minimal tensor product. 
If $\cl M$ and $\cl N$ are von Neumann algebras, we denote by $\cl M\bar\otimes\cl N$ 
their weak* spatial tensor product. Following established notation, 
we write $\cl R^{\omega}$ for the ultrapower of the hyperfinite II$_1$ factor $\cl R$ (here, $\omega$ refers to an ultrafilter over $\bb{N}$).
Given a finite set $X$, we let $\cl D_X$ be the 
algebra of all diagonal matrices over $X\times X$ and 
$\{\delta_x\}_{x\in X}$ be the canonical basis of 
(diagonal) matrix units; 
we write ${\rm tr}_{X}$ for the normalised trace of $\cl D_X$, 
$I_X$ for the identity matrix in $\cl D_X$, 
and often abbreviate $\cl D_{XX} = \cl D_X\otimes\cl D_X$. 
For a Hilbert space $H$, we let $\cl B(H)$ be the algebra of 
all bounded linear operators on $H$, and $\cl S_1(H)$ be 
its trace class ideal; note that $\cl S_1(H)^* = \cl B(H)$ via 
the trace duality $\langle \rho, T\rangle = {\rm Tr}(\rho T)$, 
$\rho\in \cl S_1(H)$, $T\in \cl B(H)$. 
If $\xi,\eta$ are vectors in $H$, we write 
$|\xi\rangle\langle \eta|$ for the rank one operator, given by 
$|\xi\rangle\langle \eta| (\zeta) = \langle \zeta,\eta\rangle\xi$.
For a compact topological space $\Omega$, we write as usual $C(\Omega)$ for the 
C*-algebra of all continuous functions on $\Omega$; we denote by $M(\Omega)$
the space of all complex regular Borel measures on $\Omega$ (and recall the canonical Banach space 
duality $C(\Omega)^* = M(\Omega)$).


\medskip

\noindent 
{\bf Use of Artificial Intelligence.}  
The first complete draft of the paper was produced in October 2026 by the human authors without the use of AI tools. The draft was 
fed into Claude Opus 5.5, which returned a list of comments and suggestions
that fell into one of the following categories: 
(a) suggestions for provision of more mathematical details in the proofs; 
(b) corrections of minor mathematical omissions, and 
(c) corrections of inconsistent notation. 
All suggestions were carefully examined by the authors, and the vast majority were implemented. The authors take the responsibilities for the content 
of the paper in its entirety.

\smallskip

\noindent 
{\bf Acknowledgements. } 
The first named author was supported by the National Research Foundation of Korea (NRF) grant funded by the Korea government (MSIT) (No.RS-2025-00561391, No.RS-2024-00413957 and RS-2022-NR069971). The second named author was supported by the National Science Center Grant OPUS-29 UMO-2025/57/B/ST1/00057. The third named author was supported by NSF grants 
CCF-2115071, DMS-2154459 and DMS-2554987. The fourth named author was supported by the Swedish Research Council project grant 2023-04555 and GS Magnusons Fond MF2022-0006. 
The authors are grateful to BIRS for funding and hospitality during a 
\lq\lq Research in Teams'' stay in September 2025, 
during which part of this work was conducted. This material is based upon work supported by the Swedish Research Council under grant no. 2021-06594 while the  authors were in residence at Institut Mittag-Leffler in Djursholm, Sweden during the Spring semester of 2026. 

\smallskip

\noindent 
{\bf Data Availability Statement.} No datasets were generated or analysed during the current study.


\section{Graded and Cantor correlations}\label{s_gradedcor}

This section contains basic definitions, introduces 
graded Cantor correlations, and establishes some preliminary facts 
that will be used later.
Let $S$ and $U$ be finite sets. 
We recall that a \emph{no-signalling correlation}
over $(S,U)$ is a family 
\begin{equation}\label{eq_deffinc}
p = \{(p(u,v|s,t))_{u,v\in U} : s,t\in S\}
\end{equation}
of conditional probability distributions, one for each pair $(s,t)\in S\times S$, satisfying the conditions
$$\sum_{w\in U} p(u,w|s,t) = \sum_{w\in U} p(u,w|s,t')
\ \mbox{ and } \ 
\sum_{w\in U} p(w,v|s,t) = \sum_{w\in U} p(w,v|s',t)$$
for all $s,s',t,t'\in S$ and all $u,v\in U$. 
Such correlations will often be referred to as \emph{finite} due to the finiteness of the sets $U$ and $S$. 

We recall that a \emph{positive operator-valued measure (POVM)} is a (finite) family $(E_i)_{i=1}^d$ of positive operators acting on a 
Hilbert space, such that $\sum_{i=1}^d E_i = I$.
A no-signalling correlation $p$ over $(S,U)$ is called 
\emph{quantum commuting} if it has the form
\begin{equation}\label{eq_defqcty}
p\left(u,v|s,t\right) = \langle P_{s,u} Q_{t,v}\xi ,\xi\rangle,
\end{equation}
where
$\xi$ is a unit vector in a Hilbert space $H$, and 
$\left(P_{s,u}\right) _{s\in S}$ and $\left(Q_{t,v}\right) _{t\in S}$, $u,v\in U$, are
POVM's on $H$ such that $P_{s,u} Q_{t,v} = Q_{t,v} P_{s,u}$ for all $s,t\in S$, $u,v\in U$.
The correlation $p$ is called \emph{quantum spatial} 
if the Hilbert space in the representation (\ref{eq_defqcty}) can be chosen 
of the form $H = H_A\otimes H_B$ for some Hilbert spaces $H_A$ and $H_B$, 
and $P_{s,u}$ (resp. $Q_{t,v}$) 
has the form $P_{s,u} = P_{s,u}'\otimes I_{H_B}$ 
(resp. $Q_{t,v} = I_{H_A}\otimes Q_{t,v}'$); 
if $H_A$ and $H_B$ are finite dimensional, we refer to $p$ being 
\emph{quantum}. 
We further say that $p$ is an \emph{approximately quantum} correlation if 
$p$ is the limit (in the vector space $\bb{R}^S\times \bb{R}^S\times \bb{R}^U\times\bb{R}^U$) of quantum spatial correlations. 
We denote by 
$\cl C_{\rm ns}(S,U)$ 
(resp. $\cl C_{\rm qc}(S,U)$, $\cl C_{\rm qa}(S,U)$, 
$\cl C_{\rm qs}(S,U)$, $\cl C_{\rm q}(S,U)$) the set 
of all no-signalling (resp. quantum commuting, approximately quantum, 
quantum spatial, quantum) correlations over the quadruple $(S,U)$. 
Finally, a correlation $p$ over $(S,U)$ is called \emph{local}
if it is a convex combination of correlations of the form 
$p^{(1)}(u|s)p^{(2)}(v|t)$, where $p^{(i)}(\cdot|s)$ is a probability 
distribution over $U$ for each $s\in S$, $i = 1,2$. We denote 
by $\cl C_{\rm loc}(S,U)$ the set of all local correlations 
over $(S,U)$;  
we use $\cl C_t$ instead of $\cl C_t(S,U)$ when $S$ and $U$ are 
clear from the context.


For $d\in \bb{N}$, we write $[d] = \{1,2,\dots,d\}$ and 
call a family $X = (X_n)_{n\in \bb{N}}$ of finite sets 
\emph{inductive} if $X_{n+1} = X_n\times[d_n]$ for some 
$d_n\in \bb{N}$, $n\in \bb{N}$. 
Let $X = (X_n)_{n\in \bb{N}}$ and $A = (A_n)_{n\in \bb{N}}$
be inductive families of finite sets, where
$X_{n+1} = X_n\times[d_n]$ and 
$A_{n+1} = A_n\times[c_n]$ for some $c_n, d_n\in \bb{N}$, 
(for the sake of avoiding trivialities) 
all assumed bigger than $1$. We set $d_0 = |X_1|$ and 
$c_0 = |A_1|$; thus, 
the inductive sequence $(\cl D_{X_n},\iota^X_{n,n+1})_n$
(resp.~$(\cl D_{A_n},\iota^A_{n,n+1})_n$), where 
$\iota^X_{n,n+1}(T) = T\otimes 1_{d_n}$ (resp.~$\iota^A_{n,n+1}(S) = S\otimes 1_{c_n}$), 
has inductive limit 
$C(\Omega_X)$ (resp.~$C(\Omega_A)$) in the C*-algebra category. 
In the sequel we will often suppress the superscripts $A$ and $X$, when the context is clear; thus, $\iota_{n,\infty}$ will denote both the canonical inclusion 
of $\cl D_{X_n}$ in $C(\Omega_X)$ and that of $\cl D_{A_n}$ in $C(\Omega_A)$.
We set $\iota^{(2)}_{n,\infty} = \iota_{n,\infty}\otimes \iota_{n,\infty}$
and $\iota^{(2)}_{n,n+1} = \iota_{n,n+1}\otimes \iota_{n,n+1}$, $n\in \bb{N}$.
Further, we equip $\Omega_X$ (resp.~$\Omega_A$) with the product 
measure $\mu_X = \otimes_{n\geq 0} \mu_{d_n}$
(resp.~$\mu_A = \otimes_{n\geq 0} \mu_{c_n}$), 
where $\mu_d$ denotes the uniform measure on the set $[d]$, $d\in \bb{N}$, and write $L^{\infty}(\Omega_X) = L^{\infty}(\Omega_X,\mu_X)$ (resp.~$L^{\infty}(\Omega_A) = L^{\infty}(\Omega_A,\mu_A)$). 
Identifying $\cl D_{X_n}$ 
(resp.~$\cl D_{A_n}$) with its image 
inside $L^{\infty}(\Omega_X)$ (resp.~$L^{\infty}(\Omega_A)$) under the map $\iota_{n,\infty}$, 
we let $\cl E_n : L^{\infty}(\Omega_X) \to \cl D_{X_n}$
be the canonical conditional expectation, and set 
$\cl E_n^{(2)} = \cl E_n\otimes \cl E_n$; 
without a risk of confusion, 
we will use the same notation for the 
expectations from $L^{\infty}(\Omega_A)$ onto $\cl D_{A_n}$.
We further write $\cl E_{n+1,n} : \cl D_{X_{n+1}}\to \cl D_{X_n}$ for the canonical conditional expectation, 
and set $\cl E_{n+1,n}^{(2)} = \cl E_{n+1,n}\otimes \cl E_{n+1,n}$, $n\in \bb{N}$. 
\par A \emph{word} is a string 
$x = x_k x_{k+1}\dots x_{m}$, where $x_i\in [d_i]$, 
$i = k,\dots,m$; 
we write $|x| = m-k+1$ for its length.
We call the word $x$ \emph{initial} if $k = 0$.
We will use notation $xx'$ for the concatenation of 
words $x$ and $x'$ (here, if $x\in \prod_{i=l}^k [d_i]$ for some $l\leq k$  
then $x'\in \prod_{i=k+1}^m [d_i]$ for some $m > k$, or $x'\in \prod_{i=k+1}^{\infty} [d_i]$). 
In the next definition, which 
extends the definition of finite correlation types 
to the Cantor setup, we introduce the main concepts examined in this paper.

\begin{definition} \rm
Let $X = (X_n)_{n\in \bb{N}}$ and $A = (A_n)_{n\in \bb{N}}$
be inductive families of finite sets. A family of scalars
\[
p=\{p(a,b|x,y) : n\in \mathbb N,\ |x|=|y|=|a|=|b|=n\},
\]
where $x,y,a,b$ are initial words, 
is called a:
\begin{enumerate}
\item[(i)] \emph{no-signalling correlation} over $(X,A)$, if for each $n\in \mathbb N$, the family
\[
p_n := \{p(a,b|x,y) : |x|=|y|=|a|=|b|=n\}
\]
(called the \emph{$n$-th component} of $p$) is a no-signalling correlation over  $(X_n,A_n)$;

\item[(ii)] \emph{graded no-signalling correlation} over $(X,A)$ if it is a no-signalling correlation and, for all initial words $x,y,a,b$ of equal length and all
words $x',y'$ of length $1$, we have
\begin{equation}\label{eq:graded_ns}
\sum_{|a'|=|b'|=1} p(aa',bb' \mid xx',yy') = p(a,b \mid x,y).
\end{equation}
\end{enumerate}
\end{definition}
For brevity we will say that $ p$ is an \emph{ns-graded correlation} over $(X,A)$. Condition \eqref{eq:graded_ns}  is equivalent by induction to requiring that
\[
\sum_{|a'|=|b'|=m} p(aa',bb' \mid xx',yy') = p(a,b \mid x,y)
\]
for all $m\in \mathbb N$ and all words $x',y'$ of length $m$.


In the case of finite correlations, there is an equivalent formulation of the no-signalling condition in the language of unital completely positive maps. This is possible  since finite ns-correlations $p$ over the pair $(S,U)$ correspond precisely to the 
unital (completely) positive linear maps 
$\Theta : \cl D_{UU} \to \cl D_{SS}$
such that 
\begin{equation}\label{eq_Thetan}
\Theta(\cl D_U\otimes 1) \subseteq \cl D_S\otimes 1
\ \mbox{ and } \ \Theta(1\otimes\cl D_U) \subseteq  1\otimes \cl D_S
\end{equation}
via the equations
$$\langle \Theta(\delta_u\otimes \delta_v),
\delta_s\otimes \delta_t\rangle = |S|^2 p(u,v|s,t), 
\ \ \ s,t\in S, u,v\in U;$$
we thus call a map $\Theta : \cl D_{UU} \to \cl D_{SS}$
satisfying (\ref{eq_Thetan}) \emph{no-signalling}
(the pairing is with respect to normalised trace).   
As was shown in \cite{gb,bctt}, such a correspondence holds in the Cantor setting, that is, in the case where the finite sets $U$ and $S$ are replaced by Cantor spaces $\Omega_X$ and $\Omega_A$ arising from two inductive families of  sets. In fact, graded no-signalling correlations over $(X,A)$ naturally fit in the broader context of Cantor no-signalling correlations and subclasses thereof, developed in the prequel \cite{bctt} of the current paper. At this point, we would like to recall a few definitions and results which will be used in the forthcoming sections. 
\par Recall that a \emph{Cantor correlation} over $(X,A)$
\cite{bctt}
is a unital completely positive map 
$\Gamma : C(\Omega_A\times \Omega_A) \to L^{\infty}(\Omega_X\times \Omega_X)$ such that 
$$\Gamma(C(\Omega_A)\otimes 1_A) \subseteq 
L^{\infty}(\Omega_X)\otimes 1_X \ \mbox{ and } \ 
\Gamma(1_A \otimes C(\Omega_A)) \subseteq 
1_X \otimes L^{\infty}(\Omega_X).$$
We note the duality $L^\infty(\Omega_X)\bar\otimes \cl{B}(H)\simeq (L^1(\Omega_X)\hat\otimes \cl{T}(H))^*$, where $\cl{T}(H)$ is the space of all trace class operators on $H$ and $\hat \otimes$ stands for the projective tensor product of Banach spaces \cite{ryan}. 
Consequently, $L^\infty(\Omega_X)\bar\otimes \cl{B}(H)$ is a dual Banach space and thus the space $\operatorname{UCP}(C(\Omega_A),L^\infty(\Omega_X)\bar\otimes \cl{B}(H))$ of all unital completely positive maps from $C(\Omega_A)$ into $L^\infty(\Omega_X)\bar\otimes \cl{B}(H)$ can  be equipped with Arveson's BW topology introduced in \cite{arveson-acta} (see also \cite{gb}), 
coinciding on bounded sets with the point weak*-topology. Given unital completely positive maps $\phi,\psi:C(\Omega_A)\rightarrow L^\infty(\Omega_X)\bar\otimes\cl{B}(H)$, we say that $\phi$ and $\psi$ 
\emph{have commuting ranges} if their values commute on the $\cl{B}(H)$ leg. 
If the latter requirement is satisfied, the dot product 
$\phi\cdot\psi$ is the canonically defined unital completely positive map from $C(\Omega_A\times \Omega_A)$ into $L^\infty(\Omega_X\times \Omega_X)\bar\otimes\cl{B}(H)$; thus, 
after viewing $\phi$ (resp.~$\psi$) as a map with target $L^\infty(\Omega_X\times \Omega_X)\bar\otimes\cl{B}(H)$, depending only on the first (resp.~second) variable, 
$(\phi\cdot\psi)(S\otimes T) = \phi(S)\Psi(T)$, where we have identified
$C(\Omega_A\times \Omega_A)$ with the tensor product $C(\Omega_A)\otimes C(\Omega_A)$.
In the case where $\phi\in \operatorname{UCP}(C(\Omega_A),L^\infty(\Omega_X)\bar\otimes \cl{B}(H))$ and $\psi\in \operatorname{UCP}(C(\Omega_A),L^\infty(\Omega_X)\bar\otimes \cl{B}(K))$ for 
(possibly different) Hilbert spaces $H$ and $K$, the 
tensor product $\phi\otimes\psi : C(\Omega_A\times \Omega_A) \to L^\infty(\Omega_X\times \Omega_X)\bar\otimes\cl{B}(H\otimes K)$ is defined similarly.

We recall the definitions of some subclasses of Cantor no-signalling correlations that will be used widely in the sequel. A unital completely positive map $\Gamma:C(\Omega_A)\otimes C(\Omega_A)\rightarrow L^\infty(\Omega_X)\bar\otimes L^\infty(\Omega_X)$ is called 

\begin{itemize}
    \item[(i)] \emph{quantum spatial}, if there exist Hilbert spaces $H$, $K$, unital completely positive maps $\phi:C(\Omega_A)\rightarrow L^\infty(\Omega_X)\bar\otimes \cl{B}(H)$ and $\psi:C(\Omega_A)\rightarrow L^\infty(\Omega_X)\bar\otimes \cl{B}(K)$ and a unit vector $\xi\in H\otimes K$ such that $$\langle \Gamma(g),h\rangle=\langle (\phi\otimes \psi)(g),h\otimes \xi\xi^*\rangle, \ \ g\in C(\Omega_A\times \Omega_A),\  h\in L^1(\Omega_X\times \Omega_X).$$
    \item[(ii)] \emph{quantum approximate}, if it is a BW limit point of quantum spatial correlations. 
    \item[(iii)] \emph{quantum commuting}, if there exists a Hilbert space $H$, unital completely positive maps 
    with commuting ranges 
 $\phi:C(\Omega_A)\rightarrow L^\infty(\Omega_X)\bar\otimes \cl{B}(H)$ and $\psi:C(\Omega_A)\rightarrow L^\infty(\Omega_X)\bar\otimes \cl{B}(H)$ and a unit vector $\xi\in H$ such that $$\langle \Gamma(g),h\rangle=\langle (\phi\cdot\psi)(g),h\otimes\xi\xi^*\rangle, \ \ g\in C(\Omega_A\times \Omega_A),\  h\in L^1(\Omega_X\times \Omega_X).$$ 
\end{itemize}

We denote by $\cl{C}_{\rm ns}(X,A)$ (resp. $\cl{C}_{\rm qs}(X,A)$, $\cl{C}_{\rm qa}(X,A)$, $\cl{C}_{\rm qc}(X,A)$) the set of all Cantor no-signalling (resp. quantum spatial, quantum approximate, quantum commuting) correlations over $(X,A)$; we omit the mention of the pair $(X,A)$ when the latter is clear from the context.

\begin{remark}\rm 
Cantor correlations are a special case of measurable correlations defined in \cite{btt} in the 
situation where the Cantor spaces are replaced with general compact metrisable spaces. 
The definitions of the subclasses of measurable no-signalling correlations given in \cite{btt} use 
the language of operator-valued information channels \cite{btt,gb}. In view of \cite[Theorem 3.11]{gb}, which provides a bijective correspondence between operator-valued information channels and 
respective unital completely positive maps, we have here followed a shorter equivalent route.
\end{remark}

There is a natural correspondence between Cantor correlations and inductive sequences of finite correlations, pointed out in \cite{bctt}. To each Cantor correlation 
$\Gamma: C(\Omega_A\times \Omega_A) \to L^{\infty}(\Omega_X\times \Omega_X)$, one associates a family of 
(unital completely positive) maps
$\Gamma_n : \cl D_{A_n\times A_n} \to \cl D_{X_n\times X_n}$ by letting 
\begin{align}\label{eq_ind_fam}
    \Gamma_n = \cl E_n^{(2)} \circ \Gamma \circ \iota^{(2)}_{n,\infty},\ \ \ n\in \bb{N}.
\end{align}
To motivate our further development, we recall the following combinations of results from \cite{bctt}:

\begin{theorem}[\cite{bctt}]\label{p_fornse}
Let $t\in\rm\{qa,qc,ns\}$ and $\Gamma : C(\Omega_A\times \Omega_A) \to L^{\infty}(\Omega_X\times \Omega_X)$ be unital and completely positive.  Then $\Gamma\in\cl{C}_{\rm t}(X,A)$ if and only if 
$\Gamma_n\in\cl{C}_{\rm t}(X_n,A_n)$, for every $n\in \bb{N}$. 
\end{theorem}

We say that the Cantor correlation 
$\Gamma : C(\Omega_A\times \Omega_A) \to L^{\infty}(\Omega_X\times \Omega_X)$ is \emph{graded} 
if $\Gamma(\iota^{(2)}_{n,\infty}(\cl D_{A_nA_n}))\subseteq \iota^{(2)}_{n,\infty}(\cl D_{X_nX_n})$ for every $n\in \bb{N}$. 
We note that $\Gamma$ is graded 
if and only if 
\begin{equation}\label{eq_iotasg}
\iota_{n,n+1}^{(2)}\circ \Gamma_n = \Gamma_{n+1}\circ \iota_{n,n+1}^{(2)}, \ \ \ n\in \bb{N};
\end{equation}
indeed, $\Gamma$ being graded is equivalent to the inclusions 
${\rm ran}(\Gamma_{n+1}\circ \iota^{(2)}_{n,n+1})\subseteq {\rm ran}(\iota^{(2)}_{n,n+1})$, $n\in \bb{N}$, 
that is, to the equalities
$$\Gamma_{n+1}\circ \iota_{n,n+1}^{(2)} = \iota_{n,n+1}^{(2)}\circ \cl E_{n+1,n}^{(2)}\circ \Gamma_{n+1}\circ \iota_{n,n+1}^{(2)}, \ \ \ n\in \mathbb{N},$$
which together with the fact that 
$\cl E_{n+1,n}^{(2)}\circ \Gamma_{n+1} \circ \iota_{n,n+1}^{(2)} = \Gamma_n$
implies the equivalence with (\ref{eq_iotasg}).

\begin{remark}\label{r_ch_mem}
\rm 
We note that, given a graded Cantor correlation $\Gamma :C(\Omega_A\times\Omega_A)\rightarrow L^\infty(\Omega_X\times\Omega_X)$, the range is contained in 
$C(\Omega_X\times \Omega_X)$. Indeed, by assigning the family $(\Gamma_n)_{n\in\bb{N}}$ defined in (\ref{eq_ind_fam}) to $\Gamma$, observe that $(\Gamma_n)_{n\in\bb{N}}$ satisfies the inductive relations (\ref{eq_iotasg}) and by the universal property of the inductive limit one obtains that 
$\Gamma(C(\Omega_A\times\Omega_A))\subseteq C(\Omega_X\times\Omega_X)$. Uniqueness follows from the fact $\{\iota_{n,\infty}(\delta_a)\otimes \iota_{n,\infty}(\delta_b) : a,b\in A_n, n\in \bb{N}\}$
(resp. 
$\{\iota_{n,\infty}(\delta_x)\otimes \iota_{n,\infty}(\delta_y) : x,y\in X_n, n\in \bb{N}\}$) has dense span in 
$C(\Omega_A\times\Omega_A)$ (resp. $C(\Omega_X\times\Omega_X)$).
In the language of information channels \cite{kakihara} this implies that graded Cantor correlations correspond to \emph{continuous channels}. 
\end{remark}

In the following, we relate the gradedness of no-signalling correlations to that of Cantor correlations. 

\begin{proposition}\label{p_samescal}
A no-signalling correlation $p = (p(a,b|x,y))_{x,y,a,b}$ over $(X,A)$ is graded if and only if there exists a unique graded
Cantor correlation 
$\Gamma : C(\Omega_A\times \Omega_A) \to C(\Omega_X\times \Omega_X)$ such that 
\begin{equation}\label{eq_deff0}
p(a,b|x,y) = |X_n|^2 \langle \iota_{n,\infty}(\delta_x)\otimes \iota_{n,\infty}(\delta_y), \Gamma(\iota_{n,\infty}(\delta_a)\otimes \iota_{n,\infty}(\delta_b)\rangle
\end{equation}
for all initial words $x,y,a,b$ of length $n$, $n\in \bb{N}$. 
\end{proposition}

\begin{proof}
Suppose that $\Gamma$ is a graded Cantor correlation, and set 
\begin{equation}\label{eq_deff}
p(a,b|x,y) = \lvert X_n\rvert^2\left\langle \delta_x\otimes \delta_y,\Gamma_n(\delta_a\otimes \delta_b)\right\rangle, \ \ |x| = |y| = |a| = |b| = n.
\end{equation}
Then 
\begin{eqnarray*}
& & 
\sum_{|a'| = |b'| = 1} p(aa',bb'|xx',yy')
= 
\lvert X_{n+1}\rvert^2 \sum_{|a'| = |b'| = 1} 
\left\langle \delta_{xx'}\otimes \delta_{yy'},\Gamma_{n+1}(\delta_{aa'}\otimes \delta_{bb'})\right\rangle\\
& = & 
\lvert X_{n}\rvert^2 d_n^2 \left\langle \delta_{xx'}\otimes \delta_{yy'},\Gamma_{n+1}(\iota_{n,n+1}(\delta_a)\otimes \iota_{n,n+1}(\delta_b))\right\rangle\\
& = & 
\lvert X_{n}\rvert^2 d_n^2 \left\langle \delta_{xx'}\otimes \delta_{yy'},\iota_{n,n+1}^{(2)}(\Gamma_{n}(\delta_a\otimes \delta_b))\right\rangle\\
& = & 
\lvert X_{n}\rvert^2 d_n^2 
\left\langle \delta_{xx'}\otimes \delta_{yy'},
\sum_{|x''| = |y''| = 1} 
(\Gamma_{n}(\delta_a\otimes \delta_b))\otimes \delta_{x''}\otimes \delta_{y''}\right\rangle\\
& = & 
\lvert X_{n}\rvert^2 d_n^2 
\left\langle \delta_{x}\otimes \delta_{y},
\Gamma_{n}(\delta_a\otimes \delta_b)\right\rangle
\sum_{|x''| = |y''| = 1} ({\rm tr}\hspace{-0.02cm}\mbox{}_{d_n}\otimes {\rm tr}\hspace{-0.02cm}\mbox{}_{d_n})((\delta_{x'}\otimes\delta_{y'})
(\delta_{x''}\otimes \delta_{y''}))\\
& = & 
\lvert X_{n}\rvert^2  
\left\langle \delta_{x}\otimes \delta_{y},
(\Gamma_{n}(\delta_a\otimes \delta_b))\right\rangle 
= p(a,b|x,y).
\end{eqnarray*}
Thus, $p$ is a graded no-signalling correlation. 
The converse follows from Remark \ref{r_ch_mem} and the preceding calculation.
\end{proof}

Given an ns-graded correlation $p$, we write $\Gamma_p$ for the Cantor correlation associated to it 
via Proposition \ref{p_samescal}. 
Theorem \ref{p_fornse} justifies the following definition.

\begin{definition}\label{def_locC}
\rm 
Let $X$ and $A$ be inductive families of sets, 
$p$ be an ns-graded correlation over $(X,A)$, and 
${\rm t}\in \{{\rm loc}, {\rm q}, {\rm qa}, {\rm qc}\}$. 
We call $p$ a \emph{t-graded correlation} if $p_n$ is of type ${\rm t}$ for every $n\in \bb{N}$.
\end{definition}

Given inductive families of sets $X$ and $A$, we write 
$\cl C_{\rm tg}(X,A)$ (or simply $\cl C_{\rm tg}$ when 
$X$ and $A$ are clear from the context) for the set 
of all t-graded correlations. 
Next, we address local graded correlations and note:

\begin{proposition}\label{p_indforlo}
Let $X$ and $A$ be inductive families of sets and 
$p$ be an ns-graded correlation over $(X,A)$.
The following are equivalent:
\begin{itemize}
    \item[(i)] $p\in \cl C_{\rm locg}$; 
    \item[(ii)] $\Gamma_p$ belongs to the BW closed 
convex hull of Cantor correlations of the form 
$\Phi\otimes\Psi$, where $\Phi, \Psi : C(\Omega_A)\to C(\Omega_X)$ are unital completely positive maps.
\end{itemize}
\end{proposition}

\begin{proof} 
Write $\Gamma = \Gamma_p$.

(i)$\Rightarrow$(ii) 
Assume that $p_n\in \cl{C}_{\rm loc}$ for every $n\in\bb{N}$; thus, there exist conditional probability distributions $\{p_{k,n}(a|x): x\in X_n,a\in A_n\}$, $\{p_{k,n}'(b|y): y\in X_n,b\in A_n\}$
such that $$p_{n}(a,b|x,y) = \sum_{k=1}^{\ell_n}\lambda_{k,n}p_{k,n}(a|x)p'_{k,n}(b|y)$$
as a convex combination. 
Moreover, setting $\tilde{\Gamma}_n = (\iota_{n,\infty}\otimes\iota_{n,\infty})
\circ\Gamma_{p_n}\circ(\cl{E}_{n}\otimes \cl{E}_{n})$, we have 
$$\tilde{\Gamma}_n
=
\sum_{k=1}^{\ell_n}\lambda_{k,n}(\iota_{n,\infty}\circ \Phi_{p_{k,n}}\circ\cl{E}_{n})\otimes (\iota_{n,\infty}\circ \Phi_{p'_{k,n}}\circ\cl{E}_{n}),$$
where $\Phi_{p_{k,n}}: \cl D_{A_n}\to \cl D_{X_n}$
(resp.~$\Phi_{p_{k,n}'} : \cl D_{A_n}\to \cl  D_{X_n}$)  are the  
unital completely positive maps corresponding to $p_{k,n}$ (resp.~$p'_{k,n}$). 
Now (ii) follows from the fact that 
$\tilde{\Gamma}_n\rightarrow \Gamma$ in the BW topology
(see \cite[Remark 5.3]{bctt}).

(ii)$\Rightarrow$(i) 
Since the unit ball is metrisable for the BW topology \cite[Theorem V.5.1]{dun_sch}, 
there exists a sequence $(\Theta_k)_{k\in\bb{N}}\subseteq \conv\{\Phi\otimes \Psi:\Phi,\Psi\in\operatorname{UCP}(C(\Omega_A),C(\Omega_X))\}$ with $\Theta_k\rightarrow_{k\to\infty} \Gamma$ in the BW topology. 
Since $(\Theta_k)_n = \cl{E}_n^{(2)}\circ \Theta_k\circ \iota^{(2)}_{n,\infty}$, we have that 
$(\Theta_k)_n \in \cl C_{\rm loc}$. 
On the other hand, since the map 
$\Theta\mapsto\cl{E}_n^{(2)}\circ \Theta\circ \iota^{(2)}_{n,\infty}$  
is BW continuous, $(\Theta_k)_n\rightarrow_{k\to \infty} \Gamma_n$ for each $n\in\bb{N}$. 
Since  the set of all local correlations over finite sets  
is closed, we conclude that $p_n$ is of local type.   
\end{proof}

\begin{remark}\label{r_closures}
\rm By Proposition \ref{p_indforlo} (resp.~\cite[Corollary 5.9]{bctt}), the set $\cl C_{\rm locg}$ (resp.~$\cl C_{\rm qcg}$) is closed in the BW topology. The example, given in 
the proof of \cite[Theorem 5.16]{bctt}, shows that this is not the case for the set $\cl C_{\rm qg}$. 
\end{remark}


\section{Synchronous correlations}\label{s_syn}

In this section, we define and examine synchronous graded correlations and identify 
a suitable C*-algebra that acts as a universal encoding object thereof via its tracial state space. 
We fix throughout the section inductive families of sets $X = (X_n)_{n\in \bb{N}}$ and 
$A = (A_n)_{n\in \bb{N}}$; we assume that 
$X_{n+1} = X_n\times [d_n]$ and $A_{n+1} = A_n\times [c_n]$, where 
$c_n, d_n > 1$, $n\in \bb{N}$.


\subsection{A universal C*-algebra for synchronicity}\label{ss_univa}

As a means of both motivation and preparation, we recall that if $S$ and $U$ are finite sets, 
a universal C*-algebra $\cl A_{S,U}$ for operator-valued information channels from $S$ to $U$ is 
defined by letting 
$$\cl A_{S,U} := \underbrace{\cl D_U\ast_1\cdots \ast_1 \cl D_U}_{|S|\,\mbox{\tiny \rm times}},$$
where the free products are amalgamated over the units (see e.g. \cite{lmprsstw}).
It is clear that $\cl A_{S,U}$ is the universal C*-algebra, 
generated by families $\{e_{s,u}\}_{u\in U}$ of orthogonal projections satisfying the relations 
$\sum_{u\in U} e_{s,u} = 1$, $s\in S$. 
In this subsection, we construct a similar universal C*-algebra in the case of 
(infinite) graded channels. As we will shortly see, it will arise as a C*-algebraic inductive limit of 
sequences of C*-algebras of the form $\cl A_{S,U}$.  

Let $\cl B_{X_1,A_1} = \cl A_{X_1,A_1}$ and, for a 
given $n > 1$, let $\cl B_{X_n,A_n}$ be the universal C*-algebra, 
generated by a family $\{e_{x,a} : x\in X_n, a\in A_n\}$ of orthogonal projections, subject to the relations

\vspace{0.1cm}

\begin{itemize}
\item[(PVM)] $\sum_{a\in A_n} e_{x,a} = 1$ for all $x\in X_n$, and 

\vspace{0.05cm}

\item[(COH)] $\sum_{|a'| = n-k} e_{xx',aa'} = \sum_{|a'| = n-k} e_{xx'',aa'}$ for all $x',x''$ with $|x'| = |x''| = n- k$ and all $k \in [n-1]$. 
\end{itemize}
In what follows we often refer to relations (COH) as \emph{(row) coherence relations}.  

It is clear that $\cl B_{X_n,A_n}$ is a canonical 
quotient of $\cl A_{X_n,A_n}$; let $q_n : \cl A_{X_n,A_n}\to \cl B_{X_n,A_n}$ be the quotient map. 
We have that the subset 
$\{\hat{e}_{x,a} : |x| = |a| = n\}$ of $\cl B_{X_{n+1},A_{n+1}}$, where 
$\hat{e}_{x,a} = \sum_{|a'| = 1} e_{xx',aa'}$, 
satisfies the properties (PVM) and (COH) for any $x'$ of length 1; by universality, 
there exists a unital *-homomorphism 
$\pi_n : \cl B_{X_n,A_n}\to \cl B_{X_{n+1},A_{n+1}}$, 
such  that $\pi_n(e_{x,a}) = \hat{e}_{x,a}$, 
$x\in X_n$, $a\in A_n$. 
We thus obtain an inductive sequence 
$\left(\cl B_{X_n,A_n},\pi_n\right)_{n\in \bb{N}}$:

\[\begin{tikzcd}
	{\mathcal{A}_{X_1,A_1}} & {\mathcal{A}_{X_2,A_2}} & \cdots & {\mathcal{A}_{X_n,A_n}} & \cdots \\
	{\mathcal{B}_{X_1,A_1}} & {\mathcal{B}_{X_2,A_2}} & \cdots & {\mathcal{B}_{X_n,A_n}} & \cdots
	\arrow["{\operatorname{id}}", from=1-1, to=2-1]
	\arrow["{q_2}", from=1-2, to=2-2]
	\arrow["{q_n}", from=1-4, to=2-4]
	\arrow["{\pi_1}", from=2-1, to=2-2]
	\arrow["{\pi_2}", from=2-2, to=2-3]
	\arrow["{\pi_{n-1}}", from=2-3, to=2-4]
	\arrow["{\pi_n}", from=2-4, to=2-5]
\end{tikzcd}\]
and let 
\begin{equation}\label{eq_defofBXA}
\cl B_{X,A} := \varinjlim (\cl B_{X_n,A_n},\pi_n)_{n\in \bb{N}}
\end{equation}
be the corresponding inductive limit in the C*-algebra category. 
The next proposition provides a concrete, free product,
description of the terms of the inductive sequence. 
If $\cl A_1,\dots,\cl A_n$ are unital C*-algebras and 
$\cl D\subseteq \cl A_i$ as a unital C*-subalgebra, $i\in [n]$, we 
denote by 
$\mathcal A_1\ast_{\;\cl D}\cdots\ast_{\;\mathcal D}\mathcal A_n$
the free product of the family $\{\cl A_i\}_{i=1}^n$ in the category of C*-algebras, amalgamated over $\cl D$. 

\begin{proposition}\label{p:freeproduct_model}
Let $n\in\mathbb N$.  
There exists a sequence $\left(\mathcal C^{(n)}_k\right)_{k=0}^n$ of C*-algebras, defined 
recursively as follows:
\begin{equation}\label{eq_dnfreepr}
\mathcal C^{(n)}_n
:=
\underbrace{
\mathcal D_{A_{n+1}}
\;\ast_{\;\mathcal D_{A_n}\otimes 1_{c_n}}\;
\cdots
\;\ast_{\;\mathcal D_{A_n}\otimes 1_{c_n}}\;
\mathcal D_{A_{n+1}}
}_{\text{$d_n$ times}},
\end{equation}
and, for $k=n-1,\dots,0$,
\[
\mathcal C^{(n)}_k
:=
\underbrace{
\mathcal C^{(n)}_{k+1}
\;\ast_{\;\mathcal D_{A_k}\otimes 1_{c_k}\otimes\cdots\otimes 1_{c_n}}\;
\cdots
\;\ast_{\;\mathcal D_{A_k}\otimes 1_{c_k}\otimes\cdots\otimes 1_{c_n}}\;
\mathcal C^{(n)}_{k+1}
}_{\text{$d_k$ times}},
\]
where $\mathcal D_{A_0}=\mathbb C$.
Furthermore, there exists a canonical *-isomorphism
$$\mathcal B_{X_{n+1},A_{n+1}} \;\cong\; \mathcal C^{(n)}_0.$$
\end{proposition}

\begin{proof}
    We note that 
$$\cl D_{A_{k+1}}\otimes 1_{c_{k+1}}\otimes\cdots \otimes 1_{c_n}\subseteq \cl C^{(n)}_{k+1}$$
canonically, and since
$$\cl D_{A_k}\otimes 1_{c_k}\otimes 1_{c_{k+1}}\otimes\cdots \otimes 1_{c_n}\subseteq 
\cl D_{A_{k+1}}\otimes 1_{c_{k+1}}\otimes\cdots \otimes 1_{c_n}, \ \ \ k\in [n-1],$$
the amalgamated free products are well-defined.

For a word $a = a_0a_1\dots a_{n}$, where 
$a_i\in [c_i]$, $i = 0,1,\dots,n$, we consider
$\otimes_{i=0}^n\delta_{a_i}$ as an element of $\cl D_{A_{n+1}}$; without a risk for confusion, we also write 
$\delta_{a'}$, where $a'$ is any word. 
Fix $x = x_0x_1\dots x_{n}$, where 
$x_i\in [d_i]$, $i = 0,1,\dots,n$. 
We let 
$\iota_{x_n} : \cl D_{A_{n+1}}\to \cl C^{(n)}_n$
be the unital *-homomorphism, placing 
an element $T\in \cl D_{A_{n+1}}$ at the $x_n$-th 
position in the definition (\ref{eq_dnfreepr}) of $ \cl C_{n}^{(n)}$. 
Further, we let 
$\iota_{x_{k}} : \cl C^{(n)}_{k+1} \to \cl C^{(n)}_{k}$
be the map, placing an element $T\in \cl C^{(n)}_{k+1}$
at the $x_k$-th position in the definition of  $ \cl C_{k}^{(n)}$. 
Define 
$\theta_n : \cl A_{X_{n+1},A_{n+1}} \to \cl C_0^{(n)}$ by letting 
$$\theta_n(e_{x_0x_1\dots x_n,a_0a_1\dots a_n}) = 
(\iota_{x_0}\circ \iota_{x_1}\circ\cdots\circ
\iota_{x_n})\left(\otimes_{i=0}^n\delta_{a_i}\right).$$
It will be convenient to abbreviate 
$\iota_{y} = \iota_{y_i}\circ \iota_{y_{i+1}}\circ\cdots\circ
\iota_{y_j}$, where $y = y_i y_{i+1}\dots y_j$, $i\leq j$, and $1_{c_k\cdots c_n}=1_{c_k}\otimes\cdots\otimes 1_{c_n}$. 
For $k\in [n]$, $|x| = |a| = k$ and 
$|y| = |y'| = n-k+1$, we have 
\begin{eqnarray*}
& & 
\theta_n\left(\sum_{|b| = n-k+1} 
e_{xy,ab} - \sum_{|b| = n-k+1} e_{xy',ab}\right)\\ 
& & 
\hspace{2cm} = \sum_{|b| = n-k+1} 
\left((\iota_x\circ \iota_y)(\delta_a\otimes \delta_b) - 
(\iota_x\circ \iota_{y'})(\delta_{a}\otimes \delta_b)\right)\\
& & 
\hspace{2cm} = \iota_x\left(\iota_y (\delta_a\otimes 1_{c_k\cdots c_n})\right) - \iota_x\left(\iota_{y'}(\delta_a\otimes 1_{c_k\cdots c_n})\right)\\
& & 
\hspace{2cm} = \iota_x\left(\iota_y(\delta_a\otimes 1_{c_k\cdots c_n}) - \iota_{y'} (\delta_a\otimes 1_{c_k\cdots c_n})\right) = 0,
\end{eqnarray*}
as 
$$\iota_y(\delta_a\otimes 1_{c_k\cdots c_n}) = 
\iota_{y'} (\delta_a\otimes 1_{c_k\cdots c_n}).$$
It follows that 
$\theta_n$ descends to a unital *-homomorphism 
(denoted in the same way) 
$\theta_n : \cl B_{X_{n+1},A_{n+1}}\to \cl C_0^{(n)}$. 

To construct the inverse map $\rho_n : \cl C_0^{(n)}\to \cl B_{X_{n+1},A_{n+1}}$, we use the universal property of the amalgamated free product. We proceed recursively.
For each word $x \in X_{n+1}$,  define a unital *-homomorphism 
$\rho^{(x)}: \cl D_{A_{n+1}}^{} \to \cl B_{X_{n+1},A_{n+1}}$  by setting $\rho^{(x)}(\delta_a) = e_{x,a}$ for all $a \in A_{n+1}$. We now construct a family of maps $\rho^{(w)}_k : \cl C_k^{(n)} \to \cl B_{X_{n+1},A_{n+1}}$ for each $k \in \{n, n-1, \dots, 0\}$ and each  $w \in X_k$.

At step $k=n$, for  $w \in X_n$, the maps $\{\rho^{(wy)}\}_{y \in [d_n]}$ must agree on the subalgebra $\cl D_{A_n} \otimes 1_{c_n}$. We have that 
\( 
\rho^{(wy)} (\sum_{|v|=1} \delta_{uv}) = \sum_{|v|=1} e_{wy, uv}.
\)
By the coherence relations in $\cl B_{X_{n+1},A_{n+1}}$, this sum is independent of $y$. Thus, these maps agree on elements of the form $ \delta_u \otimes 1_{c_n}=\sum_{|v|=1} \delta_{uv}$ and induce, by taking free products, a well-defined *-homomorphism $\rho^{(w)}_n : \cl C_n^{(n)} \to \cl B_{X_{n+1},A_{n+1}}$.

Set $X_0 = \emptyset$. 
Assuming inductively that the  maps $\rho^{(w')}_{k+1} : \cl C_{k+1}^{(n)} \to \cl B_{X_{n+1},A_{n+1}}$ are well-defined for all $w' \in X_{k+1}$, we fix  $w \in X_k$.  The maps $\{\rho^{(wy)}_{k+1}\}_{y \in [d_k]}$ evaluated on a basis element $\delta_u \otimes 1_{c_k\cdots c_n}$, $|u|=k$  yield
\( 
\rho^{(wy)}_{k+1}(\sum_{|v|=n-k+1} \delta_{uv}) = \sum_{|v|=n-k+1} e_{wyz, uv}
\)
for any $z$ with $|z| = n-k$. 
Once again, the coherence relations of $\cl B_{X_{n+1},A_{n+1}}$ guarantee this sum is independent of $y$. By the universal property, these maps induce a *-homomorphism $\rho^{(w)}_k : \cl C_k^{(n)} \to \cl B_{X_{n+1},A_{n+1}}$.

For $k=0$, assume that we have defined maps $\rho^{(y)}_{1} : \cl C_{1}^{(n)} \to \cl B_{X_{n+1},A_{n+1}}$  for all $y \in [d_0]$. By the universal property of the free product amalgamated over the unit, there exists a *-homomorphism 
\[
\rho_n : \cl C_0^{(n)} \to \cl B_{X_{n+1},A_{n+1}}.
\]
By construction, we see that $\rho_n$ and $\theta_n$ are mutually inverse, establishing the canonical *-isomorphism 
$\cl B_{X_{n+1},A_{n+1}} \cong \cl C^{(n)}_0$.
\end{proof}

\begin{lemma}\label{l_inject}
For every $n\in\mathbb N$, the *-homomorphism
\[
\pi_n : \mathcal B_{X_n,A_n} \longrightarrow \mathcal B_{X_{n+1},A_{n+1}},
\qquad
\pi_n(e_{x,a}) := \sum_{|a'|=1} e_{xx',aa'},
\]
is injective.
\end{lemma}

\begin{proof}
Using Proposition~\ref{p:freeproduct_model}, identify  
\[
\mathcal B_{X_n,A_n}\cong\mathcal C^{(n-1)}_0
\ \ \mbox{ and } \ \ 
\mathcal B_{X_{n+1},A_{n+1}}\cong\mathcal C^{(n)}_0.
\]
Under this identification, 
$\pi_n$ corresponds to a *-homomorphism
$\tilde{\pi}_n : \cl C^{(n-1)}_0\to \cl C^{(n)}_0$. 
We construct injective *-homomorphisms  
\( 
\gamma_k:\mathcal C^{(n-1)}_k\to\mathcal C^{(n)}_k\), \(k=n-1,\dots,0,
\)
fixing the common amalgamating subalgebras. As the ground step, 
we note that the canonical map 
$$\cl D_{A_n}\rightarrow \cl D_{A_n} \otimes 1_{c_n}
\rightarrow \cl C^{(n)}_{n}$$
is injective.
Now recall that
\[
\mathcal C^{(n-1)}_{\,n-1}
\;=\;
\underbrace{
\mathcal D_{A_n}
\;\ast_{\;\mathcal D_{A_{n-1}}\otimes 1_{c_{\,n-1}}}\;
\cdots
\;\ast_{\;\mathcal D_{A_{n-1}}\otimes 1_{c_{\,n-1}}}\;
\mathcal D_{A_n}
}_{d_{\,n-1}\ \text{times}} .
\]
and 
\[
\mathcal C^{(n)}_{\,n-1}
\;=\;
\underbrace{
\mathcal C^{(n)}_{\,n}
\;\ast_{\;\mathcal D_{A_{n-1}}\otimes 1_{c_{\,n-1}}\otimes 1_{c_{\,n}}}\;
\cdots
\;\ast_{\;\mathcal D_{A_{n-1}}\otimes 1_{c_{\,n-1}}\otimes 1_{c_{\,n}}}\;
\mathcal C^{(n)}_{\,n}
}_{d_{\,n-1}\ \text{times}} .
\]

The identification $\cl D_{A_{n-1}}\otimes 1_{c_{n-1}} \equiv \cl D_{A_{n-1}}\otimes 1_{c_{n-1}} \otimes 1_{c_n}$ allows us to view $\cl D_{A_{n-1}}\otimes 1_{c_{n-1}}$
as a unital C*-subalgebra of $\cl C_{n}^{(n)}$. 
By \cite[Theorem 4.2]{pedersen}, 
there exists a canonical injective map 
$\gamma_{n-1} : \cl C_{n-1}^{(n-1)}\to \cl C_{n-1}^{(n)}$.
Assume that an injective map 
$\gamma_{n-l} : \cl C_{n-l}^{(n-1)}\to \cl C_{n-l}^{(n)}$,
fixing the common C*-subalgebra 
$$\cl D_{A_{n-l}}\otimes 1_{c_{n-l}}\otimes 1_{c_{n-l+1}}\otimes\cdots \otimes 1_{c_n} \equiv \cl D_{A_{n-l}}\otimes 1_{c_{n-l}}$$
has been defined. 
The map 
$\gamma_{n-l-1} : \cl C_{n-l-1}^{(n-1)}\to \cl C_{n-l-1}^{(n)}$
is then defined by letting 
$$\gamma_{n-l-1} = 
\underbrace{\gamma_{n-l}\ast \gamma_{n-l}\ast
\cdots \ast\gamma_{n-l}}_{d_{n-l-1} \mbox{ \tiny times}}.$$
Proceeding inductively, we conclude that 
$\tilde{\pi}_n = \gamma_0$ is injective. 
\end{proof}



In view of Lemma \ref{l_inject}, we consider 
$\cl B_{X_n,A_n}$ as a unital C*-subalgebra of $\cl B_{X_m,A_m}$
whenever $n < m$, and $\cl B_{X_n,A_n}$ as a unital C*-subalgebra of $\cl B_{X,A}$
(recall the definition (\ref{eq_defofBXA})). 
The map $\pi_n$ is thus the inclusion of $\cl B_{X_n,A_n}$ into 
$\cl B_{X_{n+1},A_{n+1}}$. For $n < m$, we write 
$\pi_{n,m}$ for the inclusion of $\cl B_{X_n,A_n}$ into 
$\cl B_{X_{m},A_{m}}$, and $\pi_{n,\infty}$ for 
the inclusion of $\cl B_{X_n,A_n}$ into $\cl B_{X,A}$. 
If $|x| = |a| = n$, we use $e_{x,a}$ to denote the 
corresponding element in any of the C*-algebras 
$\cl B_{X,A}$ or $\cl B_{X_m,A_m}$, $m > n$.


\subsection{Character space}\label{ss_charactersp}

We next describe the character space of $\cl B_{X,A}$, which will be used subsequently in Section \ref{s_cantorgames}.
Let $f_n : X_n\to A_n$ be a map, $n\in \bb{N}$. 
We call the family $f = (f_n)_{n\in \bb{N}}$ a \emph{graded map} from $X$ to $A$ if, for each $n \geq 1$ and for each $x_n\in X_n$, there exists 
a map $g_{x_n} : [d_n]\to [c_n]$, such that 
\begin{equation}\label{eq_fn+1}
f_{n+1}((x_n,i)) = (f_n(x_n),g_{x_n}(i)), \ \ \ 
x_n\in X_n, i\in [d_n].
\end{equation}
Let ${\rm GM}(X,A)$ be the set of all graded maps 
from $X$ to $A$. 

The inclusion $\iota_{n,\infty} : \cl D_{X_n}\to C(\Omega_X)$
dualises to a surjection 
$\iota_{n,\infty}^* : M(\Omega_X) \to \cl D_{X_n}$, whose 
restriction to the set of point mass measures 
is the canonical projection (denoted in the same way) 
$\iota_{n,\infty}^* : \Omega_X \to X_n$. 
Similarly, we have the projection 
$\iota_{n,n+1}^* : X_{n+1} \to X_n$, $n\in \bb{N}$ (the same notation will be used for the projection 
$A_{n+1} \to A_n$, $n\in \bb{N}$).

\begin{proposition}\label{p_charactBXA}
Let $X = (X_n)_{n\in \bb{N}}$ and $A = (A_n)_{n\in \bb{N}}$
be two inductive families of sets. 
\begin{itemize}
\item[(i)] 
Every $f\in {\rm GM}(X,A)$ gives rise to a unique 
continuous map $\tilde{f} : \Omega_X\to \Omega_A$ such that 
$\iota_{n,\infty}^*(\tilde{f}(x)) = f_n(\iota_{n,\infty}^*(x))$, 
$x\in \Omega_X$, $n\in \bb{N}$. 

\item[(ii)] 
The characters $\chi : \cl B_{X,A}\to \bb{C}$ 
are in bijective correspondence 
with the elements $f = (f_n)_{n\in \bb{N}}\in {\rm GM}(X,A)$
via the assignment 
\begin{equation}\label{eq_chionexa}
\chi(e_{x,a}) = \delta_{f_n(x),a}, \ \ \ 
x\in X_n, a\in A_n, n\in \bb{N}.
\end{equation}
\item[(iii)] 
The map $\chi\to f$ from (ii) is a homeomorphism 
when ${\rm GM}(X,A)$ is equipped, via (i), 
with the topology of pointwise convergence. 
\end{itemize}
\end{proposition}

\begin{proof}
(i) The graded map $f$ gives rise to a map 
$\tilde{f} : \Omega_X\to \Omega_A$, defined by 
$$\tilde{f}((x_1,i_1,i_2,\dots)) = 
(f_1(x_1),g_{x_1}(i_1),g_{x_1i_1}(i_2),\dots), 
\ \ \ x_1\in X_1, i_k\in [d_k], k\in \bb{N}.$$
The relations $\iota_{n,\infty}^*(\tilde{f}(x)) = f_n(\iota_{n,\infty}^*(x))$,
$x\in \Omega_X$, $n\in \bb{N}$, hold by definition.

Given $n\in \bb{N}$ and $a\in A_n$, let $[a]\subseteq \Omega_A$ be the cylinder,
corresponding to $a$. Then 
$\tilde{f}^{-1}([a]) = \cup_{x\in f_n^{-1}(a)} [x]$, 
showing that $\tilde{f}^{-1}([a])$ is a (closed and) open 
subset of $\Omega_X$. Since the set of all cylinders of 
the form $[a]$, $a\in A_n$, $n\in\bb{N}$, is a subbase of the 
topology of $\Omega_A$, the map $\tilde{f}$ is continuous. 

(ii) 
Let $\chi : \cl B_{X,A}\to \bb{C}$ be a character. 
Using the paragraph after Lemma \ref{l_inject}, 
let $\chi_n : \cl B_{X_n,A_n}\to \bb{C}$ be the 
restriction of $\chi$, $n\in \bb{N}$. 
The character $\chi_1 : \cl A_{X_1,A_1}\to \bb{C}$ yields 
a family $(\chi_1^{(x)})_{x\in X_1}$ of 
*-homomorphisms, where $\chi_1^{(x)} : \cl D_{A_1}\to \bb{C}$, $x\in X_1$, thus giving rise to a 
map $f_1 : X_1\to A_1$ with the property 
$\chi_1^{(x)} (\delta_a) = \delta_{f_1(x),a}$, $a\in A_1$,
$x\in X_1$. 

Assume that the family $(f_k)_{k=1}^n$ has been defined, 
where $f_k : X_k\to A_k$ and, for each $k\in [n-1]$ and for each $x_k\in X_k$ there exists $g_{x_k} : [d_k]\to [c_k]$ such that $f_{k+1}((x_k,i)) = (f_k(x_k),g_{x_k}(i))$, $i\in [d_k]$. 
Fix $x_{k+1}\in X_{k+1}$. Since 
$e_{x_{k+1},a}$ is a projection for each $a\in A_{k+1}$
and $\sum_{a\in A_{k+1}} e_{x_{k+1},a} = 1$, 
there exists a unique $f_{k+1}(x_{k+1})\in A_{k+1}$ such that 
\begin{equation}\label{eq_fk+1}
\chi(e_{x_{k+1},f_{k+1}(x_{k+1})}) = 1;
\end{equation}
we have thus defined a map $f_{k+1} : X_{k+1}\to A_{k+1}$. 
By definition, 
$$e_{x_k,f_k(x_k)} = \sum_{j\in [c_k]} e_{x_k i,f_k(x_k) j}$$
for any $i\in [d_k]$. 
Since the elements $e_{x_k i,f_k(x_k) j}$ are projections
and $\chi(e_{x_k,f_k(x_k)}) = 1$, for every $i\in [d_k]$
there exists $g_{x_k}(i)\in [c_k]$ such that 
$\chi(e_{x_k i,f_k(x_k) g_{x_k}(i)}) = 1$. 
Comparing with (\ref{eq_fk+1}), we conclude that 
$f_{k+1}((x_k,i)) = (f_k(x_k) g_{x_k}(i))$.

(iii) 
Suppose that $f, f^{(k)}\in {\rm GM}(X,A)$, $k\in \bb{N}$.
We claim that the pointwise convergence $\tilde{f}^{(k)}\to_{k\to\infty} \tilde{f}$
is equivalent to the convergence 
$f^{(k)}_n(x)\to_{k\to\infty} f_n(x)$ for every 
$x\in X_n$ and every $n\in \bb{N}$.
To see this, equip $\Omega_X$ with its canonical 
metric; if $x = (i_0i_1i_2\dots)$ and 
$y = (j_0j_1j_2\dots)$ then 
$$d(x,y) = \sum_{n=0}^{\infty} \frac{1}{d_{n}^n} |i_n - j_n|.$$
Note that if $x$ and $y$ agree at the first $m$ coordinates ($m\geq 3$) then 
$$d(x,y) = \sum_{n=m}^{\infty} \frac{1}{d_{n}^n} |i_n - j_n|
\leq \sum_{n=m}^{\infty} \frac{2}{d_{n}^{n-1}} 
\leq \sum_{n=m}^{\infty} \frac{1}{2^{n-2}} = \frac{1}{2^{m-3}}.$$
We equip $\Omega_A$ with the analogous metric. 
Fix $x = (i_l)_{l\geq 0} \in \Omega_X$ and let $\epsilon > 0$. 
Let $m\in \bb{N}$ be such that $\frac{1}{2^{m-1}} < \epsilon$. 
Fix any $j_l\in [c_l]$, $l\geq m$. Then, writing $x = x_mx'$ with $|x_m| = m$, 
we have 
\begin{eqnarray*}
d(\tilde{f}^{(k)}(x),\tilde{f}(x))
& \leq & 
d(\tilde{f}^{(k)}(x),(f_m^{(k)}(x_m)j_{m}\dots)) \\
& + & 
d((f_m^{(k)}(x_m)j_{m}\dots), (f_m(x_m)j_{m}\dots))\\
& + & 
d(\tilde{f}(x),(f_m(x_m)j_{m}\dots))\\
& \leq &  
2\epsilon + d((f_m^{(k)}(x_m)j_{m}\dots), (f_m(x_m)j_{m}\dots)).
\end{eqnarray*}
The claim is now clear. 

Let $\chi, \chi^{(k)}$, $k\in \bb{N}$, 
be the characters corresponding to $f$ and $f^{(k)}$ according to (ii). 
The convergence 
$f^{(k)}_n(x)\to_{k\to\infty} f_n(x)$ for every 
$x\in X_n$ translates into the convergence 
$\chi^{(k)}(e_{x,a})\to_{k\to\infty} \chi(e_{x,a})$ for every 
$x\in X_n, a\in A_n$, $n\in \bb{N}$. Taking into account the 
previous paragraph, the proof is complete. 
\end{proof}


\subsection{Tracial characterisations} \label{s:charactersyn}

Recall \cite{psstw} that, if $S$ and $U$ are finite sets,  
a no-signalling correlation $p$ over $(S,U)$ is called \emph{synchronous} 
if $p(u,v|s,s) = 0$ whenever $u\neq v$, for every $s\in S$. 
By \cite[Theorem 5.5]{psstw}, the C*-algebra $\cl{A}_{S,U}$ serves as an encoding object for synchronous correlations, in the sense that every synchronous correlation $p$ of quantum commuting type over the pair $(S,U)$
arises from a trace on $\cl{A}_{S,U}$ via the equalities
$$p(u,v|s,t)=\tau(e_{s,u}e_{t,v}),\ \ \ s, t\in S,\ u,v\in U.$$ 
In this subsection we define synchronous graded correlations and
establish similar trace characterisation results to those in \cite{psstw, kps} in the Cantor setting, 
using the C*-algebra $\cl{B}_{X,A}$ constructed in Subsection \ref{ss_univa}.

\begin{definition}
\rm 
Let $X$ and $A$ be inductive families of sets and 
$p$ be an ns-graded correlation over $(X,A)$.
We call $p$ \emph{synchronous} if $p_n$ is synchronous 
for every $n\in \bb{N}$. 
\end{definition}

We write $\cl C_{\rm tg}^{\rm s}(X,A)$ for the set of all synchronous 
t-graded correlations over $(X,A)$ (or simply $\cl C_{\rm tg}^{\rm s}$ 
if $X$ and $A$ are clear from the context).

\begin{lemma}\label{l_partupton}
Let 
$$p = \{p(a,b|x,y) :  |x| = |y| = |a| = |b| \leq n, \ n\in \bb{N}\}$$
be a family of $n$ synchronous correlations of quantum commuting type, satisfying 
the conditions 
$$\sum_{|a'| = |b'| = k} p(aa',bb'|xx',yy') = p(a,b|x,y),$$
for all $x,y,a,b$ of the same length, 
where all concatenations have length $n$.
Then there exists a tracial state 
$\tau : \cl B_{X_n,A_n}\to \bb{C}$, such that 
\begin{equation}\label{eq_repforf} 
p(a,b|x,y) = \tau(e_{x,a}e_{y,b})
\end{equation}
for all initial words $x,y,a,b$ of the same length $k\leq n$. 
\end{lemma}

\begin{proof}
Since the correlation 
$p_n := \{p(a,b|x,y) : |x| = |y| = |a| = |b| = n\}$
is synchronous, by \cite{psstw} there exists a tracial 
state $\tau : \cl A_{X_n,A_n}\to \bb{C}$, such that 
$p(a,b|x,y) = \tau(\tilde{e}_{x,a}\tilde{e}_{y,b})$
for all $x,y\in X_n$, $a,b\in A_n$ (here 
$\{\tilde{e}_{x,a} : x\in X_n, a\in A_n\}$ is the set of canonical generators of $\cl A_{X_n,A_n}$). 
Fix $x\in X_k$, $x',x''\in \prod_{i=k}^{n-1} [d_i]$ and $a\in A_k$, where $k < n$, and let 
$u = \sum_{|b| = n-k} \tilde{e}_{xx',ab} - \tilde{e}_{xx'',ab}$. 
We have
$$u^2 = \sum_{|b| = |c| = n-k} 
\tilde{e}_{xx',ab}\tilde{e}_{xx',ac} - \tilde{e}_{xx',ab}\tilde{e}_{xx'',ac}
- \tilde{e}_{xx'',ab} \tilde{e}_{xx',ac} + \tilde{e}_{xx'',ab} \tilde{e}_{xx'',ac}$$
and hence 
\begin{eqnarray}\label{eq_u^2}
\tau(u^2)
& = & 
\sum_{|b| = |c| = n-k} 
(p(ab,ac|xx',xx') - p(ab,ac|xx',xx'') \nonumber\\
& - &  p(ab,ac|xx'',xx') + p(ab,ac|xx'',xx''))\\
& = & 
2 p(a,a|x,x) - 2 p(a,a|x,x) = 0. \nonumber
\end{eqnarray}
It follows, by the Cauchy-Schwarz inequality, that 
$\tau$ annihilates the ideal generated by the 
elements $u$, and hence descends to a tracial state 
on the quotient $\cl B_{X_n,A_n}$ of the C*-algebra $\cl A_{X_n,A_n}$. 
It remains to observe that the obtained tracial state, denoted in the same way, satisfies (\ref{eq_repforf}).
\end{proof}

We recall \cite[Section 11.5]{pisier} that
a \emph{matricial ultraproduct} is a 
von Neumann algebra of the form 
$M_{\omega} := \left(\oplus_{n\in \bb{N}} M_{k_n}\right)/\cl J_{\omega}$, where $\omega$ is an ultrafilter on $\bb{N}$ 
and 
$$\cl J_{\omega} = \left\{(w_n)_{n\in \bb{N}} \in \oplus_{n\in \bb{N}} M_{k_n} : \lim\hspace{-0.05cm}\mbox{}_{\omega} \|w_n\|_2 = 0\right\}$$
is the canonical ideal of 
$\oplus_{n\in \bb{N}} M_{k_n}$ associated with $\omega$. 
Such a matricial ultraproduct comes equipped with a 
canonical trace $\tau_{\omega}$, given by 
$\tau_{\omega}([w_n]_{\omega}) = \lim_{\omega} {\rm tr}_{M_{k_n}}(w_n)$, where $[w_n]_{\omega}$ denotes the 
coset of $(w_n)_{n\in \bb{N}}$ in $M_{\omega}$. 
A unital *-homomorphism $\pi : \cl B_{X,A}\to M_{\omega}$ will be called \emph{approximately finite dimensional} if
there exist unital *-homomorphisms $\pi_n : \cl B_{X_n,A_n}\to M_{k_n}$, $n\in \bb{N}$, such that 
$\pi(u) = [\pi_n(u)]_{\omega}$ whenever $u\in \cup_{m\in \bb{N}} \cl B_{X_m,A_m}$
(note that, if $u\in \cl B_{X_m,A_m}$ then the sequence 
$(\pi_n(u))_{n}$ is only eventually defined (whenever $n\geq m$) but this does not affect its coset with respect to $\omega$). Given an approximately finite dimensional 
*-homomorphism $\pi : \cl B_{X,A}\to M_{\omega}$, 
the trace $\tau_{\omega}$ will be called \emph{stable (relative to $\pi$)}
if the sequence $({\rm tr}_{M_{k_n}}(\pi_n(u)))_{n\in \bb{N}}$
is eventually constant whenever $u$ is a word of length two 
on the canonical generators $\{e_{x,a}\}$ of $\cl B_{X,A}$. 

\par 
Recall \cite{bctt} 
that the Cantor correlations 
$\Gamma : C(\Omega_A\times \Omega_A) \to L^{\infty}(\Omega_X\times \Omega_X)$ are in bijective correspondence 
with the corresponding associated inductive families $(\Gamma_n)_{n\in \bb{N}}$ of 
no-signalling correlations 
$\Gamma_n : \cl D_{A_nA_n}\to \cl D_{X_nX_n}$, $n\in \bb{N}$.
For $x\in X_n$, let 
$[x] = \{xy : y \in \prod_{i\geq n} [d_i]\}$ be the cylinder in $\Omega_X$ of all words commencing with $x$.
Given a function $h\in C(\Omega_X\times \Omega_X)$, 
let 
$$\Delta_{X_n}(h) = |X_n|^2 \sum_{x\in X_n} \left(\int_{[x]\times [x]} h d(\mu_X\times \mu_X)\right) \chi_{[x] \times [x]}.$$
We have that $\Delta_{X_n}: C(\Omega_X\times \Omega_X)\to C(\Omega_X\times \Omega_X)$ is an idempotent with 
range ${\rm span}\{\chi_{[x]\times [x]} : x\in X_n\}$; indeed, the fact that 
${\rm ran}(\Delta_{X_n}) \subseteq {\rm span}\{\chi_{[x]\times [x]} : x\in X_n\}$ 
is trivial, while the fact that 
$\chi_{[x] \times [x]} = \Delta_{X_n}(\chi_{[x] \times [x]})$
implies the reverse inclusion.
Similarly, given a function $h\in C(\Omega_X\times \Omega_X)$, 
let 
$$\Theta_{X_n}(h) = |X_n|^2 \sum_{x,y\in X_n} \left(\int_{[x]\times [y]} h d(\mu_X\times \mu_X)\right) \chi_{[x] \times [y]}.$$
We have in particular, that $\Theta_{X_n}(\chi_{[x]\times[y]})=\chi_{[x]\times[y]}$ for $x,y\in X_n$.
We define $\Delta_{A_n}$ and $\Theta_{A_n}$ similarly.

\begin{theorem}\label{th_trBXA}
Let $X$ and $A$ be inductive families of sets, and
$p\in \cl C_{\rm nsg}$. 
The following are equivalent: 
\begin{itemize}
\item[(i)] $p\in \cl C_{\rm qcg}^{\rm s}$; 

\item[(ii)] the associated
Cantor correlation
$\Gamma_p : C(\Omega_A\times \Omega_A) \to C(\Omega_X\times \Omega_X)$ is quantum commuting and satisfies the identities
\begin{equation} \label{eq_diagvanish}
\Delta_{X_n}\circ \Gamma_p \circ \Theta_{A_n} = 
\Delta_{X_n}\circ \Gamma_p \circ \Delta_{A_n}, 
\ \ \  n\in \bb{N};
\end{equation}

\item[(iii)] there exists a tracial state $\tau : \cl B_{X,A}\to \bb{C}$, such that 
\begin{equation}\label{eq_pabxy=tau}
p(a,b|x,y) = \tau(e_{x,a}e_{y,b})
\end{equation}
for all initial words $x,y,a,b$ of the same length.
\end{itemize}
Moreover, 
\begin{itemize}
\item[(a)] $p\in \cl C_{\rm qag}^{\rm s}$ if and only if the trace $\tau$ in (\ref{eq_pabxy=tau}) can be chosen to be amenable; 

\item[(b)] $p\in \cl C_{\rm qg}^{\rm s}$ if and only if 
there exists an approximately finite dimensional unital *-homomorphism 
$\pi : \cl B_{X,A}\to M_{\omega}$ into a matricial ultraproduct, relative to which the canonical trace is stable
and satisfies (\ref{eq_pabxy=tau});

\item[(c)] $p\in \cl C_{\rm locg}^{\rm s}$ if and only if the trace $\tau$ in (\ref{eq_pabxy=tau}) can be chosen to 
factor through an abelian representation of $\cl B_{X,A}$. 
\end{itemize}
\end{theorem}

\begin{proof}
(iii)$\Rightarrow$(i) 
Fixing $x\in X_n$ and $a\in A_n$, we have 
\begin{eqnarray*}
& & 
\sum_{|a'| = |b'| = 1} p(aa',bb'|xx',yy')
= 
\sum_{|a'| = |b'| = 1}  \tau(e_{xx',aa'}e_{yy',bb'})\\
& = & 
\tau\left(\left(\sum_{|a'| = 1}e_{xx',aa'}\right) \left(\sum_{|b'| = 1} e_{yy',bb'}\right)\right)
= \tau(e_{x,a}e_{y,b}) = 
p(a,b|x,y),
\end{eqnarray*}
showing that $p$ is graded. Synchronicity is 
immediate in view of conditions (PVM). 

To show that $p$ is of quantum commuting type, 
note that (iii) implies, via 
\cite{psstw}
that the $n$-th component $p_n$ of 
$p$ is of quantum commuting type, and hence so is $\Gamma_n$ for each $n$. Now \cite[Theorem 5.8]{bctt} implies that $\Gamma$, and hence $p$, is of quantum commuting type.

(i)$\Rightarrow$(iii)
By 
Lemma \ref{l_partupton}, for each $n\in \bb{N}$
there exists a tracial state $\tau_n : \cl B_{X_n,A_n}
\to \bb{C}$, such  that (\ref{eq_pabxy=tau}) holds for 
all words $x,y,a,b$ with 
$|x| = |y| = |a| = |b| \leq n$. 
Using Lemma \ref{l_inject}, 
we consider $\cl B_{X_n,A_n}$ as a unital C*-subalgebra of 
$\cl B_{X,A}$, and extend 
$\tau_n$ to a state $f_n : \cl B_{X,A}\to \bb{C}$. 
Without loss of generality, 
assume that $f_n\to_{n\to \infty} \tau$ in the weak* topology, for some state $\tau : \cl B_{X,A}\to \bb{C}$. 
We claim that $\tau$ is tracial. 
In fact, suppose that $u,v\in \cl B_{X_n,A_n}$; then 
\begin{eqnarray*}
& & \tau(\pi_{n,\infty}(u)\pi_{n,\infty}(v)) 
= 
\lim_{m\to \infty} f_m(\pi_{n,m}(u)\pi_{n,m}(v))
= 
\lim_{m\to \infty} \tau_m(\pi_{n,m}(u)\pi_{n,m}(v))\\
& = &  
\lim_{m\to \infty} \tau_m(\pi_{n,m}(v)\pi_{n,m}(u))
= 
\lim_{m\to \infty} f_m(\pi_{n,m}(v)\pi_{n,m}(u))
= 
\tau(\pi_{n,\infty}(v)\pi_{n,\infty}(u)).
\end{eqnarray*}
Thus, $\tau$ is tracial and condition (\ref{eq_pabxy=tau}) readily follows. 

(i)$\Leftrightarrow$(ii)
Condition (\ref{eq_diagvanish}) is  equivalent to the condition 
\begin{eqnarray*}
\left\langle \Gamma_p(\chi_{[a]\times [b]}),\chi_{[x]\times [x]}\right\rangle 
& = & \delta_{a,b}
\left\langle \Gamma_p(\chi_{[a]\times [a]}),\chi_{[x]\times [x]}\right\rangle\\
& = & 
\delta_{a,b}
\left\langle (\Gamma_p\circ\Delta_{A_n})(\chi_{[a]\times [b]}),\chi_{[x]\times [x]}\right\rangle, 
\ \ \ x\in X_n, n\in \bb{N},
\end{eqnarray*}
that is, to the condition 
$$
\left\langle (\Delta_{X_n}\circ \Gamma_p\circ\Theta_{A_n})(h), \chi_{[x]\times [y]}\right\rangle = 
\left\langle (\Delta_{X_n}\circ \Gamma_p\circ \Delta_{A_n})(h),\chi_{[x]\times [y]}\right\rangle,$$
$x,y\in X_n, n\in \bb{N}$, $h\in C(\Omega_A\times\Omega_A)$,
which is equivalent to $\Gamma_p$ satisfying the displayed equality. 

\smallskip

(a) Suppose that $\tau : \cl B_{X,A}\to \bb{C}$ is an 
amenable trace satisfying (\ref{eq_pabxy=tau}). 
Letting $\tau_n = \tau|_{\cl B_{X_n,A_n}}$, we have that 
$\tau_n$ is an amenable trace on $\cl B_{X_n,A_n}$ such that 
$p_n(a,b|x,y) = \tau_n(e_{x,a}e_{y,b})$ for all $x,y\in X_n$ and all $a,b\in A_n$. By \cite{kps}, $p_n$ is of approximately quantum type for each $n\in \bb{N}$; by Theorem \ref{p_fornse}, $p\in \cl C^{\rm s}_{\rm qag}$. 

Conversely, suppose that 
$p\in \cl C^{\rm s}_{\rm qag}$. By Theorem \ref{p_fornse},
$p_n$ is of approximately quantum 
type for each $n$. Using \cite[Theorem 3.6]{kps}, let 
$\tilde{\tau}_n : \cl A_{X_n,A_n}\to \bb{C}$ be an amenable trace
satisfying (\ref{eq_pabxy=tau}) for all words of length $n$. 
Letting $\cl J$ be the ideal of $\cl A_{X_n,A_n}$ for which 
$\cl B_{X_n,A_n} = \cl A_{X_n,A_n}/\cl J$, the gradedness conditions satisfied by 
$p_n$ imply that $\tilde{\tau}_n$ annihilates $\cl J$ (see (\ref{eq_u^2})). 
By \cite[Proposition 6.3.5]{bo}, the induced trace $\tau_n : \cl B_{X_n,A_n} \to \bb{C}$ is amenable.

Assume that $\cl B_{X,A}\subseteq \cl B(H)$ as a unital C*-subalgebra. 
Let $f_n$ be an extension of $\tau_n$ to a state of $\cl B(H)$
with the property that $f_n(uTu^*) = f_n(T)$ for every $T\in \cl B(H)$ and every unitary $u\in \cl B_{X_n,A_n}$, $n\in \bb{N}$. 
Let $(f_{n_i})_{i\in \bb{I}}$ be a subnet of 
$(f_n)_{n\in \bb{N}}$, and $f : \cl B(H)\to \bb{C}$ be a state,
such that $f_{n_i}\to_{i\in \bb{I}} f$ in the weak* topology.
Fix $k\in \bb{N}$ and a unitary $u\in \cl B_{X_k,A_k}$. Whenever $n_i\geq k$ we have that 
$$f(uTu^*) = \lim_{i\in \bb{I}} f_{n_i}(uTu^*)
= \lim_{i\in \bb{I}} f_{n_i}(T) = f(T).$$
On the other hand, if $u$ is a unitary in $\cl B_{X,A}$ and $\epsilon > 0$
then there exists $k\in \bb{N}$ and a unitary $u'\in \cl B_{X_k,A_k}$ such that $\|u - u'\| < \epsilon$. The latter follows from the density of the set of the selfadjoint elements $\cup_k\cl B_{X_k,A_k}^{\rm sa}$ in $\cl B_{X,A}^{\rm sa}$ and continuity of the continuous functional calculus 
(letting $u = e^{ih}$, where $h$ is a selfadjoint element of $\cl B_{X,A}$,  we can find a selfadjoint 
element $h'$ of $\cl B_{X_k,A_k}$ such that $\|u-e^{ih'}\|<\epsilon$ and as $\cl B_{X_k,A_k}$ is a $C^*$-subalgebra of $\cl B_{X,A}$, we have that $u'=e^{ih'}\in \cl B_{X_k,A_k}$).  
It follows that $f(uTu^*) = f(T)$ for all $T\in \cl B(H)$ and all 
unitaries $u\in \cl B_{X,A}$. 

Let $\tau = f|_{\cl B_{X,A}}$; then $\tau$ is an amenable 
trace on $\cl B_{X,A}$ \cite[Section 6.2]{bo}. In addition, 
if $x,y\in X_n$ and $a,b\in A_n$ then
$$\tau(e_{x,a} e_{y,b}) = 
\lim_{i\to\infty} f_{n_i}(e_{x,a} e_{y,b}) = 
\lim_{i\to\infty} \tau_{n_i}(e_{x,a} e_{y,b}) = \lim_{i\to\infty} p_{n_i}(a,b|x,y) = p(a,b|x,y).$$

(b) 
Let $M_{\omega}$ be a matricial ultraproduct and
$\pi : \cl B_{X,A} \to M_{\omega}$ be an  
approximately finite dimensional *-representation
for which $\tau_{\omega}$ is a stable trace 
satisfying (\ref{eq_pabxy=tau}). 
For a fixed $n$, we have that 
$p_n(a,b|x,y) = {\rm tr}_{M_{k_n}}(\pi_n(e_{x,a})\pi_n(e_{y,b}))$ for all $x,y\in X_n$ and 
all $a,b\in A_n$, and by \cite{hmps} $p_n$ is of quantum type.

Conversely, suppose that $p_n$ is of quantum type
for every $n\in \bb{N}$. Using \cite{hmps} let 
$k_n\in \bb{N}$ and $\pi_n : \cl A_{X_n,A_n}\to M_{k_n}$
be a unital *-homomorphism, such that 
$p_n(a,b|x,y) = {\rm tr}_{M_{k_n}}\circ\pi_n (\tilde{e}_{x,a} \tilde{e}_{y,b})$ whenever $x,y\in X_n$ and 
$a,b\in A_n$, where $\tilde{e}_{z,c}$ denote 
the canonical generators of $\cl A_{X_n,A_n}$. 
As in the calculations (\ref{eq_u^2}), 
the faithfulness of ${\rm tr}_{M_{k_n}}$ implies that 
$\pi_n$ induces a *-homomorphism (denoted in the same way) 
$\pi_n : \cl B_{X_n,A_n}\to M_{k_n}$, such that 
$p_n(a,b|x,y) = {\rm tr}_{M_{k_n}}\circ\pi_n (e_{x,a}e_{y,b})$
whenever $x,y\in X_n$ and $a,b\in A_n$. 

Let $\omega$ be an ultrafilter on $\bb{N}$, 
$M_{\omega}$ be the matricial ultraproduct along $\omega$, 
arising from the sequence $(k_n)_{n\in \bb{N}}$, 
and 
$\pi_0 : \cup_{n\in \bb{N}} \cl B_{X_n,A_n}\to M_{\omega}$
be the *-homomorphism, given by 
$$\pi_0(u) = [(\pi_m(u))_{m\geq n}]_{\omega}, \ \ \ 
u\in \cl B_{X_n,A_n}.$$
As $\pi_0$ is bounded, it possesses a 
(unique) extension $\pi : \cl B_{X,A}\to M_{\omega}$ to a unital *-homomorphism; it is clear that $\pi$ is 
approximately finite dimensional. 
Fix $m\in \bb{N}$ and note that, if 
$x,y\in X_m$ and $a,b\in A_m$ then 
$$p_m(a,b|x,y) = p(a,b|x,y) = {\rm tr}\hspace{-0.05cm}\mbox{}_{M_{k_n}}(\pi_n(e_{x,a})\pi_n(e_{y,b})), \ \ \ n\geq m.$$
It follows that $\tau_{\omega}$ is stable and 
satisfies (\ref{eq_pabxy=tau}). 

(c) Suppose that $\cl D$ is an abelian C*-algebra, 
$f : \cl D\to \bb{C}$ a state, and 
$\pi : \cl B_{X,A}\to \cl D$ is a unital *-homomorphism, such that (\ref{eq_pabxy=tau}) holds for the trace 
$\tau := f\circ \pi$. 
Set $\tilde{\tau}_n = \tau\circ \phi_n \circ q_n$, 
where $\phi_n : \cl B_{X_n,A_n}\to \cl B_{X,A}$
temporarily denotes the inclusion map; we have that 
$\tilde{\tau}_n : \cl A_{X_n,A_n}\to \bb{C}$
is a trace that yields the correlation $p_n$. 
By \cite{hmps}, $p_n$ is of local type, $n\in \bb{N}$.

Conversely, assume that $p_n$ is of local type 
for each $n$. Following the proof of (b), we 
may now replace the matrix algebras $M_{k_n}$ with 
finite dimensional abelian C*-algebras $\cl D_n$, $n\in \bb{N}$. The statement now follows from the fact 
that the arising matricial ultraproduct $M_{\omega}$ is abelian. 
\end{proof}


\section{Bisynchronous correlations}\label{s_cqpg}

In this section, we 
define bisynchronous graded correlations and 
characterise them in terms 
of tracial states on universal C*-algebras. 
We will see that the latter are continuous function algebras
on projective limits of compact quantum groups first studied
in \cite{bgs}. 

As a means of preparation, recall \cite{paulsen-rahaman} that, if $S$ is a finite set,  
a synchronous no-signalling correlation $p$ over $(S,S)$ is called \emph{bisynchronous} 
if $p(u,u|s,t) = 0$ whenever $s\neq t$, for every $u\in S$. 
Recall, further, that a \emph{magic unitary} (or a \emph{quantum permutation}) over $S$ 
on a Hilbert space $H$
is a matrix $(P_{s,t})_{s,t\in S}$ with entries in $\cl B(H)$, such that the families 
$(P_{s,t'})_{t'\in S}$ and $(P_{s',t})_{s'\in S}$ are PVM's for all $s,t\in S$. 
The \emph{quantum permutation group} over $S$ is the universal C*-algebra 
$C({\rm S}_{S}^+)$ of a magic unitary $(\hat{p}_{s,t})_{s,t\in S}$, that is, the C*-algebra 
generated by a family $\{\hat{p}_{s,t}\}_{s,t\in S}$ of orthogonal projections subject to the relations 
$\sum_{t'\in S} \hat{p}_{s,t'} = \sum_{s'\in S} \hat{p}_{s',t} = 1$
for all $s,t\in S$
(\cite{wang}, see also \cite{bigalois}); thus, unital *-representations $\pi : C({\rm S}_{S}^+)\to \cl B(H)$ on 
a Hilbert space $H$ correspond to magic unitaries $(P_{s,t})_{s,t\in S}$ in $\cl B(H)$ via the 
assignments $\pi(\hat{p}_{s,t}) = P_{s,t}$. 
Further, the tracial states $\tau$ of $C({\rm S}_{S}^+)$ exhaust all bisynchronous no-signalling 
correlations $p$ over $S$ via the assignment (\cite{paulsen-rahaman})
$$p(u,v|s,t)=\tau(\hat{p}_{s,u}\hat{p}_{t,v}),\ \ \ s,t,u,v\in S.$$ 


\subsection{A universal C*-algebra for bisynchronicity}\label{ss_biscstar}

In this subsection, we present a construction of a C*-algebra that will play
the role of the quantum permutation group for Cantor spaces as opposed to finite sets. 
We note that the latter C*-algebra is a compact quantum group already studied in 
\cite{bgs} and that it can be viewed as a projective limit in the category of compact quantum groups 
\cite{ss};  
for our purposes (see Subsection \ref{ss_tracialchar}), we follow an approach that relates it to the C*-algebras of the form $\cl B_{X,A}$ introduced in Subsection \ref{ss_univa}.
We discuss in detail the connection with quantum group theory in 
Subsection \ref{subs_projlims}.

Let $X=(X_n)_{n\in \bb{N}}$ be an inductive family. For $n\in \bb{N}$, we set 
\[
\mathcal B_{X_n}
:=
C({\rm S}_{X_n}^+)\big/ \left\langle
\sum_{|b|=n-k} \hat{p}_{xy',ab}
-
\sum_{|b|=n-k} \hat{p}_{xy'',ab}, 
\sum_{|y|=n-k} \hat{p}_{xy,ab'}
-
\sum_{|y|=n-k} \hat{p}_{xy,ab''}
\right\rangle,
\]
where the ideal, denoted $\cl J_{X_n}$, is generated by the specified elements for all $k\in [n-1]$,
$|x|=|a|=k$, and $|y'|=|y''|=|b'|=|b''|=n-k$.

\begin{remark}\label{rmk:welldef} 
\rm
Inside $\mathcal B_{X_{n+1}}$, for every $x,a\in X_n$ we define
\( 
p_{x,a}':=\sum_{|b|=1} \hat p_{xy',ab},
\)
where $y'\in[d_n]$ is arbitrary. By the first family of coherence relations at level
$n+1$ with $k=n$, this is independent of the choice of $y'$. Let also
\( 
 p_{x,a}'' = \sum_{|y|=1}\hat p_{xy,ab'},
\)
which by the second family of coherence relations  is independent of $b'$. Hence, summing over all one-letter
extensions, 
\[
d_n p_{x,a}'=\sum_{y,b\in[d_n]}\hat p_{xy,ab}=d_n p_{x,a}'',
\]
so that $  p_{x,a}' = p_{x,a}''$. The PVM relations for the elements
$ p_{x,a}'$ follow from the corresponding relations at level
$n+1$, while the level-$(n+1)$ coherence relations imply that these
elements satisfy both coherence relations at level $n$. 
\end{remark}

We write $p_{x,a}$ for the image of $\hat{p}_{x,a}$ under the quotient map. It is straightforward to verify that 
there exists a (unital) *-homomorphism 
$\rho_n : C({\rm S}_{X_{n}}^+)\to \cl B_{X_{n+1}}$, such that 
$\rho_n(\hat{p}_{x,a}) = \sum_{|a'| = 1} p_{xx',aa'}$
where $x'$ is any word of length $1$ (the expression being independent of the choice of $x'$).
Moreover, by Remark \ref{rmk:welldef}, $\rho_n$ vanishes on $\cl J_{X_n}$, and hence
descends to a unital *-homomorphism (denoted in the same way)
$\rho_n : \cl B_{X_n}\to \cl B_{X_{n+1}}$. 
We let 
\[
\cl B_{X} := \varinjlim (\cl B_{X_n},\rho_n)_{n\in \bb{N}}
\]
be the inductive limit of the 
sequence $(\cl B_{X_n},\rho_n)_{n\in \bb{N}}$ in the category of C*-algebras. 

\begin{remark}\label{r_qX}
\rm 
By the universal property of the inductive limit and of the 
finite quantum permutation groups, the C*-algebra $\cl B_{X}$ is the universal C*-algebra, 
generated by a family $\{p_{x,a} : x,a\in X_n, n\in \bb{N}\}$ of projections such that $(p_{x,a})_{x,a\in X_n}$ is a 
quantum permutation for every $n$, and satisfying the relations 
$$p_{x,a} = \sum_{|b|=1} p_{xy',ab} = 
\sum_{|b|=1} p_{xy'',ab} \ \mbox{ and } \  
p_{x,a} = \sum_{|y|=1} p_{xy,ab'} =  
\sum_{|y|=1} p_{xy,ab''}
$$
for all initial words $x$ and $a$ of equal length and  
all $y', y'', b', b''$ of length 1.
Thus, 
\begin{itemize}
\item[(i)] there exists a *-epimorphism 
$q_X : \cl B_{X,X}\to \cl B_X$ such that $q_X(e_{x,a}) = p_{x,a}$ for all initial words $x$ and $a$ of equal length; 

\item[(ii)]
for a Hilbert space $H$, the unital 
*-representations $\rho : \cl B_X\to \cl B(H)$ correspond precisely to
a \emph{Cantor spectral bi-measure} over $X = (X_n)_{n\in \bb{N}}$ acting on $X$, that is, 
families $P = \{(P_{x,a})_{x,a\in X_n} : n\in \bb{N}\}$ of projections on $H$, satisfying the defining relations of $\cl B_X$
concretely.
\end{itemize}
\end{remark}

Let $f_n : X_n\to X_n$ be a map, $n\in \bb{N}$, 
such that the family $f = (f_n)_{n\in \bb{N}}$ is a graded map 
from $X$ to $X$, with associated families of maps 
$g_{x_n} : [d_n]\to [d_n]$, $x_n\in X_n$, $n\in \bb{N}$
(see equation (\ref{eq_fn+1})).
We call $f$ a \emph{graded bijection} on $X$ if 
$f_1$ is a bijection, and 
$g_{x_n}$ is a 
bijection for all $x_n\in X_n$ and all $n\in \bb{N}$.
We let 
${\rm GB}(X)$ be the set of all graded bijections on $X$. 
Clearly, ${\rm GB}(X)\subseteq {\rm GM}(X,X)$; 
thus, every $f\in {\rm GB}(X)$ gives rise, 
via Proposition \ref{p_charactBXA}, to a 
continuous map $\tilde{f} : \Omega_X\to \Omega_X$.

\begin{proposition}\label{p_cohhome}
Let $X = (X_n)_{n\in \bb{N}}$ be an inductive family of sets. 
\begin{itemize}
\item[(i)] 
If $f\in {\rm GB}(X)$ then $\tilde{f} : \Omega_X\to \Omega_X$ 
is a homeomorphism.

\item[(ii)] 
The characters $\chi : \cl B_{X}\to \bb{C}$ 
are in bijective correspondence 
with the elements $f = (f_n)_{n\in \bb{N}}\in {\rm GB}(X)$
via the assignment 
\begin{equation}\label{eq_chionexabij}
\chi(p_{x,a}) = \delta_{f_n(x),a}, \ \ \ 
x\in X_n, a\in A_n, n\in \bb{N}.
\end{equation}
\item[(iii)] 
The map $\chi\mapsto f$ from (ii) is a homeomorphism 
when ${\rm GB}(X)$ is equipped, via (i), 
with the topology of pointwise convergence. 
\end{itemize}
\end{proposition}

\begin{proof}
(i) Given $f = (f_n)_{n\in \bb{N}} \in {\rm GB}(X)$, 
let $f^- := (f_n^{-})_{n\in \bb{N}}$, where $f_1^- := f_1^{-1}$ 
and 
$f^-_{n+1}((x_n,i)) = (f_n^{-1}(x_n),g_{f_n^{-1}(x_n)}^{-1}(i))$, 
$x_n\in X_n$, $i\in [d_n]$, $n\in \bb{N}$. 
It is straightforward to see that $f^-\in {\rm GB}(X)$ and 
that 
$$\widetilde{f^-}\circ \tilde{f} 
= \tilde{f} \circ\widetilde{f^-} = {\rm id}.$$
Thus, $\tilde{f}$ is a homeomorphism.

(ii) follows similar lines to those of Proposition \ref{p_charactBXA} (ii), and 
uses the fact that the restriction of the character of $\cl B_{X,X}$ defined therein to $\cl B_{X_n,X_n}$ annihilates the generators of the kernel of the canonical 
quotient map from $\cl B_{X_n,X_n}$ onto $\cl B_{X_n}$, 
and thus descends to a character $\chi_n$ of $\cl B_{X_n}$.

(iii) follows in a fashion similar to Proposition \ref{p_charactBXA} (iii).
\end{proof}


In view of Proposition \ref{p_cohhome}, 
$\cl B_{X}$ can be thought of as the quantum graded homeomorphism group of $\Omega_X$; note that, in particular, the latter proposition implies 
that $\cl B_{X}$ is non-trivial.


\subsection{Tracial characterisation}\label{ss_tracialchar}

In this subsection, we extend bisynchronicity to the graded setting and obtain a characterisation of graded bisynchronous correlations via tracial states
on the C*-algebra $\cl B_X$ introduced in Subsection \ref{ss_biscstar}. 

\begin{definition} \rm
Let $(X_n)_{n\in \mathbb N}$ be an inductive family of sets, and set $A_n = X_n$, 
$n\in \mathbb N$.
A graded no-signalling correlation
\[
p=\{p(a,b|x,y) : n\in \mathbb N,\ |x|=|y|=|a|=|b|=n\}
\]
over $(X,A)$ is called a:
\begin{itemize}
    \item[(i)] \emph{graded bicorrelation} if the transposed family
\[
\tilde p=\{\tilde p(x,y|a,b) : n\in \mathbb N,\ |x|=|y|=|a|=|b|=n\},
\]
where $\tilde p(x,y|a,b):=p(a,b|x,y)$, is a graded ns-correlation over $(A,X)$;
\item[(ii)]  \emph{bisynchronous graded correlation} if $p$ is a graded bicorrelation such that 
$p$ and $\tilde p$ are synchronous.
\end{itemize}
\end{definition}

We write $\cl C_{\rm tg}^{\rm bs}(X)$ for the set of all bisynchronous ${\rm t}$-graded correlations over $(X,X)$ (or simply $\cl C_{\rm tg}^{\rm bs}$ if $X$ is clear from the context), $\rm t\in\{loc, q, qa, qc, ns\}$. 

\begin{remark}
\rm 
Suppose that $p = (p(a,b|x,y))_{x,y,a,b}$ is a graded
bicorrelation. Equation (\ref{eq_deff0}) yields a 
Cantor correlation 
$\Gamma : C(\Omega_X\times \Omega_X)\to C(\Omega_X\times \Omega_X)$, canonically associated with $p$. 

\begin{itemize}
\item[(i)]
    
If $(\Gamma_n)_{n\in \bb{N}}$ is the sequence of unital completely positive maps, $\Gamma_n : \cl D_{X_n X_n}\to \cl D_{X_n X_n}$, canonically associated with $\Gamma$ then 
$\Gamma_n$ is in addition trace-preserving; thus, its 
dual $\Gamma_n^* : \cl D_{X_n X_n}\to \cl D_{X_n X_n}$
is unital. 

\item[(ii)] The Cantor correlation $\tilde{\Gamma}$, 
associated with $\tilde{p}$ via formula (\ref{eq_deff0}) 
satisfies the conditions
$$\left\langle \Gamma(g),h\right\rangle = \langle g,\tilde{\Gamma}(h)\rangle, \ \ \ g,h\in C(\Omega_X\times \Omega_X).$$
\end{itemize}
\end{remark}

\begin{theorem}\label{th_trBX}
Let $p = \{p(a,b|x,y) : n\in \bb{N}, |x| = |y| = |a| = |b| = n\}$ be a no-signalling correlation. 
The following are equivalent: 
\begin{itemize}
\item[(i)] $p$ is graded, bisynchronous and of quantum commuting type; 

\item[(ii)] 
the graded Cantor correlation
$\Gamma_p : C(\Omega_X\times \Omega_X) \to C(\Omega_X\times \Omega_X)$ is of quantum commuting type, and 
$\Gamma_p$ and $\tilde{\Gamma}_p$ satisfy (\ref{eq_diagvanish});

\item[(iii)] there exists a tracial state 
$\tau : \cl B_{X}\to \bb{C}$, such that 
\begin{equation}\label{eq_pabxy=taubis}
p(a,b|x,y) = \tau(p_{x,a}p_{y,b})
\end{equation}
for all initial words $x,y,a,b$ of the same length.
\end{itemize}
Moreover, 
\begin{itemize}
    \item[(a)] $p\in \cl C_{\rm qag}^{\rm bs}$ if and only if  there exists a unital *-homomorphism $\pi: \cl B_X\to \cl R^\omega$ such that 
    $p(a,b|x,y) = \tau_\omega(\pi(p_{x,a}p_{y,b}))$, where $\tau_\omega$ is the canonical trace on $\cl R^\omega$; 

\item[(b)] $p\in \cl C_{\rm qg}^{\rm bs}$ if and only if 
there exist an approximately finite dimensional unital *-homomorphism 
$\pi : \cl B_{X}\to M_{\omega}$ into a matricial ultraproduct, with respect to which the canonical trace is stable
and satisfies $p(a,b|x,y) = \tau(\pi(p_{x,a}p_{y,b}))$;

\item[(c)] $p\in \cl C_{\rm locg}^{\rm bs}$ if and only if the trace $\tau$ in (\ref{eq_pabxy=taubis}) can be chosen to 
factor through an abelian representation of $\cl B_{X}$. 
\end{itemize}
\end{theorem}

\begin{proof}
(iii)$\Rightarrow$(i) follows similarly to the implication 
(iii)$\Rightarrow$(i) in Theorem \ref{th_trBXA}. 

(i)$\Rightarrow$(iii)
Let $p_n$ be the $n$-th component of $p$, $n\in \bb{N}$. 
By Theorem \ref{th_trBXA}, there exists a tracial state 
$\tau : \cl B_{X,X}\to \bb{C}$, such that 
\begin{equation}\label{eq_pabtau}
p(a,b|x,y) = \tau(e_{x,a}e_{y,b}) 
\end{equation}
for all initial words 
$x,y,a,b$ of equal length. 
Let $\tau_n : \cl B_{X_n,X_n}\to \bb{C}$ be the tracial state, 
arising canonically from $\tau$ in view of the inductive 
sequence $(\cl B_{X_n,X_n},\pi_n)_{n\in \bb{N}}$. 
Then $p_n$ is a bisynchronous correlation, for which 
$p_n(a,b|x,y) = \tau_n(e_{x,a}e_{y,b})$, when $\cl B_{X_n,X_n}$ is viewed as a C*-subalgebra of $\cl B_{X,X}$. 

We claim that $\tau$ annihilates the ideal
$\cl I := \ker(q_X)$ (see Remark \ref{r_qX}). 
To see this, note that $\cl I$ is generated by the set 
$$\left\{\sum_{z\in X_n} e_{z,b}\hspace{-0.1cm}-\hspace{-0.1cm}1,\sum_{|x'| = k} e_{xx',aa'} \hspace{-0.1cm}-\hspace{-0.1cm} \sum_{|x'| = k}e_{xx',aa''} :b\in X_n, a'\neq a'', |x| \hspace{-0.1cm}=\hspace{-0.1cm} |a| \hspace{-0.1cm}=\hspace{-0.1cm} n-k, n\in \bb{N}\right\}.$$
The fact that $\tau_n$ annihilates generators of the first type follows from bisynchronicity. We verify the second ones. Fixing $|x| = |a| = n-k$ and $a'\neq a''$ with $|a'| = |a''| = k$, we have 
\begin{eqnarray*}
& & 
\hspace{-2cm}
\tau_n\left(\left(\sum_{|x'| = k} e_{xx',aa'} - \sum_{|x'| = k}e_{xx',aa''}\right)^2\right)\\
& = & 
\sum_{|x'| = |y'| = k} \tau_n(e_{xx',aa'}e_{xy',aa'}) 
- 
\sum_{|x'| = |y'| = k} \tau_n(e_{xx',aa'}e_{xy',aa''})\\
& - &  
\sum_{|x'| = |y'| = k} \tau_n(e_{xx',aa''}e_{xy',aa'})
+ 
\sum_{|x'| = |y'| = k} \tau_n(e_{xx',aa''}e_{xy',aa''})\\
& = & 
\sum_{|x'| = |y'| = k} p(aa',aa'|xx',xy')
- 
\sum_{|x'| = |y'| = k} p(aa',aa''|xx',xy')\\
& - &  
\sum_{|x'| = |y'| = k} p(aa'',aa'|xx',xy')
+ 
\sum_{|x'| = |y'| = k} p(aa'',aa''|xx',xy')\\
& = & 
2p(a,a|x,x) - 2p(a,a|x,x) = 0.
\end{eqnarray*}
By the Cauchy-Schwarz inequality, $\tau_n$ induces a 
tracial state $\tilde{\tau}_n$ on the C*-algebra $\cl B_{X_n}$. 

We note that the sequence $(\tilde{\tau}_n)_{n\in \bb{N}}$ is inductive; to this end, note first that 
$$q_{X_{n+1}}\circ \pi_n = \rho_n \circ q_{X_n}, \ \ \ n\in \bb{N},$$
where $q_{X_n}: \cl B_{X_n,X_n}\to\cl B_{X_n}$ is the quotient map and $\pi_n:\cl B_{X_n,X_n}\to\cl B_{X_{n+1},X_{n+1}}$ are the connecting maps defining the inductive limit $\cl B_{X,X}=\varinjlim (\cl B_{X_n,X_n}, \pi_n)_{n\in \bb{N}}$. For $\tilde{u}\in \cl B_{X_n}$, let $u\in \cl B_{X_n,X_n}$ be an element with $q_{X_n}(u) = \tilde{u}$. Then 
\begin{eqnarray*}
(\tilde{\tau}_{n+1}\circ\rho_n)(\tilde{u})
& = & 
(\tilde{\tau}_{n+1}\circ\rho_n\circ q_{X_n})(u)
= 
(\tilde{\tau}_{n+1}\circ q_{X_{n+1}}\circ \pi_n)(u)\\
& = & 
\tau_{n+1}(\pi_n(u)) = \tau_n(u) = \tilde{\tau}_n(\tilde{u}).
\end{eqnarray*}
Hence $(\tilde{\tau}_n)_{n\in \bb{N}}$ induces a 
tracial state $\tilde{\tau} : \cl B_X\to \bb{C}$. 
It is now straightforward that 
$\tilde{\tau}\circ q_X = \tau$, and hence (\ref{eq_pabtau}) 
implies that $p(a,b|x,y) = \tilde{\tau}(p_{x,a}p_{y,b})$ 
for all initial finite words $x,y,a,b$ of equal length. 

(a) If $p(a,b|x,y)=\tau_\omega( \pi(p_{x,a}p_{y,b}))$, then $p_n$, $n\in\mathbb N$, are of approximately quantum type by \cite[Theorem 2.2]{paulsen-rahaman}. Conversely, if $p\in\cl C_{\rm qag}^{\rm bs}$ then by Theorem \ref{th_trBXA} there is an amenable trace $\tau: \cl B_{X,X}\to\mathbb C$ 
such that $p(a,b|x,y)=\tau(e_{x,a}e_{y,b})$. 
By a result of E. Kirchberg  in \cite[Proposition 3.2]{kirchberg}, there exists a *-homomorphism $\pi: \cl B_{X,X}\to\cl R^\omega$ such that $\tau=\tau_\omega\circ\pi$. 
As in the proof of the implication (i)$\Rightarrow$(iii),
$\tau$  annihilates the ideal $\cl I$. By faithfulness of $\tau_\omega$, the representation 
$\pi$ annihilates the ideal $\cl I$ and hence determines a *-representation $\pi: \cl B_X\to\cl R^\omega$. 

The proofs of (b) and  (c) are similar, following the arguments in the proof of (b) and (c) in Theorem \ref{th_trBXA}. 
 \end{proof}


\section{Factorisable maps from bisynchronous correlations}\label{s_fmsc}

In this section, we show that graded bisynchronous correlations give rise to factorisable maps over 
infinite dimensional Hilbert spaces. We start with some definitions.
Let $H$ be a Hilbert space. 
Following \cite{hm1} (see also \cite{ctt}), we will call 
a linear map $\Phi:\cl B(H)\to\cl B(H)$ \emph{factorisable} 
if there exist a tracial von Neumann algebra $(\cl N,\tau)$ and a unitary $D\in\cl B(H)\bar\otimes \cl N$ such that $$ \Phi(T)=(\id\otimes\tau)(D^*(T\otimes 1)D)$$ for any $T\in\cl B(H)$; the algebra $\cl N$ is called an \emph{ancilla}
(we point out that in \cite{ctt} these maps were called \emph{exact factorisable}, 
but we have decided to shorten it for the sake of brevity). We remark that both the ancilla $\cl N$ and the unitary operator $D$ are not unique. 
We note also that every map that has the latter 
form is necessarily normal, unital, completely positive and trace preserving in that 
$\Tr(\Phi(T))=\Tr(T)$ for all $T\geq 0$, where we have set,
as usual, $\Tr(T)=\infty$ in the case where $T$ is not an element of the trace class $\cl S_1(H)$. 

In \cite{ctt} different types of factorisable maps have been defined according to the type of admissible ancillas 
(see \cite{paulsen-rahaman} for the finite-dimensional case). 
More precisely, we say that $\Phi:\cl B(H)\to\cl B(H)$ is 
\begin{itemize}
    \item[(i)] \emph{locally} factorisable if it admits an abelian ancilla;
    \item[(ii)] \emph{quantum} factorisable if it admits a finite-dimensional ancilla;
    \item[(iii)] \emph{approximately quantum} factorisable if it admits an ancilla that can be embedded in a matricial ultraproduct $\cl M^{\frak \omega}$ in a trace preserving manner. 
\end{itemize}
    We will say $\Phi:\cl B(H)\to\cl B(H)$ is \emph{stably approximately quantum} ($\rm qas$) factorisable if there exists a trace preserving  embedding of the ancilla $\cl N$ into $\cl M^{\frak \omega}$ such that $\tau_{\frak \omega}$ is stable on the image of $\cl N$.
For $\rm t\in\{ loc, q, qas, qa\}$, we thus refer to $\Phi$ as being $\rm t$-factorisable, and we say that 
$\Phi$ is $\rm qc$-factorisable if it is is factorisable. 

Let $X=(X_n)_{n\in\mathbb N}$ be an inductive family of sets, where $X_{n+1}=X_n\times[d_n]$, $n\in\mathbb N$, and write
$\{\epsilon_{s,t}:s,t\in S\}$ for the canonical matrix units in $M_S$. In particular, the matrices in $M_S$ are viewed as operators on $ L^{2}(S,\mu_S)$ equipped with the uniform measure via the identification $ \epsilon_{s,t} \mapsto |e_s\rangle\langle e_t|$, where $ e_s = |S|^{1/2} \delta_s$.

Consider the inductive system
\[
M_{X_1}\stackrel{\jmath_{1,2}}{\longrightarrow} 
M_{X_2}\stackrel{\jmath_{2,3}}{\longrightarrow} \cdots
\stackrel{\jmath_{n-1,n}}{\longrightarrow}
M_{X_n}\stackrel{\jmath_{n,n+1}}{\longrightarrow}\cdots,
\]
where, under the identification $M_{X_{n+1}}\simeq M_{X_n}\otimes M_{d_n}$,
the connecting maps are given by
\[
\jmath_{n,n+1}(u)=u\otimes q_{d_n}, \ \mbox{ where } 
q_{d_n}=\frac{1}{d_n}\sum_{x,y\in[d_n]}\epsilon_{x,y}
= |1_{[d_n]}\rangle\langle 1_{[d_n]}|,
\]
with $1_{[d_n]} = (1,\dots,1)\in L^2([d_n],\mu_{[d_n]})$.
Equivalently,
\[
\jmath_{n,n+1}(\epsilon_{x,y})
=
\frac{1}{d_n}\sum_{x',y'\in[d_n]}\epsilon_{xx',yy'}; 
\]
we note that $\jmath_{n,n+1}$ is an injective *-homomorphism. 
For brevity, we denote the identity matrix in $M_{X_n}$ by $I_n$.

\begin{lemma}
Let $X=(X_n)_{n\in\mathbb N}$ be an inductive family of sets. 
Then
\[
\varinjlim (M_{X_n},\jmath_{n,n+1})_{n\in \bb N}
\cong
\mathcal K(L^2(\Omega_X)).
\]
\end{lemma}

\begin{proof} 
Let $\Omega_n=\prod_{k\geq n} [d_k]$ and  $\mu_n=\otimes_{k\geq n}\mu_{d_k}$. Under the identification
\( 
L^2(\Omega_{X})\simeq L^2(X_n,\mu_{X_n})\otimes L^2(\Omega_n,\mu_{n}),
\) 
we have
\( 
\jmath_{n,\infty}(u)=u\otimes \tilde{q}_n\), with  
\(\tilde{q}_n=|1_n\rangle\langle 1_n|\)
where $1_n$ denotes the constant function in $L^2(\Omega_n,\mu_n)$.  Hence each
$\jmath_{n,\infty}(u)$ is a finite-rank operator, so the inductive limit is contained in
$\mathcal K(L^2(\Omega_X))$. Conversely, functions depending only on finitely many coordinates are dense in
$L^2(\Omega_X)$. It follows that every rank-one operator on $L^2(\Omega_X)$ is a norm limit
of operators of the form
\( 
|\xi\rangle\langle \eta|\otimes \tilde{q}_n
=
\jmath_{n,\infty}(|\xi\rangle\langle \eta|)\),
\( \xi,\eta\in L^2(X_n,\mu_{X_n}),
\)
and hence every finite-rank operator belongs to the norm closure of
$\bigcup_n \jmath_{n,\infty}(M_{X_n})$. Since the compact operators are the norm closure of the
finite-rank operators, the result follows.
\end{proof}

\subsection{Factorisable maps via Cantor quantum permutations}

Let $(\Gamma_n)_{n\in\mathbb N}$ be an inductive family of bisynchronous correlations that arise from a Cantor correlation $\Gamma: C(\Omega_A\times\Omega_A)\to C(\Omega_X\times \Omega_X)$.
Let $p$ be the bisynchronous correlation with $\Gamma = \Gamma_p$, and $\tau$ be the tracial state $\tau: \cl B_X\to\mathbb C$ associated to $p$ via Theorem \ref{th_trBX}. In particular,
$p(a,b|x,y)=\tau(p_{x,a}p_{y,b})$
for all initial words $x$, $y$, $a$, $b$ of length $n$. 
Following \cite{paulsen-rahaman}, to each $\Gamma_n$ we associate the map $\Phi_n:M_{X_n}\to M_{X_n}$, given by
$$\Phi_n(\epsilon_{x,y})=\sum_{a,b\in X_n} p(a,b|x,y) \epsilon_{a,b},  \ \ \ x,y\in X_n.$$
As $\sum_{x,y\in X_n}p(a,b|x,y)=\sum_{a,b\in X_n}p(a,b|x,y)=1$ for all $x$, $y$, $a$, $b\in X_n$, a 
straightforward verification shows that $\Phi_n$ is unital and trace preserving (see \cite{paulsen-rahaman}). 
The operator $D_n := [p_{x,a}]_{x,a\in X_n}$, 
considered naturally as an element of 
$M_{X_n}(\pi_{\tau}(\cl B_{X_n})'')$ (where $\pi_{\tau}$ is the 
GNS representation arising from $\tau$) is unitary and
$$\Phi_n(T)=(\id\otimes \tau)(D_n^*(T\otimes 1)D_n), \ \ \  T\in M_{X_n},$$
showing that $\Phi_n$ is a factorisable map.

\begin{lemma}\label{l_inductiverelPhi}
The family $(\Phi_n)_{n\in\mathbb N}$ of maps is inductive
in that
$$\Phi_{n+1}\circ \jmath_{n,n+1}=\jmath_{n,n+1}\circ\Phi_n,
\ \ \ n\in\mathbb N.$$
\end{lemma}
\begin{proof} 
We have that 
\begin{eqnarray*}
&&(\Phi_{n+1}\circ\jmath_{n,n+1})(\epsilon_{x,y})=\frac{1}{d_n}\Phi_{n+1}\left(\sum_{x',y'\in [d_n]} \epsilon_{xx',yy'}\right)\\
&&=\frac{1}{d_n}\sum_{x',y'\in [d_n]}\sum_{aa', bb'\in X_{n+1}}\tau(p_{xx',aa'}p_{yy',bb'})\epsilon_{aa',bb'}\\
&&=\frac{1}{d_n} \hspace{-0.05cm}
\sum_{aa',bb'\in X_{n+1}}\tau(p_{x,a}p_{y,b}) \epsilon_{aa',bb'}
=\sum_{a,b\in X_n}\tau(p_{x,a}p_{y,b})\jmath_{n,n+1}(\epsilon_{a,b})\\
&&=\jmath_{n,n+1}(\Phi_n(\epsilon_{x,y})).
    \end{eqnarray*}
\end{proof}

Let $\Phi:\mathcal K(L^2(\Omega_X))\to \mathcal K(L^2(\Omega_X))$ be the canonical map induced by the inductive sequence $(\Phi_n)_{n\in\mathbb N}$; thus, 
\begin{equation} \label{eq:def_factor}
\Phi\circ \jmath_{n,\infty}=\jmath_{n,\infty}\circ \Phi_n, \quad n \in \bb N.
\end{equation}
Set $\Phi_\Gamma:=\Phi^{**}$ to be the extension to the bidual; we have that $\Phi_\Gamma$ is a normal completely positive map on $\mathcal B(L^2(\Omega_X))$.

For $n\in\mathbb N$ and $x,y\in X_n$, set
\[
\mathcal E_{x,y}:=\jmath_{n,\infty}(\epsilon_{x,y})\in \mathcal K(L^2(\Omega_X)).
\]
Then each $\mathcal E_{x,y}$ has rank one, their linear span,
over all $n$ and all $x,y\in X_n$, is dense in $\mathcal K(L^2(\Omega_X))$, and
\[
\mathcal P_n:=\sum_{x\in X_n}\mathcal E_{x,x}=\jmath_{n,\infty}(I_{X_n})
\] 
is a projection. It is clear that the sequence $(\mathcal P_n)_n$ is increasing and $\bigvee_n \mathcal P_n=I$.

By virtue of Remark \ref{r_qX}, we write
\( 
D_n=\sum_{x,a\in X_n}\epsilon_{x,a}\otimes p_{x,a},
\)
considered as a unitary element of $M_{X_n}\otimes \cl B_{X}$, $n\in \bb{N}$.



\begin{lemma}\label{L:compat} 
The family $(D_n)_{n\in \bb{N}}$ satisfies the conditions
\begin{equation}\label{eq:compat}
D_{n+1}(\jmath_{n,n+1}(I_{X_n})\otimes 1)=(\jmath_{n,n+1}\otimes \id)(D_n)
=(\jmath_{n,n+1}(I_{X_n})\otimes 1)D_{n+1}, 
\end{equation}
for all $ n\in \bb{N}$.
\end{lemma}

\begin{proof}
Fix $n\in \bb{N}$. Using the identity 
$\jmath_{n,n+1}(I_{X_n})
= \frac{1}{d_n} \sum_{x\in X_n} \sum_{x',y'\in [d_n]} \epsilon _{xx',xy'}$ and the definition of $D_{n+1}$, we obtain
\begin{align*}
D_{n+1}(\jmath_{n,n+1}(I_{X_n})\otimes 1)
&=
\frac{1}{d_n} \sum_{x\in X_n} \sum_{x',y'\in [d_n]} D_{n+1}(\epsilon _{xx',xy'}\otimes 1) \\
&=
\frac{1}{d_n} \sum_{x,z\in X_n} \sum_{x',y',z'\in [d_n]} \epsilon _{zz',xy'}\otimes p_{zz',xx'} \\
&=
\sum_{x,z \in X_n} \jmath_{n,n+1}(\epsilon_{z,x})\otimes p_{z,x} 
=
(\jmath_{n,n+1}\otimes \id)(D_n).
\end{align*}
On the other hand, expanding \(D_{n+1}\) and using the defining relations of
\(\mathcal B_{X}\), we have
\[
(\jmath_{n,n+1}(I_{X_n})\otimes 1)D_{n+1}
=
\frac{1}{d_n}
\sum_{x,a\in X_n}\sum_{x',a'\in[d_n]}
\epsilon_{xx',aa'}\otimes
\sum_{y'\in[d_n]}p_{xy',aa'}.
\]
The last sum $\sum_{y'\in[d_n]}p_{xy',aa'}$ is independent of \(a'\) and equals 
$p_{x,a}$. Hence the expression on the previous line coincides with 
\[
\frac{1}{d_n}
\sum_{x,a\in X_n}\sum_{x',a'\in[d_n]}
\epsilon_{xx',aa'}\otimes p_{x,a}
=
(\jmath_{n,n+1}\otimes \id)(D_n).
\]
\end{proof}

Let $\cl B_X$ be faithfully represented on a Hilbert space $H$.
For $\xi\in \operatorname{Ran}(\mathcal P_n)\otimes H$, define $ D : L^2(\Omega_X)\otimes H \to L^2(\Omega_X)\otimes H$ by letting
\begin{equation}\label{eq:defD}
D\xi:=(\jmath_{n,\infty}\otimes \id)(D_n)\,\xi.
\end{equation}
By Lemma~\ref{L:compat}, this definition is independent of $n$, hence $D$ is well-defined on $\bigcup_{n\in\mathbb N}\operatorname{Ran}(\mathcal P_n)\otimes H$. Moreover, since each $D_n$ is unitary, we have 
that $\|D\xi\|=\|\xi\|$ for all $\xi$ in the latter dense subspace, and therefore $D$ extends uniquely to a bounded operator on $L^2(\Omega_X)\otimes H$. Finally, by (\ref{eq:defD}), 
\begin{equation}\label{eq:D_on_Pn}
D(\mathcal P_n\otimes 1)
=
(\jmath_{n,\infty}\otimes \id)(D_n),
\qquad n\in\mathbb N.
\end{equation}
We say that the sequence $(D_n)_{n\in \bb{N}}$ is \emph{associated with} $D$. 

For a $C^*$-algebra $A$ write $M(A)$ for the multiplier algebra of $A$.
\begin{lemma}\label{L:cantor_perm} 
The operator $D$ belongs to $M(\cl K(L^2(\Omega_X))\otimes \cl B_X)$, is unitary, and satisfies the identities
\[
D(\mathcal P_n\otimes 1)=(\mathcal P_n\otimes 1)D,
\qquad n\in\mathbb N.
\]
\end{lemma}

\begin{proof}

We prove first the commutation relation. Applying \(\jmath_{n+1,\infty}\otimes\id\) to the second identity in
\eqref{eq:compat}, we get
\[
(\mathcal P_n\otimes 1)(\jmath_{n+1,\infty}\otimes \id)(D_{n+1})
=
(\jmath_{n,\infty}\otimes \id)(D_n).
\]
Iterating this identity gives
\begin{equation}\label{iter}
(\mathcal P_n\otimes 1)(\jmath_{m,\infty}\otimes \id)(D_m)
=
(\jmath_{n,\infty}\otimes \id)(D_n), \ \ \ m\ge n.
\end{equation}
Since 
\( 
D(\mathcal P_m\otimes 1)
=
(\jmath_{m,\infty}\otimes \id)(D_m),
\)
we obtain for all $m\ge n$ that
\[
(\mathcal P_n\otimes 1)D(\mathcal P_m\otimes 1)
=
(\jmath_{n,\infty}\otimes \id)(D_n)
=
D(\mathcal P_n\otimes 1).
\]
Letting \(m\to\infty\), and using the fact that \(\mathcal P_m\uparrow 1\) strongly, 
we obtain
$(\mathcal P_n\otimes 1)D=D(\mathcal P_n\otimes 1)$.

To see the first claim it is enough to show that for $T\in\cl K(L^2(\Omega_X))$, $D(T\otimes 1)$ and $(T\otimes 1)D\in \cl K(L^2(\Omega_X))\otimes\cl B_X$. 
As $\cl P_n\to_{n\to\infty} 1$ strongly, and hence $\cl P_n T\to_{n\to\infty} T$  and 
$T\cl P_n\to_{n\to\infty} T$ in norm, we obtain that 
$$D(T\otimes 1)=\lim_{n\to\infty} D(\cl P_n\otimes 1)(T\otimes 1) = \lim_{n\to\infty} ((\jmath_{n,\infty}\otimes \id)(D_n)(T\otimes 1),$$
where the limits are in the norm topology. It follows that 
$D(T\otimes 1)\in \cl K(L^2(\Omega_X))\otimes\cl B_X$;
similarly, $(T\otimes 1)D\in \cl K(L^2(\Omega_X))\otimes\cl B_X$.
Finally, since $D_n$ is unitary for each $n$, we have
\[
D^*D(\mathcal P_n\otimes 1)
=
(\jmath_{n,\infty}\otimes \id)(D_n^*D_n)
=
\mathcal P_n\otimes 1,
\]
and similarly $DD^*(\mathcal P_n\otimes 1)=\mathcal P_n\otimes 1$. As
$\bigvee_n \mathcal P_n=1$, it follows that $D$ is unitary.
\end{proof}

The unitary operator $D \in  M(\cl K(L^2(\Omega_X))\otimes \cl B_X)$ defined in \ref{eq:defD} will 
be called \emph{the universal magic unitary over $X$}. 
We recall that, given C*-algebras $\cl A$ and $\cl B$, where $\cl B$ is unital, any non-degenerate *-homomorphism $\rho : \cl A\to \cl B$ extends uniquely to a (unital) *-homomorphism (denoted in the same way) $\rho : M(\cl A) \to \cl B$ where, here and in the sequel, $M(\cl A)$ stands for the 
multiplier algebra of $\cl A$ (see e.g. \cite[Proposition 2.1]{lance}). 

\begin{definition} \rm
Let $X$ be an inductive family of sets and $D\in M(\cl K(L^2(\Omega_X))\otimes\cl B_X)$ be the universal magic unitary over $X$. Let $\cl N $ be a von Neumann algebra acting on $H$ and $\rho: \cl B_X\to\cl N$ be a $*$-representation of $\cl B_X$. The unitary operator 
$(\id\otimes\rho)(D)$ will be called a \emph{quantum permutation on the Cantor space} $\Omega_X$  (associated to $\rho$) or simply a \emph{Cantor quantum permutation}. 
\end{definition}
In what follows we shall often use the same letters $D$  and $D_n$ for a Cantor quantum permutation $(\id\otimes\rho)(D)$ and for $(\id_{M_{X_n}}\otimes\rho)(D_n)$ and call the latter sequence $(D_n)_{n\in \mathbb N}$ for a sequence of unitary operators associated to $D$.

\begin{remark}\label{r_aboutD} \rm 
\begin{itemize}
\item[(i)] 
If $D$ is a Cantor quantum permutation and $(D_n)_{n\in \mathbb N}$ is the sequence of unitary operators associated to $D$ then   $D= \lim_{n\to\infty}\,D_n$
in the strong operator topology. 
\item[(ii)] It is straightforward to check that $D$ is a Cantor quantum permutation associated to $\rho$ if and only if 
the identities
$(\cl E_{s,t}\otimes 1)D(\cl E_{x,y}\otimes 1)=\cl E_{s,y}\otimes \rho(p_{t,x})$  hold for all initial words $x$, $y$, $s$, $t$ of length $n$.
\end{itemize}
\end{remark}

Let $ \rm t\in \{loc,q,qa,qas, qc\}$. 
We say that a map $ \Phi: \cl B(L^{2}(\Omega_X)) \to  \cl B(L^{2}(\Omega_X))$ is \emph{$\rm t$-factorisable via a quantum permutation on $\Omega_X$} 
if $\Phi$ is $\rm t$-factorisable and the unitary $D \in \cl B(L^2(\Omega_X))\bar \otimes \cl N$ can be chosen to be  a Cantor quantum permutation.

\begin{theorem}\label{factormap}
Let $\rm t\in\{ loc, qa, qc\}$ and $p$ be a 
graded bisynchronous correlation of type ${\rm t}$.
Then $\Phi_{\Gamma_p}$ is a $ \rm t$-factorisable map via 
a quantum permutation $D$ on $\Omega_X$. 
If $\Gamma_p$
is of $\rm q$-type, then  $\Phi_{\Gamma_p}$ is $\rm qas$-factorisable. 

Conversely,
if $\Phi: \cl B(L^2(\Omega_X))\to\cl B(L^2(\Omega_X))$ is $\rm t$-factorisable via a quantum permutation $D$ on $\Omega_X$ and we write
$\Phi(T)=(\id\otimes\tau)(D^*(T\otimes 1)D)$, then
$p(a,b|x,y) := \tau(p_{x,a}p_{y,b})$ for all initial words $x$, $y$, $a$, $b$ of the same length, is a graded bisynchronous 
correlation of type ${\rm t}$
such that $\Phi=\Phi_{\Gamma_p}$; further, if $\Phi$ is $\rm qas$-factorisable then  
$\Gamma_p$ is a Cantor $\rm q$-correlation.
\end{theorem}

\begin{proof}
We treat the $\rm qc$-case; the other cases are similar. 
Let $\Gamma$ be a bisynchronous Cantor quantum commuting correlation and let $\tau$ be the associated trace on $\cl B_X$ via Theorem \ref{th_trBX}. 
Applying the GNS construction to $\tau$, we obtain a Hilbert space $H$, 
a *-representation $\pi_{\tau} : \cl B_X\to \cl B(H)$ and 
a tracial von Neumann algebra $(\cl N,\tau_{\cl N})$, canonically arising from 
$(\cl B_X,\tau)$ (where $\cl N = \pi_{\tau}(\cl B_X)''$). 
We continue to denote the canonical generators of $\cl N$ by 
$p_{x,a}$, and note that $p(a,b|x,y)=\tau_{\cl N}(p_{x,a}p_{y,b})$ for all $x,y,a, b$ of the same length. Similarly, we continue to denote by $D$ its image under 
${\rm id}\otimes\pi_{\tau}$. 
Taking $x$, $y$, $s$, $t$ of length $n$, we obtain
$$D^*(\cl E_{x,y}\otimes 1)D= \sum_{|s| = |t|= n}\cl E_{s,t}\otimes p_{x,s}p_{y,t}$$
and hence
$$(\id\otimes\tau_{\cl N})(D^*(\cl E_{x,y}\otimes 1)D)=\sum_{|s| = |t|= n}\tau_{\cl N}(p_{x,s}p_{y,t})\cl E_{s,t}.$$
On the other hand,
\begin{eqnarray*}
\Phi_\Gamma(\cl E_{x,y})
& = &  
\jmath_{n,\infty}(\Phi_n(\epsilon_{x,y}))\\
& = & \jmath_{n,\infty}\left(\sum_{|s| = |t|= n}\tau_{\cl N}(p_{x,s}p_{y,t})\epsilon_{s,t} \right)
= \sum_{|s| = |t|= n}\tau_{\cl N}(p_{x,s}p_{y,t})\cl E_{s,t},
\end{eqnarray*}
giving 
\begin{equation}\label{fact}
\Phi_\Gamma(\cl E_{x,y})= (\id\otimes\tau_{\cl N})(D^*(\cl E_{x,y}\otimes 1)D),
\end{equation}
that is, $\Phi_\Gamma$ is qc-factorisable. 

Conversely, let the map 
$\Phi:\cl B(L^2(\Omega_X))\to \cl B(L^2(\Omega_X))$ be $\rm qc$-factorisable via a quantum permutation on $\Omega_X$; thus, 
there exists a tracial von Neumann   algebra ($\cl N,\tau_{\cl N})$ and projections $\tilde p_{x,y}\in\cl N$, $x,y\in X_n$, $n\in\mathbb N$, such that the assignment $\rho(p_{x,y})=\tilde p_{x,y}$, $x,y\in X_n$, $n\in \bb{N}$, extends to a *-representation of $\cl B_X$ into $\cl N$ and $\Phi(T)=(\id\otimes \tau_{\cl N})(\tilde{D}^*(T\otimes 1)\tilde{D})$, where 
$\tilde{D}\in\cl B(L^2(\Omega_X))\bar\otimes \cl N$ is a unitary which satisfies 
$(\cl E_{s,t}\otimes 1)\tilde{D}(\cl E_{x,y}\otimes 1)=\cl E_{s,y}\otimes \tilde{p}_{t,x}$ for all initial words $x$, $y$, $s$, $t$ of equal length (see Remark \ref{r_aboutD} (ii)). 
Letting $p(a,b|x,y)=\tau_{\cl N}(\tilde p_{x,a}\tilde p_{y,b})$ for all $x$, $y$, $a$, $b\in X_n$, we obtain a quantum commuting correlation $\Gamma_n: \cl D_{X_n,X_n}\to\cl D_{X_n,X_n}$. Moreover, $(\Gamma_n)_{n\in\mathbb N}$ is graded: 
\begin{eqnarray*}
    &&\Gamma_{n+1}(\iota_{n,n+1}^{(2)}(\delta_a\otimes\delta_b))
    = \Gamma_{n+1}\left(\sum_{|a'| = |b'| = 1}\delta_{aa'}\otimes\delta_{bb'}\right)\\&&
 = \sum_{|x| = |y| = n} \sum_{|x'| = |y'| = 1} \sum_{|a'| = |b'| = 1}\tau(\tilde p_{xx',aa'}\tilde p_{yy',bb'})\delta_{xx'}\otimes\delta_{yy'}\\
 && = \sum_{|x| = |y| = n} \sum_{|x'| = |y'| = 1}\tau(\tilde p_{x,a}\tilde p_{y,b})\delta_{xx'}\otimes\delta_{yy'}
 =\iota_{n,n+1}^{(2)}(\Gamma_n(\delta_a\otimes\delta_b)).
\end{eqnarray*}
Let $\Gamma : C(\Omega_X\times\Omega_X)\to C(\Omega_X\times\Omega_X)$ be the associated quantum commuting correlation. 
By Theorem \ref{th_trBX}, $\Gamma$ is bisynchronous. 
After reversing the previous arguments we obtain $\Phi_\Gamma=\Phi$.
\end{proof}


\subsection{Integral formulas}\label{ss_intformu}

Given a graded bisynchronous correlation $p$ over the pair $(X,X)$, 
the unital channel $\Phi_{\Gamma_p}$ was defined via (\ref{eq:def_factor}) as a consequence of Lemma \ref{l_inductiverelPhi}, using a canonical coordinatisation of $\cl K(L^2(\Omega_X))$. 
In the remainder of this section, we show that $\Phi_{\Gamma_p}$ can be defined in a coordinate-free manner 
via an explicit integral formula. 
Fix a graded bisynchronous correlation $p$ over $(X,X)$ and recall from \cite{gb} that the 
unital completely positive map $\Gamma_p : C(\Omega_X\times\Omega_X)\to L^{\infty}(\Omega_X\times\Omega_X)$ is canonically and bijectively associated with a bipartite information channel 
on $\Omega_X$, that is, a family $p = (p_{x,y})_{x,y\in \Omega_X}$ of regular Borel probability 
measures on $\Omega_X\times\Omega_X$ which is measurable in that, for every Borel set $E\subseteq \Omega_X\times\Omega_X$, the scalar-valued function $(x,y)\mapsto p_{x,y}(E)$ is measurable
(we refer the reader to \cite{kakihara} for details on information channels in the measurable setting). 
The correspondence between $\Gamma_p$ and $p$ is given via the identities
\begin{align}\label{eq_chop}
\langle \Gamma_p(g),f\rangle =
\int_{\Omega_X\times\Omega_X}\int_{\Omega_X\times\Omega_X}
g(a,b)f(x,y)dp_{x,y}(a,b)d\mu_X(x)d\mu_X(y),
\end{align}
where 
$g\in C(\Omega_X\times\Omega_X)$ and $f\in L^1(\Omega_X\times\Omega_X)$, 
and the duality in the left hand side of (\ref{eq_chop}) is that between 
$L^1(\Omega_X\times\Omega_X)$ and its dual space $L^{\infty}(\Omega_X\times\Omega_X)$. 
Identity (\ref{eq_deff0}) easily implies that, if $[z]$ denotes the cylinder in $\Omega_X$ based in the finite 
word $z$, then 
\begin{equation}\label{eq_sigmawithpn}
p_n(a,b|x,y) = |X_n|^2 \int_{[x]\times [y]} p_{s,t}([a]\times [b])d\mu_X(s) d\mu_X(t),
\end{equation}
for all initial words $x,y,a,b$ of length $n$. 

For $k\in L^2(\Omega_X\times\Omega_X)$, let $T_k \in \cl B(L^2(\Omega_X))$ be the 
Hilbert-Schmidt operator, given by 
$$(T_k\xi)(x) = \int_{\Omega_X} k(x,y)\xi(y)d\mu_X(y), \ \ \ \xi\in L^2(\Omega_X), x\in \Omega_X.$$
and for Hilbert-Schmidt operators $T$, $S$ on $L^2(\Omega_X)$ set $\langle T, S\rangle:=\Tr(TS)$. 
Fix $n\in \bb{N}$ and let $k,h\in L^2(\Omega_X\times\Omega_X)$ 
have the form 
$$k = \iota_{n,\infty}(\delta_a)\otimes \iota_{n,\infty}(\delta_b) \ 
\mbox{ and } \ 
h = \iota_{n,\infty}(\delta_x)\otimes \iota_{n,\infty}(\delta_y)$$
for some $x,y,a,b\in X_n$. 
Noting  that if $x,y\in X_n$ then
$\iota_{n,\infty}(\delta_x)=\chi_{[x]}\in C(\Omega_X)$ and 
$$j_{n,\infty}(\epsilon_{x,y})=\epsilon_{x,y}\otimes |1_n\rangle\langle 1_n|=|X_n||\chi_{[x]}\rangle\langle\chi_{[y]}|,$$ 
where $|1_n\rangle$ is the constant function 1 on $\Omega_n=\prod_{k=n}^\infty[d_k]$, we have $j_{n,\infty}(\epsilon_{x,y})=|X_n|T_k$ and $\Tr(j_{n,\infty}(\epsilon_{x,x}))=|X_n|\|\chi_{[x]}\|_2^2=1$.

As $\Phi_{\Gamma_p}(\jmath_{n,\infty}(\epsilon_{a,b}))=\sum_{x,y\in X_n}p_n(a,b|x,y)\jmath_{n,\infty}(\epsilon_{x,y})$,
$$\langle \Phi_{\Gamma_p}(\jmath_{n,\infty}(\epsilon_{a,b})), \jmath_{n,\infty}(\epsilon_{x,y})^*\rangle=p_n(a,b|x,y)\Tr(j_{n,\infty}(\epsilon_{x,x}))=p_n(a,b|x,y).$$
Therefore, 
\begin{align*}
\langle \Gamma_p(k),h\rangle 
&= 
\frac{1}{|X_n|^2} 
p_n(a,b|x,y)\\
& = \frac{1}{|X_n|^2}\left\langle\Phi_{\Gamma_p}(\jmath_{n,\infty}(\epsilon_{a,b})),
\jmath_{n,\infty}(\epsilon_{x,y})^*\right\rangle
= \left\langle\Phi_{\Gamma_p}(T_k),T_{h^*} \right\rangle,
\end{align*}
where $h^*(x,y)=h(y,x)$, 
and hence, by (\ref{eq_chop}), 
\begin{equation}\label{eq_intfor}
\left\langle\Phi_{\Gamma_p}(T_k),T_{h^*} \right\rangle = 
\int_{\Omega_X\times\Omega_X}\int_{\Omega_X\times\Omega_X} k(a,b)h(x,y)dp_{x,y}(a,b)d\mu_X(x) d\mu_X(y).
\end{equation}

In the remainder of this section we show that (\ref{eq_intfor}) holds for arbitrary $k,h\in L^2(\Omega_X\times \Omega_X)$. 
\begin{lemma}\label{lem_bic_l2}
    If $p$ is a graded Cantor bicorrelation, then the associated map $\Gamma_p$ is $\|\cdot\|_{2\rightarrow 2}$-contractive.  
\end{lemma} 
\begin{proof}
    First note that for each $n\in\bb{N}$ the associated channel $\Gamma_n$ of $p_n$ is $\|\cdot\|_{2\rightarrow 2}$-contractive. Indeed, for any $f\in L^2(X_n\times X_n)$
    \begin{align*}
        \|\Gamma_n(f)\|_2^2 & =
        \sum_{x,y}\Bigg\vert\sum_{a,b}p(a,b|x,y) f(a,b)\Bigg\rvert^2\\
        & \leq\sum_{x,y}\sum_{a,b}p(a,b|x,y)\lvert f(a,b)\rvert^2=\sum_{a,b}\lvert f(a,b)\rvert^2=\|f\|_2^2.
    \end{align*}
    Now since $\Gamma_n$ are all $\|\cdot\|_2$-contractions and the inclusions are trace-preserving one obtains that $\|\Gamma(f)\|_2\leq\|f\|_2$ for every $f\in C(\Omega_X\times \Omega_X)$. 
\end{proof}

\begin{lemma}\label{lem_fact_2cont}
    If $\Phi:\cl{B}(L^2(\Omega_X))\rightarrow\cl{B}(L^2(\Omega_X))$ is factorisable, then it is a contraction in the Hilbert-Schmidt norm. 
\end{lemma}
\begin{proof}
     Set $\Phi(u)=(\id\otimes\tau)(D^*(u\otimes 1)D)$, $u\in\cl{B}(L^2(\Omega_X))$. 
     Using the Schwarz inequality in the third line of the subsequent chain,
     for any $k\in L^2(\Omega_X\times \Omega_X)$ we have 
    \begin{align*}
        \|\Phi(T_k)\|_2^2 & = \|(\id\otimes\tau)(D^*(T_k\otimes1)D)\|^2_2\\
        & =\Tr((\id\otimes\tau)(D^*(T_k\otimes1)D)^*(\id\otimes\tau)(D^*(T_k\otimes1)D))\\
        & \leq \Tr((\id\otimes \tau)(D^*(T_k^*\otimes1)DD^*(T_k\otimes1)D)\\
        & =(\Tr\otimes\tau)(D^*(T_k^*T_k\otimes1)D)=(\Tr\otimes\tau)(T_k^*T_k\otimes1)
        =\Tr(T_k^*T_k)=\|T_k\|_2^2.
    \end{align*}
\end{proof}

\begin{proposition}
    If $p$ is a graded Cantor bicorrelation, then $$\langle\Phi_{\Gamma_p}(T_k),T_{h^*}\rangle= \int_{\Omega_X\times\Omega_X}\int_{\Omega_X\times\Omega_X} k(a,b)h(x,y)dp_{x,y}(a,b)d\mu_X(x) d\mu_X(y)$$
    for every $h,k\in L^2(\Omega_X\times \Omega_X)$.
\end{proposition}

\begin{proof}
First we show that (\ref{eq_chop}) holds for arbitrary $h,k\in L^2(\Omega_X\times \Omega_X)$. Let $h, k\in L^2(\Omega_X\times \Omega_X)$ and  $(k_n)_{n\in\bb{N}}$ be a sequence in $C(\Omega_X\times \Omega_X)$ converging to $k$ in $\|\cdot\|_2$-norm. 
Without loss of generality, we assume that $h$ takes non-negative values. 
By Lemma \ref{lem_bic_l2}, $\Gamma_p$ is 2-contractive and hence for every $n,m\in\bb{N}$, 
\begin{align*}
& \int_{\Omega_X\times \Omega_X}\int_{\Omega_X\times \Omega_X}\lvert k_n-k_m\rvert(a,b)dp(a,b|x,y)h(x,y)d\mu_X(x)d\mu_X(y)\\
& =\int_{\Omega_X\times \Omega_X} \Gamma_p(\lvert k_n-k_m\rvert)(x,y)h(x,y)d\mu_X(x)d\mu_X(y))\\
& = \langle \Gamma_p(\lvert k_n-k_m\rvert),h\rangle
\leq \|\Gamma_p(\lvert k_n-k_m\rvert)\|_2\| h\|_2\leq \|k_n-k_m\|_2\|h\|_2.
\end{align*}
It follows that the sequence $(k_n)_{n\in\bb{N}}$ is Cauchy in $L^1(\Omega_X\times \Omega_X,\nu)$, where $\nu$ is the Borel measure on $\Omega_X\times \Omega_X$, defined by letting
$$\nu(\alpha)=\int_{\Omega_X\times\Omega_X}
p(\alpha|x,y)h(x,y)d\mu_X(x)d\mu_X(y)$$
(note that the measure $\nu$ is finite due to the finiteness
of $\mu_X$ and the square integrability of $h$). 
 Write $g\in L^1(\Omega_X\times \Omega_X,\nu)$ for the limit of $(k_n)_{n\in\bb{N}}$ and note that there exists a subsequence $k_{n_j}$ converging pointwise to $k$ $\nu$-almost everywhere. On the other hand $(k_{n_j})$ converges to $k$ in $L^2(\Omega_X\times \Omega_X,\mu_X\otimes \mu_X)$ and hence there exists a further subsequence converging pointwise to $k$ $\mu_X\otimes \mu_X$-almost everywhere. Since $\nu\ll\mu_X\otimes\mu_X$ we obtain $g = k$ $\nu$-almost everywhere.
 
Next set $$\cl{T}=\{f\in L^2(\Omega_X\times \Omega_X):T_f\in \cup_{n\in\bb{N}}\jmath_{n,\infty}(M_{X_n})\},$$
and note that $\cl{T}$ is dense in $C(\Omega_X\times \Omega_X)$, and hence 
in $L^2(\Omega_X\times \Omega_X)$. 
As was shown before Lemma \ref{lem_bic_l2}, equation (\ref{eq_intfor}) 
holds for all functions $h,k\in \cl T$. 
By Lemma \ref{lem_fact_2cont}, (\ref{eq_intfor}) holds for all 
$h,k \in L^2(\Omega_X\times \Omega_X)$. 
The proof is complete in view of the first paragraph. 
\end{proof}


\subsection{Projective limit of free wreath products} \label{subs_projlims}

In the present subsection, we show that the universal C*-algebra for bisynchronous graded correlations arises as a projective limit of free wreath products of compact quantum groups \cite{bgs} and discuss the associated coaction on the algebra of continuous functions on the Cantor space as well as the implemented co-representation.

Before proceeding, we briefly present basic concepts and results 
from compact quantum group theory; we refer the reader to \cite{nesh} for a more detailed exposition.

A \textit{compact quantum group (CQG)} $\bb{G}$ is a pair $(C(\bb{G}), \Delta)$, where $C(\bb{G})$ is a unital ${\rm C}^{*}$-algebra and $\Delta: C(\bb{G})\rightarrow C(\bb{G})\otimes C(\bb{G})$ is a unital *-homomorphism, called the \emph{comultiplication}, such that:
\begin{itemize}
  \item[(i)] $(\mathrm{id}\otimes \Delta)\Delta = (\Delta\otimes \mathrm{id})\Delta$;
  \item[(ii)] 
  \( 
    \overline{\mathrm{span}}\bigl\{ \Delta(C(\bb{G}))(1\otimes C(\bb{G})) \bigr\}
    = C(\bb{G})\otimes C(\bb{G})
    =
    \overline{\mathrm{span}}\bigl\{ (C(\bb{G})\otimes 1)\Delta(C(\bb{G})) \bigr\}.
  \)
\end{itemize}
Every compact quantum group admits a unique state $\varphi_{\bb{G}}: C(\bb{G})\rightarrow \bb{C}$, called the \textit{Haar state}, such that
\( 
(\mathrm{id}\otimes \varphi_{\bb{G}})\Delta(a) = \varphi_{\bb{G}}(a)\,1
=
(\varphi_{\bb{G}}\otimes \mathrm{id})\Delta(a)\),
 \(a \in C(\bb{G}).
\)
If the Haar state is tracial, then $\bb{G}$ is said to be of \textit{Kac type}.

Given two compact quantum groups $\bb{G} = (C(\bb{G}), \Delta_{\bb{G}})$ and $\bb{H} = (C(\bb{H}), \Delta_{\bb{H}})$, a unital *-homomorphism $\varphi: C(\bb{G})\rightarrow C(\bb{H})$ is a \emph{morphism of compact quantum groups} if
\(
(\varphi\otimes \varphi)\circ \Delta_{\bb{G}} = \Delta_{\bb{H}}\circ \varphi.
\)
If $\varphi$ is surjective, we write $\bb H \subseteq \bb G$ and call $\bb H$ a subgroup of $\bb G$; if $\varphi$ is injective, we call $\bb G$ a quotient of $\bb H$; and if $\varphi$ is bijective, then $\bb{G}$ and $\bb{H}$ are \emph{isomorphic as compact quantum groups}.

Let $\bb G$ be a compact quantum group and $C$ a unital $C^*$-algebra.
A \emph{coaction} of $\bb G$ on $C$ is a unital $*$-homomorphism
$\alpha:C\to C\otimes C(\bb G)$ such that
$(\alpha\otimes\id)\circ\alpha=(\id\otimes\Delta)\circ\alpha$ and the
linear span of $\alpha(C)(1\otimes C(\bb G))$ is dense in
$C\otimes C(\bb G)$.

Let $\bb{G} = (C(\bb{G}), \Delta)$  be a compact quantum group and  $H$ be a Hilbert space. A \emph{co-representation} of $ \bb G$ on $ H$ is an invertible element $ U \in M(\cl K(H) \otimes C(\bb G))$ such that $ (\id \otimes \Delta_{})(U) = U_{12}U_{13}$. If moreover $ U$ is unitary, it is called a \emph{unitary co-representation}. Here we have used that $\id \otimes \Delta_{}$ is a non-degenerate *-homomorphism and thus extends to the multiplier algebra. Also  we have used the leg notation, that is   $U_{12}=U\otimes 1$ and
$U_{13}=(\mathrm{id}\otimes\sigma)(U_{12})$, $\sigma$ the flip of
$ C(\bb G)\otimes C(\bb G)$, both giving rise to non-degenerate *-homomorphisms and extend to the multiplier algebras. 

A \emph{compact matrix quantum group} is a triple $\bb G=(C(\bb G), \Delta, u)$ where $C(\bb G)$ is a unital ${\rm C}^{*}$-algebra, $\Delta: C(\bb G)\rightarrow C(\bb G)\otimes C(\bb G)$ is a unital *-homomorphism, and $u=[u_{i,j}]\in M_{n}(C(\bb G))$ is unitary such that: (i) $\Delta(u_{i,j}) = \sum_{k=1}^{n}u_{i,k}\otimes u_{k,j}$ for all $1\le i,j\le n$; (ii) $\bar u := [u_{i,j}^{*}]_{i,j}$ is invertible in $M_{n}(C(\bb G))$; (iii) the entries $\{u_{i,j}\}_{1\le i,j\le n}$ generate $C(\bb G)$ as a ${\rm C}^{*}$-algebra.
The unitary matrix $u$ is called the \emph{fundamental representation}. Every compact matrix quantum group is a compact quantum group.

A  \emph{Hopf *-algebra} is a quadruple $(\cl A,\Delta,\varepsilon,S)$ consisting of a unital *-algebra $\cl A$, a unital *-homomorphism $\Delta:\cl A\to \cl A\otimes \cl A$  such that $(\mathrm{id}\otimes \Delta)\Delta=(\Delta\otimes \mathrm{id})\Delta$ and  morphisms  $\varepsilon:\cl A\to\bb C$ (the \emph{co-unit}) and $S:\cl A\to \cl A$ (the \emph{antipode}) such that, for all $a\in \cl A$, (i) $(\varepsilon\otimes \mathrm{id})\Delta(a)=(\mathrm{id}\otimes \varepsilon)\Delta(a)=a$ and (ii) $m(S\otimes \mathrm{id})\Delta(a)=m(\mathrm{id}\otimes S)\Delta(a)=\varepsilon(a)\,1$,
where $m:\cl A\otimes \cl A\to \cl A$ is the multiplication map.

Given a compact matrix quantum group $\bb G$, the *-algebra generated by the matrix coefficients $\{u_{i,j}\}$ is denoted by $\mathcal{O}(\bb{G})$.  Then there exist a unital *-homomorphism $\varepsilon:\mathcal{O}(\bb G)\to\bb C$ and a unital anti-homomorphism $S:\mathcal{O}(\bb G)\to\mathcal{O}(\bb G)$ such that
\( 
\varepsilon(u_{i,j})=\delta_{i,j}\),  \(S(u_{i,j})=u_{ji}^{*}\),  \((1\le i,j\le n)\),
making $(\mathcal{O}(\bb G),\Delta,\varepsilon,S)$ a Hopf *-algebra.

Let $\bb{G} = (C(\bb{G}), [u_{x,y}]_{x,y\in X})$ and $\bb{H} = (C(\bb{H}), [v_{a,b}]_{a,b\in A})$ be  compact matrix quantum groups. The {\rm free product}  $\bb{G}\ast \bb{H}$ is the compact matrix quantum group denoted by $\bb{G}\ast \bb{H} $ with associated C*-algebra $C(\bb G * \bb H):= C(\bb{G})\ast C(\bb{H})$ and fundamental co-representation $u \oplus v$. 

The free wreath product of a compact quantum group $\bb G$ by the quantum permutation group $ S_{X}^{+}$ was introduced by  Bichon \cite{bichon04}. In \cite[Definition 5.5]{br_rob1} the authors extended the notion to any quantum subgroup of $ S_{X}^+$.

Let $\bb G =(C(\bb G) , u=[u_{x,y}]_{x,y\in X})$ be a compact matrix quantum group and $ \bb H = (C(\bb H), v=[v_{a,b}]_{a,b\in A})$ a subgroup of $S_{A}^+$. The \textit{free wreath product} of $ \bb G$ by $\bb H$  denoted by $ \bb G \wr_{*} \bb H $ is the compact matrix quantum group whose corresponding C*-algebra is the quotient of  $ C(\bb G^{*A}) *C(\bb H) $ by the two sided ideal generated by the elements
\[
u^{(a)}_{x,y}v_{a,b}- v_{a,b}u^{(a)}_{x,y}, \quad \quad  x,y\in X, a,b \in A,
\]
where $ u^{(a)}= [u_{x,y}^{(a)}]_{x,y\in X}$ is the $a$-th copy in $ \bb G^{*A}$ of the fundamental co-representation of $ \bb G$.  The fundamental co-representation $ [w_{ax,by}]_{(a,y),(b,y)\in A\times X}$ of $ \bb G \wr_{*} \bb H$ is given by 
\[
w_{ax,by}= u^{(a)}_{x,y}v_{a,b}, \quad \quad  x,y\in X, a,b \in A.
\]

\begin{remark} \label{re_wreath_incl}
\rm Assume that $\bb G$ is a compact matrix quantum group such that the co-unit $\varepsilon_{\bb G}$ extends to a *-homomorphism on $ C(\bb G)$. The canonical inclusion $\rho : C( \bb H) \hookrightarrow C(\bb G \wr_{*} \bb H)$, $ v_{a,b} \mapsto v_{a,b}$ is injective. Indeed, let $\pi :  C(\bb G \wr_{*} \bb H) \to C( \bb H)$ be the surjective *-homomorphism defined by $ \pi(w_{ax,by})= \delta_{x,y}v_{a,b}$. We show that $ \pi$ is a left inverse for $ \rho$. To show that  $\pi$ is well defined, consider the *-homomorphisms $ \pi_{a}: C(\bb G) \to C(\bb H); \; u_{x,y} \mapsto \delta_{x,y}1 $, $ a\in A$ that are the composition of the map $ \bb C \to C(\bb H); \; \lambda \mapsto \lambda 1$ and the co-unit and let $ *_{A}\pi_{a}: C(\bb G)^{*A} \to C(\bb H)$ be their free product. Take the free product map $ (*_{A}\pi_{a}) * \id : C(\bb G)^{*A} *_{1} C(\bb H) \to C(\bb H); \; u_{x,y}^{(c)} v_{a,b} \mapsto \delta_{a,b} v_{a,b}$ where $ a,b,c \in A$ and $ x,y \in X$. It is readily seen that this *-homomorphism annihilates the ideal generated by $ u_{x,y}^{(a)}v_{a,b}- v_{a,b} u_{x,y}^{(a)}$ and thus induces the unital *-homomorphism $\pi$. Now note that $ \pi \circ \rho = \id$ thus showing that $\rho $ is injective.   That is, $ \bb H$ is a quotient of $ \bb G \wr_{*} \bb H$. 
\end{remark}

Let $ (X_n)_{n \in \bb N}$ be an inductive family of finite sets, where 
$ X_{n+1}= X_{n}\times [d_{n}]$, $n\in \bb{N}$. We define a projective sequence of compact quantum groups. Set $ \bb W_1:= {\rm S}_{{X_1}}^{+}$ and define inductively $\bb W_{n+1}= {\rm S}_{d_n}^{+} \wr_{*} \bb W_{n}$, $ n \in \bb N$. That is, 
\[
\bb W_{n+1}=S^+_{d_n}\wr_*\Big(S^+_{d_{n-1}}\wr_*\big(\cdots\wr_*
\big(S^+_{d_1}\wr_*S^+_{X_1}\big)\cdots\big)\Big),\qquad n\in\bb N .
\]
To see that the construction above is well defined, note that if $\bb G$ and $\bb H$ are subgroups of the quantum permutation groups ${\rm S}_{N_1}^{+}$ and ${\rm S}_{N_2}^+$, respectively, then $ \bb G\wr_{*}\bb H$ is a subgroup of ${\rm S}_{N_1 N_2}^{+}$.
We have the following inductive sequence of C*-algebras
\begin{align*}
    \cdots \xrightarrow{} C(\bb W_{n}) \xrightarrow{\rho_n} C(\bb W_{n+1}) \xrightarrow[]{\rho_{n+1}} \cdots, 
\end{align*}
where the unital *-homomorphisms $ \rho_n $ take $ C(\bb W_n)$ to  its associated copy in the quotient of the free product.
Now by universality of the C*-algebra $  C({\rm S}_{d_n}^{+})$, the co-unit $ \varepsilon: \cl O({\rm S}_{d_n}^{+}) \to \bb C$ extends to a unital *-homomorphism on $  C({\rm S}_{d_n}^{+})$. By Remark \ref{re_wreath_incl}, the connecting maps $ \rho_{n}$ are unital *-monomorphisms. 
We show that the maps $ \rho_n : C(\bb W_n) \to C(\bb W_{n+1})$ are compact quantum group morphisms.  Let $ \Delta_n$ denote the comultiplication of $ \bb W_n$ and $v=(v_{x,a})_{x,a\in X_n}$ denote the fundamental representation. Then, 
noting that $\Delta_n(v_{x,a})=\sum_{z\in X_n}v_{x,a}\otimes v_{z,a}$ on the generators $v_{x,a}$, it is straightforward to see that $\Delta_{n+1}\circ \rho_n=(\rho_n\otimes \rho_n)\circ\Delta_n$.

Thus, the resulting inductive limit of the C*-algebras $ C(\bb W_n)$ gives rise to the projective limit compact quantum group (see \cite{ss} and \cite[Lemma 1.1]{bgs}) $\bb W_{\infty} := \lim\limits_{\longleftarrow}\bb W_{n}$. 
The comultiplication  $\Delta_{\infty}: C(\bb W_{\infty}) \to C(\bb W_{\infty})\otimes C(\bb W_{\infty}) $ satisfies;
\[
\Delta_{\infty} \circ \rho_{n,\infty} = (\rho_{n,\infty} \otimes \rho_{n,\infty}) \circ \Delta_n, \quad n \in \bb N.
\]

\begin{proposition}\label{p_BXSWinf}
There is a canonical *-isomorphism
$\cl B_{X_n} \cong C(\mathbb W_n)$, $n \in \bb N$. 
Moreover, $ \cl B_{X} \cong C(\bb W_{\infty}) $.
\end{proposition}

\begin{proof}
Recall that $ C(\bb W_{n})$ is the C*-algebra generated by the elements 
\[
w_{xx',aa'}=u_{x',a'}^{(x)}v_{x,a},\quad   x,a \in X_{n-1}, \;  x',a' \in [d_{n-1}]
\]
where $ u_{x',a'}^{(x)}$ are entries of the quantum permutation matrix $ u^{(x)}= [u^{(x)}_{x',a'}]_{x',a' \in [d_{n-1}]}$ of the $x$-th copy of $ C({\rm S}_{d_{n-1}}^{+})$ in the free product.

By the universal property of the C*-algebra $\cl B_{X_n}$, there exists a surjective unital *-homomorphism for each $n \in \bb N$, 
\[
\sigma_{n}: \cl B_{X_n} \to C(\bb W_{n}); \ \ p_{xx',aa'} \mapsto  u_{x',a'}^{(x)}v_{x,a},
\]
where $x,a \in X_{n-1}$, $  x',a' \in [d_{n-1}]$.

On the other hand, let $x,a\in X_{n-1}$ and $x',a'\in [d_{n-1}]$, and 
set 
$$u^{(x)}_{x',a'} := \sum_{c\in X_{n-1}}p_{xx',ca'} 
\ \mbox{ and } \ v_{x,a} := \sum _{a''\in [d_{n-1}]} p_{xx',aa''}$$ 
(note that $v_{x,a}$ is well-defined due to the coherence relations). 
We have  
\[
u^{(x)}_{x',a'} v_{x,a}=\sum _{a''\in [d_{n-1}]} \sum_{c\in X_{n-1}}p_{xx',ca'} p_{xx',aa''} = p_{xx',aa'}
\]
by the fact that each of the columns is a PVM. 
In addition, $(u^{(x)}_{x',a'})_{x',a'}$ is a magic unitary; indeed check first that 
\begin{align*}
\sum_{x'\in [d_{n-1}]} p_{xx',ca'} 
& = \frac{1}{d_{n-1}} \sum_{a' \in [d_{n-1}]}\sum_{x'\in [d_{n-1}]}p_{xx',ca'} \\ 
& = \frac{1}{d_{n-1}}\sum_{x'\in [d_{n-1}]}\sum_{a'\in [d_{n-1}]}p_{xx',ca'}
= \sum_{a'\in [d_{n-1}]}p_{xx',ca'},
\end{align*}
where in the first equality  we use independence on the variable $a'$ and in the third equality we used 
independence on the variable $x'$. Hence,
\begin{align*}
\sum_{x' \in [d_{n-1}]}u^{(x)}_{x',a'} 
& = 
\hspace{-0.15cm}\sum_{x'\in [d_{n-1}]} \sum_{c\in X_{n-1}}p_{xx',ca'}\\
& =  
\hspace{-0.15cm}\sum_{c\in X_{n-1}} \sum_{x'\in [d_{n-1}]} p_{xx',ca'}
= 
\hspace{-0.15cm}\sum_{c\in X_{n-1}} \sum_{a'\in [d_{n-1}]} p_{xx',ca'}=1.
\end{align*}
It is immediate that $ \sum_{a'\in [d_{n-1}]} u^{(x)}_{x',a'}=1$ and that $ (u^{(x)}_{x',a'})^{*}= u^{(x)}_{x',a'}=(u^{(x)}_{x',a'})^2$.  
We conclude that there exists a unital *-homomorphism $ \sigma'_n : C(\bb W_n) \to \cl B_{X_n}$ determined by the assignments $\sigma'_n(w_{xx',aa'}) = p_{xx',aa'}$. The maps $ \sigma_n$ and $\sigma'_n$ are inverses of each other and thus 
\( 
\cl B_{X_n} \cong C(\bb W_{n}).
\)
\end{proof}

For a function $f\in C(\Omega_X)$, let $M_f$ be the operator 
on $L^2(\Omega_X)$ of multiplication  by $f$. In the next theorem, we 
identify $C(\Omega_X)$ with the C*-algebra $\{M_f : f\in C(\Omega_X)\}$, 
acting on $L^2(\Omega_X)$. 

\begin{theorem}\label{thm:corepsandcoacts}
    Consider the compact quantum group $\bb W_{\infty}= (\cl B_X, \Delta_{\infty})$. Then we have the following: 
\begin{enumerate}
        \item The unitary operator  $D \in M(\cl K(L^2(\Omega_X))\otimes \cl B_X)$  defined in (\ref{eq:defD}) is a unitary co-representation of $\bb W_{\infty}$ on $ L^2(\Omega_X)$;
        \item There is a canonical coaction of the compact quantum group $ \bb W_\infty$  on $C(\Omega_X)$ denoted by 
$\alpha:C(\Omega_X)\to C(\Omega_X)\otimes\cl B_X$ and defined as
\[
\alpha(\chi_{[x]})=\sum_{a\in X_n}\chi_{[a]}\otimes p_{a,x},
\qquad x\in X_n,\ n\in\bb N;
\]

\item 
The map $\alpha$ is a coaction implemented by the unitary co-representation $D$ (see \cite{vaes}) in that 
\[
\alpha(M_f)=D(M_f\otimes 1)D^*, \quad f \in C(\Omega_X).
\]
    \end{enumerate}
\end{theorem}
\begin{proof}
    (i) Fix $n\in\bb N$. The magic unitary $ D_n=[p_{x,a}]_{x,a\in X_n}$ is the fundamental co-representation of the compact matrix quantum group $ \bb W_n$. We identify each $ p_{x,a}$  with its image in $ \cl B_X$ under $ \rho_{n,\infty}$.  For $x,a\in X_n$ we thus have
\[
(\id\otimes\Delta_\infty)(D_n)
=\sum_{x,a\in X_n}\epsilon_{x,a}\otimes
\sum_{b\in X_n}p_{x,b}\otimes p_{b,a}
=(D_{n})_{12}(D_{n})_{13}
\quad\text{in } M_{X_n}\otimes\cl B_X\otimes\cl B_X .
\]

We compute,
\begin{align*}
(\id\otimes\Delta_\infty)(D)(\cl P_n\otimes 1\otimes 1)
&=(\id\otimes\Delta_\infty)\big(D(\cl P_n\otimes 1)\big)\\
&=(\jmath_{n,\infty}\otimes\id\otimes\id)((D_{n})_{12}(D_{n})_{13})\\
&=\big((\jmath_{n,\infty}\otimes\id)(D_n)\big)_{12}
  \big((\jmath_{n,\infty}\otimes\id)(D_n)\big)_{13}\\
&=\big(D(\cl P_n\otimes 1)\big)_{12}\big(D(\cl P_n\otimes 1)\big)_{13}\\
&=D_{12}(\cl P_n\otimes 1\otimes 1)\,D_{13}(\cl P_n\otimes 1\otimes 1)\\
&=D_{12}D_{13}(\cl P_n\otimes 1\otimes 1),
\end{align*}
where in the first equality we used $(\id\otimes\Delta_\infty)(\cl P_n\otimes 1)=\cl P_n\otimes 1\otimes 1$, in the 
second equality we used equation (\ref{eq:D_on_Pn}) and $(\id\otimes\Delta_\infty)\circ(\jmath_{n,\infty}\otimes\id)
=(\jmath_{n,\infty}\otimes\id\otimes\id)\circ(\id\otimes\Delta_\infty)$ and in the last step we used the identity
$D_{13}(\cl P_n\otimes 1\otimes 1)=(\cl P_n\otimes 1\otimes 1)D_{13}$,
which follows from Lemma~\ref{L:cantor_perm}.

Now since both $ (\id\otimes\Delta_\infty)(D)$ and $D_{12}D_{13} $ are elements of the multiplier algebra $M(\cl K(L^2(\Omega_X)) \otimes \cl B_X \otimes \cl B_X)$ and  $\cl P_n\otimes 1\otimes 1\to 1$ strictly, we conclude that $ (\id\otimes\Delta_\infty)(D) = D_{12}D_{13} $, as we wanted.

(ii) Fix $n$ and define $\alpha_n:\cl D_{X_n}\to \cl D_{X_n}\otimes\cl B_X$ by $ \alpha_n (\delta_x)=\sum_{a\in X_n}\delta_a\otimes p_{a,x}$.
It is clear that  $\alpha_n$ is a unital $*$-homomorphism. 
Note that 
\[
\alpha_{n+1} \circ\iota_{n,n+1}(\delta_x) 
=\sum_{a\in X_n}\sum_{a'\in[d_n]}\delta_{aa'}\otimes
\sum_{x'\in[d_n]}p_{aa',xx'}
=(\iota_{n,n+1} \otimes \id)\circ \alpha_n(\delta_x),
\]
and thus by the properties of the inductive limits we infer the existence of a unital *-homomorphism $ \alpha : C(\Omega_X) \to C(\Omega_X) \otimes \cl B_X$ given by the displayed formula.

For $x\in X_n$ we have
$(\alpha\otimes\id)\alpha(\chi_{[x]})
=\sum_{a,b\in X_n}\chi_{[b]}\otimes p_{b,a}\otimes p_{a,x}
=(\id\otimes\Delta_\infty)\alpha(\chi_{[x]})$, and coassociativity
follows by density and continuity. By the orthogonality of the entries
of $(p_{a,x})$ along its columns, for $a\in X_n$ and $b\in\cl B_X$,
\[
\sum_{x\in X_n}\alpha(\chi_{[x]})(1\otimes p_{a,x}b)
=\sum_{c\in X_n}\chi_{[c]}\otimes\sum_{x}p_{c,x}p_{a,x}b
=\chi_{[a]}\otimes b ,
\]
which yields the density condition. 

(iii) It is straightforward to check that the *-homomorphism 
${ \alpha}$ is again a coaction of $ \bb W_{\infty}$ on the image of $  C(\Omega_X)$ in $\cl B(L^2(\Omega_X))$. We will show the unitary implementation.  Let $x\in X_n$ and let $ e_x= |X_n|^{1/2} \chi_{[x]}$. These vectors form an orthonormal basis of $ {\rm Ran}(\cl P_n)$; in particular if $ \xi \in L^{2}(\Omega_X)$, then $ \cl P_n \xi = \sum_{y\in X_n} \xi_y e_y $ and $$
M_{\chi_{[x]}}\cl P_{n} \xi = \sum_{y\in X_n} \xi_y M_{\chi_{[x]}}e_y = \sum_{y\in X_n} \xi_y \delta_{x,y} e_y=\xi_xe_x,
$$
so that $M_{\chi_{[x]}}\cl P_{n} = \cl E_{x,x}=\jmath_{n,\infty}(\epsilon_{x,x}) $ and by selfadjointness $M_{\chi_{[x]}}\cl P_{n} = \cl E_{x,x} = \cl P_{n}M_{\chi_{[x]}} $. 

Now we compute both sides:
\begin{align*}
    {\alpha}(M_{\chi_{[x]}}) (\cl P_n \otimes 1) = \sum_{a\in X_n}M_{\chi_{[a]}} \cl P_n\otimes p_{a,x} 
    = \sum_{a\in X_n}\cl E_{a,a} \otimes p_{a,x},
\end{align*}
and on the other hand,
\begin{align*}
    D(M_{\chi_{[x]}}\otimes 1)D^*(\cl P_n \otimes 1)& =  D(M_{\chi_{[x]}}\cl P_n\otimes 1)D^*\\
    &= D(\cl P_n \otimes 1) (M_{\chi_{[x]}}\cl P_n\otimes 1) (\cl P_n \otimes 1)D^*\\
    &= D(\cl P_n \otimes 1) (\cl E_{x,x}\otimes 1) (\cl P_n \otimes 1)D^*\\
    &= (\jmath_{n,\infty} \otimes \id )(D_n)(\jmath_{n,\infty} \otimes \id )(\epsilon_{x,x} \otimes 1)(\jmath_{n,\infty} \otimes \id )(D_n^*)   \\
    &= (\jmath_{n,\infty} \otimes \id ) ( D_n(\epsilon_{x.x} \otimes 1) D_n^*)\\
    &= (\jmath_{n,\infty} \otimes \id )(\sum_{a\in X_n} \epsilon_{a,a} \otimes p_{a,x} )
    = \sum_{a\in X_n}\cl E_{a,a} \otimes p_{a,x}.
\end{align*}
We conclude that 
\[
 {\alpha}(M_{\chi_{[x]}}) (\cl P_n \otimes 1) = D(M_{\chi_{[x]}}\otimes 1)D^*(\cl P_n \otimes 1)
\]
for each $ n \in \bb N$, and thus by using that $ \cl P_{n} \to_{n \to \infty}1$ strongly and the fact that the cylinders span a dense subspace of $ C(\Omega_X)$ we conclude the equality for any $ M_f$, $ f\in C(\Omega_X)$ as required. 
\end{proof}

\begin{remark}\rm
Quantum groups of the type of $\bb W_\infty$ first appeared in \cite{bgs},
where the quantum isometry group of the Christensen--Ivan spectral triple
on a commutative AF algebra is obtained as the inductive limit of the
quantum symmetry groups of the truncated Bratteli diagrams
\cite[Theorems~1.2 and~2.6]{bgs}; the special case of the algebra of
continuous functions on the middle-third Cantor set is treated explicitly
in \cite[Theorem~3.1]{bgs}. The relations linking consecutive levels in
the definition of $\cl B_X$ are those of \cite[Lemma~2.3]{bgs}, where they
are derived from the compatibility of the coactions of consecutive levels,
namely the requirement that the coaction $\alpha_{n+1}$ of
$\cl B_{X_{n+1}}$ on $\cl D_{X_{n+1}}$ satisfies
$\alpha_{n+1}(\cl D_{X_n})\subseteq \cl D_{X_n}\otimes\cl B_{X_{n+1}}$; in other
words, the quantum permutation of $X_{n+1}$ preserves the blocks over
$X_n$. By \cite[Theorem~2.6]{bgs}, the resulting level quantum groups are
the quantum automorphism groups, in the sense of Bichon, of the rooted
trees obtained by restricting the Bratteli diagram at a finite level.

A result similar to Proposition~\ref{p_BXSWinf} was obtained in
\cite[Theorem~5.7]{br_rob1}, where the quantum automorphism group of a
homogeneous rooted tree is considered. In our setting, $\bb W_\infty$ can
be thought of as  the quantum automorphism group of the
inhomogeneous rooted tree.
\end{remark}



\section{Cantor games}\label{s_cantorgames}

In this section we define the game algebra of a Cantor game and, using the results from Sections \ref{s_syn} and \ref{s_cqpg}, characterise its perfect strategies of various types via traces of corresponding types.


\subsection{Perfect strategies and rule functions}

We recall the notions of a Cantor game and its values first introduced in \cite{bctt}. 
Given a quadruple of inductive families of finite sets $X=(X_n)_{n\in\mathbb N}$, $Y=(Y_n)_{n\in \mathbb N}$, $A=(A_n)_{n\in \mathbb N}$, $B=(B_n)_{n\in \mathbb N}$, consider as before the corresponding Cantor spaces 
$\Omega_X,\Omega_Y,\Omega_A,\Omega_B$ (see Section \ref{s_gradedcor}) and denote by $\mu_{XY}=\mu_X\otimes \mu_Y$ the product measure of $\mu_X$ and $\mu_Y$. A \emph{Cantor game} \cite{bctt} is a pair $(\kappa,\mu_{XY})$ where $\kappa\subseteq \Omega_X\times \Omega_Y\times \Omega_A\times \Omega_B$ is a closed set. 
\par Given a Cantor game $\cl{G}=(\kappa,\mu_{XY})$ one can associate a sequence of finite games $\cl{G}_n=(\kappa_n,\mu_{X_nY_n})$ over $(X_n,Y_n,A_n,B_n)$, $n\in\bb{N}$, by projecting $\kappa$ on the first $n$ coordinates. The resulting sequence $(\kappa_n)_{n\in\bb{N}}$ has the property of \emph{nestedness}, namely $\kappa_{n+1}\subseteq \kappa_{n}$, $n\in\bb{N}$, when viewed as cylinder sets in $\Omega_X\times \Omega_Y\times \Omega_A\times \Omega_B$. Conversely, given a sequence of nested sets $(\kappa_n)_{n\in\bb{N}}$ one obtains the associated Cantor game by setting $\kappa=\cap_{n\in\bb{N}}\kappa_n$. 
\par The set $\kappa$ can be viewed as a generalization of the set of all winning quadruples to the measurable setup. Equivalently, we can think of the winning quadruples as the support of a rule function $\lambda$ which encodes the rules of the game. Taking advantage of this correspondence we can define the \emph{rule function} $\lambda:\Omega_X\times \Omega_Y\times \Omega_A\times \Omega_B\rightarrow\{0,1\}$ of a Cantor game $(\kappa,\mu_{XY})$ as the characteristic function of the set $\kappa$. Consequently, the nestedness condition of $(\kappa_n)_{n\in \bb{N}}$ gives rise to a sequence of rule functions 
$\lambda = (\lambda_n)_{n\in\Bbb N}$ with
\[
\lambda_n: X_n\times Y_n\times A_n\times B_n \to \{0,1\},
\]
satisfying the condition
\[
\lambda_n(x,y,a,b)=0
\ \ \Longrightarrow \ \
\lambda_{n+1}(xx',yy',aa',bb')=0
\]
for all length-one extensions $x'\in [d_n]$, $y'\in [d_n]$, 
$a'\in [c_n]$, and $b'\in [c_n]$. We will sometimes refer to the sequence $(\lambda_n)_{n\in\Bbb N}$ as being \emph{nested}. 
Conversely, given a nested sequence $(\lambda_n)_{n\in \bb{N}}$ of 
rule functions, we let 
$$\kappa_{\lambda_n} = \{(xx',yy',aa',bb') : |x| = |y| = |a| = |b| = n, 
\lambda(x,y,a,b) = 1\}, \ \ \ n\in \bb{N};$$
such a nested sequence gives rise to the Cantor game $\cap_{n=1}^{\infty} \kappa_{\lambda_n}$.
We note that Cantor games fall into the larger class of \emph{measurable games} introduced in \cite{btt} and are precisely specialisations of measurable games restricted to the Cantor setup. 

Let 
$\cl G=(\kappa,\mu_{XY})$ be a Cantor game and
\[
\Gamma : C(\Omega_A)\otimes C(\Omega_B)\longrightarrow
L^\infty(\Omega_X)\,\bar\otimes\,L^\infty(\Omega_Y)
\]
be a no-signalling correlation. Let, further, 
$p_\Gamma$ be the associated channel, that is, the $\Omega_X\times\Omega_Y$ measurable family $(p_\Gamma(\cdot,\cdot|x,y))_{(x,y)\in\Omega_X\times\Omega_Y}$ of Borel probability measures over $\Omega_A\times\Omega_B$ such that
$$\Gamma(f)(x,y)=\int_{\Omega_A\times\Omega_B}f(a,b)d p_\Gamma(a,b|x,y) \quad \text{$\mu_{XY}$-a.e.}$$
Following the framework developed in \cite{btt}, the value of a Cantor strategy $\Gamma$ is given by the 
value at $\kappa$ of the compound measure $\mu_{XY}\otimes p_\Gamma$, namely 
\begin{equation}\label{eq_compaoundm}
(\mu_{XY}\otimes p_\Gamma)(\kappa)=\int_{\Omega_X\times\Omega_Y}p_{\Gamma}(\kappa_{x,y}|x,y)d\mu_{XY}(x,y),
\end{equation}
where $\kappa_{x,y}=\{(a,b)\in\Omega_A\times\Omega_B: (x,y,a,b)\in\kappa\}$ is the $(x,y)$-section of $\kappa$.
Let $\Lambda_n = {\rm supp}(\lambda_n)$, viewed as a subset of 
$X_n\times Y_n\times A_n\times B_n$, and assume for a moment that 
$\Lambda_n = \{(x,y,a,b)\}$ for some $(x,y,a,b)\in X_n\times Y_n\times A_n\times B_n$. 
We have that 
$$(\mu_{XY}\otimes p_\Gamma)([x]\times [y]\times [a]\times [b]) = 
\frac{1}{|X_n||Y_n|} p(a,b|x,y) = (\mu_{X_nY_n}\otimes p_{\Gamma_n})(\Lambda_n)$$
(see \cite[Lemma 6.4]{bctt}).
By additivity, 
\begin{equation}\label{eq_compaoundm2}
(\mu_{XY}\otimes p_\Gamma)(\kappa_{\lambda_n})
= 
(\mu_{X_nY_n}\otimes p_{\Gamma_n})(\Lambda_n)
\end{equation}
for any $\Lambda_n\subseteq X_n\times Y_n\times A_n\times B_n$. 

We say that $\Gamma$ is a \emph{perfect strategy}  for $\cl G$ if
$(\mu_{XY}\otimes p_\Gamma)(\kappa)=1$.
We point out that the latter notion generalises the concept of a 
perfect strategy of a non-local game over finite input and output sets; 
see e.g. \cite{lmprsstw}. 

\begin{proposition} \label{p_perfect_cantor} 
Let $\cl G=(\kappa,\mu_{XY})$ be a Cantor game with rule function
$\lambda=(\lambda_n)_{n\in\Bbb N}$, $\Gamma$ be a no-signalling  correlation and $(\Gamma_n)_{n\in\Bbb N}$ the associated inductive family. 
The following are equivalent:
\begin{enumerate}
\item[(i)] 
$\Gamma$ is a perfect strategy for the Cantor game $\cl G$;
\item[(ii)] 
$p_\Gamma(\kappa_{x,y}| x,y)=1$ for $\mu_{XY}$-almost all $(x,y)$; 
\item[(iii)] 
for every $n\in\Bbb N$, the finite--level correlation $\Gamma_n$ is perfect
for $\lambda_n$, that is,
\begin{equation*}\label{eq:finite-perfect}
  \lambda_n(x_n,y_n,a_n,b_n)=0
  \ \Longrightarrow\
  p_{\Gamma_n}(a_n,b_n| x_n,y_n)=0,
\end{equation*}
for all $(x_n,y_n,a_n,b_n)\in X_n\times Y_n\times A_n\times B_n$ and every $n\in \bb{N}$. 
\end{enumerate}
\end{proposition}

\begin{proof}
(i)$\Leftrightarrow$(ii) follows immediately from the definition (\ref{eq_compaoundm}) of the compound measure.

\smallskip

(i)$\Rightarrow$(iii)
Fix $n$ and set $\Lambda_n:=\mathrm{supp}\,\lambda_n\subseteq X_n\times Y_n\times A_n\times B_n$.  
By \cite[Lemma~6.4]{bctt}, (\ref{eq_compaoundm2}) and the inclusion $\kappa\subseteq\kappa_{\lambda_n}$, 
we have 
\begin{equation}\label{eq:integral-chain}
\begin{aligned}
\int_{X_n\times Y_n}
& p_{\Gamma_n}\big((\Lambda_n)_{x,y}| x,y\big)\,d\mu_{X_nY_n}(x,y)
=(\mu_{X_nY_n}\otimes p_{\Gamma_n})(\Lambda_n)\\[2pt]
&=(\mu_{XY}\otimes p_\Gamma)(\kappa_{\lambda_n})
\ge (\mu_{XY}\otimes p_\Gamma)(\kappa)
=1.
\end{aligned}
\end{equation}
Since the integrand takes values in $[0,1]$, equality in \eqref{eq:integral-chain} forces 
\( 
p_{\Gamma_n}\big((\Lambda_n)_{x,y}| x,y\big)=1\) 
for all \( (x,y)\in X_n\times Y_n.
\)
But for each $(x,y)\in X_n\times Y_n$ we can expand
\[
p_{\Gamma_n}\big((\Lambda_n)_{x,y}| x,y\big)
=\sum_{(a,b)\in(\Lambda_n)_{x,y}} p_{\Gamma_n}(a,b| x,y).
\]
Hence
\( 
\sum_{(a,b)\in(\Lambda_n)_{x,y}} p_{\Gamma_n}(a,b| x,y)=1\) for all \( (x,y)\in X_n\times Y_n.
\)
Now, because 
$$\sum_{(a,b)\in A_n\times B_n}p_{\Gamma_n}(a,b| x,y)=1,$$ 
it follows that
$p_{\Gamma_n}(a,b| x,y)=0$ if $(x,y,a,b)\notin \Lambda_n$,
which is precisely the finite--level perfectness condition.

\smallskip

(iii)$\Rightarrow$(i) 
If a finite--level strategy is perfect, that is,
\[
\lambda_n(x,y,a,b)=0 
\ \Longrightarrow\ 
p_{\Gamma_n}(a,b| x,y)=0
\quad\text{for all }x,y,a,b,
\]
then
\begin{align*}
(\mu_{X_nY_n}\otimes p_{\Gamma_n})(\Lambda_n^{\,c})
&=\sum_{(x,y)\in X_n\times Y_n}
\mu_{X_nY_n}(\{(x,y)\})
\sum_{(a,b)\notin (\Lambda_n)_{x,y}} 
p_{\Gamma_n}(a,b| x,y)=0,
\end{align*}
and hence
\( 
(\mu_{X_nY_n}\otimes p_{\Gamma_n})(\Lambda_n)=1.
\)
By  \cite[Theorem 6.5]{bctt} and \cite[Lemma 6.4]{bctt}
\[
(\mu_{XY}\otimes p_\Gamma)(\kappa)
=\lim_{n\to\infty}(\mu_{X_nY_n}\otimes p_{\Gamma_n})(\Lambda_n)
=1.
\]
\end{proof}


\subsection{The graded synchronous game *-algebra} \label{s:graded_game_alg}

Let $X = (X_n)_{n\in \Bbb N}$ and $A = (A_n)_{n\in \Bbb N}$ be inductive families of finite sets,
and let $\lambda = (\lambda_n)_{n\in\Bbb N}$ be a rule function over the quadruple $(X,X,A,A)$. 
We call $\lambda$ \emph{synchronous} if $\lambda_n(x,x,a,b) = 0$ whenever $a\neq b$, for all 
initial words $x,y,a,b$ of length $n$, and all $n\in \bb{N}$. 
Assuming $\lambda$ is synchronous, consider, for each $n\in\Bbb N$, the synchronous non-local game
\(  
\cl G_n = (X_n,A_n,\lambda_n),
\)
and let $\cl A(\cl G_n)$ denote the game algebra of $\cl G_n$ \cite{hmps}; thus, 
$\cl A(\cl G_n)$ is the unital *-algebra generated by
symbols $\{e_{x,a} : x\in X_n,\, a\in A_n\}$,
subject to the relations, for each fixed $x\in X_n$,
\( e_{x,a}^* = e_{x,a}\), \( e_{x,a}^2 = e_{x,a}\), 
\( e_{x,a} e_{x,b} = 0\) for \(a\neq b\), 
\(\sum_{a\in A_n} e_{x,a} = 1\),
and the \emph{rule relations}
\[
\lambda_n(x,y,a,b) = 0
\ \Longrightarrow\
e_{x,a} e_{y,b} = 0.
\]
We call $\cl A(\cl G_n)$ the \emph{level-$n$ synchronous game *-algebra}.

\vspace{0.2cm}

To ensure compatibility of these algebras along the inductive chain, one would need some additional row-coherence relations to hold as in the construction of C*-algebras $\cl B_{X_n,A_n}$ in Subsection~\ref{ss_univa}. However, these become automatic in this case via the property of the nestedness and the structure of the game *-algebra.

\begin{proposition}\label{prop:auto_coh}
    Let $(X_n)_{n\in \bb N}$, $ (A_n)_{n\in \bb N}$ be inductive families of finite sets, $ (\lambda_{n})_{n\in \bb N}$ be a nested family of synchronous rule functions and $ \cl G_n=(X_n,A_n,\lambda_n)$ be the non-local games. Then, for each $ n \in \bb N$,  the generators $ e_{x,a}$,  of $ \cl A(\cl G_n)$ satisfy
    \begin{equation}\label{eq:s_cohrel}
    \sum_{|a'| = n-k} e_{x x', a a'} =
\sum_{|a'| = n-k} e_{x x'', a a'},
    \end{equation}
    for all $k\in [n-1]$, $x \in X_k$ and all $|x'|=|x''|=n-k$ and $|a|=k$.
\end{proposition}
\begin{proof}
    We consider the case $k=n-1$; the rest follow by induction. Let $T_{x,a;x'}:= \sum_{|a'|=1}e_{xx',aa'}$ and calculate
    \begin{align*}
    T_{x,a;x'}
    &= \left(\sum_{|a'|=1}e_{xx',aa'}\right) \left(\sum_{c\in A_n}e_{xx'',c}\right)\\
    &= \sum_{c\in X_n} \sum_{|a'| = 1} e_{xx',aa'}e_{xx'',c}
    =\sum_{|a'|=1,|c'|=1} e_{xx',aa'}e_{xx'',ac'},
    \end{align*}
    where we used the synchronicity at the $(n-1)$-level and nestedness. 
    Indeed, if the $(n-1)$-th level prefix $c_0$ of $ c$ is not equal to $a$, synchronicity implies that the rule function $\lambda_{n-1}$ evaluates to zero at the quadruple $(x,x,a,c_0)$, and by the fact that $(\lambda_m)_{m\in \bb{N}}$ is nested, we have that $\lambda_{n}$ evaluates at zero as well at the quadruple $(xx',xx'',aa',c)$.

    On the other hand, if we let $ T_{x,a;x''}:= \sum_{|a'|=1} e_{xx'',aa'}$, then analogous arguments yield that 
    \[
     T_{x,a;x''}=\sum_{|a'|=1,|c'|=1} e_{xx'',aa'}e_{xx',ac'}.
    \]
    By selfadjointness, $ T_{x,a;x''}=\sum_{|a'|=1,|c'|=1} e_{xx',ac'}e_{xx'',aa'}$ and if we relabel  $ a'$ with $ c'$ we obtain the desired equality.
\end{proof}

Inside $\cl A(\cl G_{n+1})$, for every $x\in X_n$ and $a\in A_n$ we define the
element
\begin{equation}\label{eq_hat_e}
\hat e_{x,a} := \sum_{|a'|=1} e_{x x', a a'},
\end{equation}
where $x'\in [d_n]$ is arbitrary.
By Proposition \ref{prop:auto_coh}, the
expression in \eqref{eq_hat_e} is independent of $x'$.
Within $\cl A(\cl G_{n+1})$, the family $\{\hat e_{x,a}\}_{a\in A_n}$ satisfies 
the same relations as $\{e_{x,a}\}_{a\in A_n}$ at level $n$:
each $\hat e_{x,a}$ is a self-adjoint idempotent, $\hat e_{x,a}\hat e_{x,b}=0$
for $a\neq b$, and $\sum_{a\in A_n}\hat e_{x,a}=1$.  Moreover, by gradedness of
$\lambda$, the rule relations are preserved:
if $\lambda_n(x,y,a,b)=0$ then
$\hat e_{x,a}\hat e_{y,b}=0$ in $\cl A(\cl G_{n+1})$.
Hence, by the universality of $\cl A(\cl G_n)$, there exists a unique unital
*-homomorphism
\[
\pi_n^{\rm s} : \cl A(\cl G_n) \longrightarrow \cl A(\cl G_{n+1}), \ \ \mbox{ such that }
\pi_n^{\rm s} (e_{x,a}) = \hat e_{x,a}, \ x\in X_n, a\in A_n;
\]
we thus obtain an inductive system of unital *-algebras  $(\cl A(\cl G_n),\pi_n^{\rm s} )_{n\in\Bbb N}$.
The \emph{synchronous game *-algebra} associated to
the synchronous Cantor game $\cl G=(\kappa,\mu_{XX})$ is defined as the inductive limit
\[
\cl A(\cl G)
 :=
\varinjlim\, (\cl A(\cl G_n), \pi_n^{\rm s} )_{n\in \bb{N}}.
\]
Using the universal property of the 
inductive limit, one can see that $\cl A(\cl G)$ is canonically *-isomorphic to 
the  *-algebra generated by symbols
\[
\{\, e_{x,a} \ : \ x\in X_n,\ a\in A_n,\ n\in\Bbb N \,\},
\]
subject to the following relations, holding for all $n$ and all
$x,y\in X_n$, $a,b\in A_n$:
\vspace{0.1cm}

\noindent
\textrm{(PVM)}   
for each fixed $x\in X_n$,
\[
e_{x,a}^* = e_{x,a}, \qquad
e_{x,a}^2 = e_{x,a}, \qquad
e_{x,a}e_{x,b}=0\ (a\neq b), \qquad
\sum_{a\in A_n} e_{x,a} = 1.
\]

\noindent
\textrm{(COH)} 
for all $k<n$, for every $x\in X_k$, all $x'\in \prod_{i=k}^{n-1}[d_i]$, and all $a\in A_k$,
\[ 
e_{x,a}= \sum_{|a'|=n-k} e_{x x',\, a a'}.
\]

\noindent
\textrm{(RULE)} 
$e_{x,a} e_{y,b} = 0$ whenever $x,y,a,b$ are initial words of length $n$ and 
$\lambda_n(x,y,a,b)$ $=$ $0$, $n\in \bb{N}$.

\begin{remark}\label{r_quotofbxa}
\rm
Conditions (PVM) and (COH) are precisely the 
defining conditions for the universal C*-algebra 
$\cl B_{X,A}$ defined in Subsection \ref{ss_univa}. 
It follows that the universal enveloping C*-algebra of $\cl A(\cl G)$ is a canonical quotient of $\cl B_{X,A}$. 
\end{remark}



\begin{theorem} \label{th_game_alg_reps}
    Let $ \cl G= (\kappa,\mu_{XX})$ be a synchronous Cantor game. The following hold: 
    \begin{enumerate}
        \item[(i)] 
        $\cl G$ has a perfect graded quantum commuting strategy if and only if there exists a unital *-homomorphism $ \cl A(\cl G) \to  \cl A$ for a C* -algebra $\cl A$ with a faithful tracial state;
        
\item[(ii)] $\cl G$ has a perfect graded quantum approximate strategy if and only if there exists a unital *-homomorphism $ \cl A(\cl G) \to \cl R^{\omega}$ for an ultrapower of the hyperfinite {\rm II}$_1$ factor $\cl R$;

\item[(iii)] 
$\cl G$ has a perfect graded local strategy if and only if the character space of $ \cl A(\cl G)$ is non-trivial.
\end{enumerate}
\end{theorem}

\begin{proof}
(i) 
We continue to denote by $ e_{x,a}$, $x\in X_n$, $ a\in A_n$,  $n\in \bb N$, the image of the generators of $ \cl A(\cl G)$ inside $\cl A$ under the *-homomorphism. Define 
\begin{equation}\label{eq_pdeftau}
p(a,b|x,y) := \tau (e_{x,a} e_{y,b}),
\end{equation}
    where $ x,y,a,b$ are of the same length and $ \tau$ is the trace in $\cl A$ and note that the (PVM) and (COH) conditions in the graded game algebra ensure that $p $ is a synchronous graded correlation as in the proof of Theorem \ref{th_trBXA}. Finally, by the (RULE) condition of the graded game algebra and Proposition \ref{p_perfect_cantor} we have that $p$ is perfect.

 Conversely, assume that $p$ is a perfect graded synchronous strategy of quantum commuting type. By Theorem \ref{th_trBXA}, there exists a tracial state $ \tau : \cl B_{X,A} \to \bb C$ such that 
(\ref{eq_pdeftau}) holds. 
    Let  $\cl  A$ be the image of $\cl B_{X,A}$ under the GNS representation with respect to $\tau$, 
    equipped with the 
    faithful tracial state arising from $\tau$. 
    The fact that $p$ is perfect ensures condition
(RULE) for the projections $e_{x,a}$ in $\cl A$. 
Using Remark \ref{r_quotofbxa}, we obtain a unital *-homomorphism from the graded game algebra into $\cl  A$.

(ii) Let $\tau : \cl B_{X,A}\to \bb{C}$ be an amenable trace
that gives rise to a perfect graded quantum approximate strategy $p$; the existence of $\tau$ is due to Theorem \ref{th_trBXA}.   
By \cite[Proposition 3.2]{kirchberg}, 
there exist an ultrafilter $\omega$ and a unital 
*-homomorphism $\pi : \cl B_{X,A}\to \cl R^{\omega}$ such that 
$\tau = \tau_{\omega}\circ \pi$. Noting that $\tau_{\omega}$ is faithful, we obtain that 
$\pi(e_{x,a}e_{y,b}) = 0$ whenever 
$|x| = |y| = |a| = |b| = n$ and $\lambda_n(x,y,a,b) = 0$. 
Since $\pi$ is a homomorphism, 
$\pi$ descends to a unital *-homomorphism (denoted in the same way) $\pi : \cl A(\cl G)\to \cl R^{\omega}$. 

Conversely, suppose that $\pi : \cl A(\cl G)\to \cl R^{\omega}$ is a unital *-homomorphism and let 
$p$ be the (graded) correlation corresponding to 
the trace $\tau_{\omega}\circ \pi$. 
It is trivial that $p$ is a perfect synchronous strategy for $\cl G$. 
Composing with the quotient and inclusion maps, let 
$\pi_n : \cl A_{X_n,A_n} \to \cl R^{\omega}$ be the 
unital *-homomorphism arising from $\pi$; we have that 
$\pi_n$ gives rise to $p_n$. 
By \cite[Theorem 3.6]{kps}, $p_n$ is of 
approximately quantum type. Thus, $p\in \cl C_{\rm qag}$.

(iii) Let $\pi:\cl{A}(\cl G)\rightarrow \bb{C}$ be a unital *-homomorphism and $\tilde{\pi}:\cl{B}_{X,A}\rightarrow \mathbb{C}$ be the character after composing with the quotient map. By Proposition \ref{p_charactBXA}  there exists $f=(f_n)_{n\in\bb{N}}\in{\rm GM}(X,A)$ given by $\tilde{\pi}(e_{x,a})=\delta_{f_n(x),a}$ for $a\in A_n$ and $x\in X_n$. By the (RULE) condition one obtains that ${\rm Gr}(f\times f)\subseteq \kappa$.  
It is straightforward that the resulting correlation $p$ is a perfect strategy. The (PVM) and (COH) conditions imply that $p$ is synchronous and graded. 

For the converse, 
recall that the deterministic strategies of a finite synchronous game 
over the pair $(S,T)$
are in a canonical correspondence with functions $g : S\to T$. 
Assume that there exists a perfect local graded strategy for the game $\cl G$. 
Denote by $\bb{D}_n$ the set of all deterministic strategies
for the game $\cl G_n$, and by $\tilde{\bb{D}}_n$ the set of all 
graded maps $f = (f_m)_{m\in \bb{N}}$ such that $f_n\in \bb{D}_n$, 
$n\in \bb{N}$. By assumption, $\tilde{\bb{D}}_n\neq \emptyset$, $n\in \bb{N}$. 
Clearly, the family $(\tilde{\bb{D}}_n)_{n\in \bb{N}}$ is nested. 
By the compactness of ${\rm GM}(X,A)$, there exists 
$f\in \cap_{n\in \bb{N}} \tilde{\bb{D}}_n$.
It is now straightforward that ${\rm Gr}(f\times f)\subseteq \kappa$; 
following the first paragraph, $f$ gives rise to a 
character of $\cl A$. 
\end{proof}

We finish this section with a discussion on how infinite parallel repetition fits in our setting.
Let $\cl G = (S,U,\lambda)$ be a finite synchronous game. Its $n$-fold parallel repetition is the synchronous game
\( 
\cl G_n = (S^n,U^n,\lambda_n),
\)
where
\[
\lambda_n(x,y,a,b)
:=
\prod_{i=1}^n \lambda(x_i,y_i,a_i,b_i),
\;
x=(x_i)_{i},y=(y_i)_{i},\ a=(a_i)_{i},b=(b_i)_{i};
\]
equivalently, $\lambda_n(x,y,a,b)=1$ if and only if $\lambda(x_i,y_i,a_i,b_i)=1$ for all $i=1,\dots,n$.
It is straightforward that the
sequence $(\lambda_n)_{n\in\Bbb N}$ is a nested family of synchronous rule functions, and hence $(\cl G_n)_{n\in\Bbb N}$ is an inductive family of synchronous games in the sense of the previous subsection. The associated graded synchronous game 
*-algebra
\( \cl A(\cl G)
=
\varinjlim \cl A(\cl G_n)
\)
therefore encodes the algebraic structure of the infinite parallel repetition of $\cl G$. The coherence relations (COH) express the compatibility of strategies under refinement, while the rule relations (RULE) enforce that every finite block satisfies the original game constraints.


\subsection{The graded bisynchronous game *-algebra}\label{s:graded_bisync_game_alg}

Let $X = A = (X_n)_{n\in\Bbb N}$ be an inductive family of finite sets and 
$\lambda=(\lambda_n)_{n\in\Bbb N}$ be a bisynchronous 
rule function in that $\lambda$ is synchronous and 
$\lambda_n(x,y,a,a) = 0$ provided $x,y,a\in X_n$ and $x\neq y$. 
For each $n\in\Bbb N$, we consider the bisynchronous non-local game
\( 
\mathcal G_n=(X_n,X_n,\lambda_n).
\)
Since each $\mathcal G_n$ is synchronous, the level-$n$ synchronous game *-algebra
\( 
\mathcal A(\mathcal G_n)
\)
is already defined as in Subsection~\ref{s:graded_game_alg}.

Define
\[
\cl A^{\rm bs}(\cl G_n):= \cl A(\cl G_n)/ \left\langle 1-\sum_{x \in X_n}e_{x,a}:a\in X_n\right\rangle, 
\]
where the quotient is taken by the corresponding two-sided ideal,
and note that by arguing as in Proposition~\ref{prop:auto_coh}, one automatically has 
both the row and column coherence relations 
\[
 \sum_{|a'|=n-k} e_{xx',aa'}=\sum_{|a'|=n-k} e_{xx'',aa'},\quad 
\
\sum_{|y|=n-k} e_{xy,ab'}=\sum_{|y|=n-k} e_{xy,ab''}
\]
for all  $k\in[n-1]$, $|x|=|a|=k$, and
\( 
|x'|=|x''|=|b'|=|b''|=n-k.
\)

By arguing as in  Remark~\ref{rmk:welldef}, we see that  the family $\{\hat e_{x,a}=\sum_{|a'|=1} e_{xx',aa'}\}_{a\in X_n}$ satisfies the same defining relations as
the generators of $\mathcal A^{\rm bs}(\mathcal G_n)$, and therefore induces a well-defined unital
*-homomorphism
\[
\pi_n^{\rm bs}:\mathcal A^{\mathrm{bs}}(\mathcal G_n)\to \mathcal A^{\mathrm{bs}}(\mathcal G_{n+1}).
\]
Consequently, \((\mathcal A^{\mathrm{bs}}(\mathcal G_n),\pi_n^{\rm bs})_{n\in\Bbb N}\) forms an
inductive system of unital *-algebras. We define the \emph{bisynchronous game *-algebra} by
letting 
\[
\mathcal A^{\mathrm{bs}}(\mathcal G)
:=
\varinjlim (\mathcal A^{\mathrm{bs}}(\mathcal G_n),\pi_n^{\rm bs})_{n\in \bb N}.
\]

\begin{lemma}\label{l_bs_sync_Cstar_reps} 
The synchronous game *-algebra $ \cl A(\cl G)$ and the  bisynchronous *-algebra $ \cl A^{\rm bs}(\cl G)$ have the same bounded *-representations. 
\end{lemma}

\begin{proof}
Using \cite[Remark 2.1]{paulsen-rahaman}, the conclusion of the lemma 
holds for the *-algebras $ \cl A(\cl G_n)$ and $ \cl A^{\rm bs}(\cl G_n)$ for every $n\in \bb{N}$. 
On the other hand, the quotient maps $q_n$ are compatible with the connecting maps in that
\( 
 q_{n+1}\circ\pi_n^{\rm s}
 =
 \pi_n^{\rm bs}\circ q_n.
\)
The claim follows using the universal property of the inductive limits.
\end{proof}

\begin{theorem} \label{th_bs_game_alg_reps}
Let $ \cl G= (\kappa,\mu_{XX})$ be a bisynchronous Cantor game. The following hold:
\begin{enumerate}
\item[(i)] 
$\cl G$ has a perfect graded quantum commuting strategy if and only if there exists a unital *-homomorphism $ \cl A^{\rm bs}_{}(\cl G) \to  \cl A$ for a C* -algebra $\cl A$ equipped with a faithful tracial state;
        
\item[(ii)] $\cl G$ has a perfect graded quantum approximate strategy if and only if there exists a unital *-homomorphism $ \cl A^{\rm bs}_{}(\cl G) \to \cl R^{\omega}$ for an ultrapower of the hyperfinite {\rm II}$_1$ factor $\cl R$;

\item[(iii)] 
$\cl G$ has a perfect local graded strategy if and only if the character space of $ \cl A^{\rm bs}_{}(\cl G)$ is non-trivial.
\end{enumerate}
\end{theorem}

\begin{proof}
By Lemma~\ref{l_bs_sync_Cstar_reps}, the graded synchronous and
graded bisynchronous game algebras have the same representations into
C$^*$-algebras. The result therefore follows directly from
Theorem~\ref{th_game_alg_reps}.
\end{proof}


\section{Quantum isomorphisms of Cantor graphs}\label{s_cicg}

In this section, we consider a particular instance of a Cantor game, namely the 
Cantor graph isomorphism game, and characterise in operator algebraic terms the 
existence of perfect strategies of different types. 
The next subsection contains the general setup and some preliminary results. 
Subsection \ref{ss_cohgamealg} examines 
bi-Galois extensions over the Hopf algebras of the 
quantum automorphism groups of the underlying Cantor graphs. Finally, Subsection \ref{ss_maincharteo} contains the main characterisation 
result, connecting perfect strategies, bi-Galois extensions and trace representations. 


\subsection{The Cantor graph setup}\label{ss_Cgsetup}

Let $X = (X_n)_{n\in\mathbb{N}}$ be an inductive family of sets, 
where $X_{n+1} = X_n\times [d_n]$, $d_n\geq 2$, $n\in \bb{N}$, and set $d_0 = |X_1|$. We set $A_n = X_n$, $n\in \bb{N}$, and $A = X$ throughout the section 
(note both $A$ and $X$ will be used in our notation for clarity). 
A \emph{Cantor graph} is a closed subset 
$\mathfrak{G}\subseteq \Omega_X\times\Omega_X$ that is reflexive in that it contains the diagonal 
$\Delta_X = \{(x,x) : x\in  \Omega_X\}$ and is symmetric in that 
$$(x,y)\in \mathfrak{G} \ \implies \ (y,x)\in \mathfrak{G}.$$
For every $n\in \mathbb{N}$, let $\mathfrak{p}_n : \Omega_X\times\Omega_X \to X_n\times X_n$ be the projection onto the first $n$ coordinates, and set $G_n := \mathfrak{p}_n(\mathfrak{G})$. 
Clearly, the subset $G_n$ is symmetric and contains the diagonal $\Delta_n$ of $X_n\times X_n$. 
Write $A_{G_n}$ for the adjacency matrix of $G_n$: for $x$, $y\in X_n$, the $x,y$-entry 
$(A_{G_n})_{x,y}=1$ if and only if 
$(x,y)\in G_n$, if and only if $([x]\times[y])\cap\mathfrak{G}\ne\emptyset$. 
Since $\mathfrak{G}$ was assumed reflexive, $G_n$ contains the diagonal in $X_n\times X_n$, 
that is, $(A_{G_n})_{x,x}=1$ for all $x\in X_n$. Let $\tilde{A}_{G_n} = A_{G_n} - I$. 

For $n\in\mathbb N$, write $\Omega_n$ for the Cantor set $\prod_{k\geq n} [d_k]$ and let $\mu_n=\otimes_{k\geq n}\mu_{d_k}$ and $\mu_{X_n}=\otimes_{k=1}^{n-1}\mu_{d_k}$.  We have $L^2(\Omega_X)\simeq L^2(X_n,\mu_{X_n})\otimes L^2(\Omega_n,\mu_n)$. 
Below, $T_{\chi_{\mathfrak{G}_n}}$ and $T_{\chi_{G_n}}$
denote the corresponding operators with (integral) kernels 
the characteristic functions $\chi_{\mathfrak{G}_n}$ and $\chi_{G_n}$. 

\begin{lemma} \label{ident}
Let $\mathfrak{G} \subseteq \Omega_X\times\Omega_X$ be a Cantor graph and let $\mathfrak{G}_n=\mathfrak{p}_n^{-1}(G_n)$. Then 
\begin{enumerate}
    \item[(i)] $T_{\chi_{\mathfrak{G}_n}}=T_{\chi_{G_n}}\otimes T_{1_{\Omega_n\times\Omega_n}} = \frac{1}{|X_n|}\jmath_{n,\infty}(A_{G_n})$;
    \item[(ii)] $\mathcal{P}_n=I_n\otimes T_{1_{\Omega_n\times\Omega_n}}=\jmath_{n,\infty}(I_n)$.
\end{enumerate}
\end{lemma}
\begin{proof}
We note that (ii) is immediate by definition. For (i), since $\mathfrak G_n=\mathfrak p_n^{-1}(G_n)$, we have
\( 
\chi_{\mathfrak G_n}\big((x,\omega),(y,\eta)\big)=\chi_{G_n}(x,y)\),
 \(x,y\in X_n,\ \omega,\eta\in \Omega_n.
\)
Thus
\( 
\chi_{\mathfrak G_n}=\chi_{G_n}\otimes 1_{\Omega_n\times \Omega_n},
\)
and as a consequence 
\[
T_{\chi_{\mathfrak G_n}}
=
T_{\chi_{G_n}}\otimes T_{1_{\Omega_n\times\Omega_n}}.
\]
On the other hand, if $ f \in L^{2}(\Omega_X)$, $ x\in X_n$, $ \omega \in \Omega_n$ then
\begin{align*}
(T_{\chi_{\mathfrak{G}_n}} f)(x, \omega) 
&= \int_{\Omega_X} \chi_{\mathfrak{G}_n}\big((x,\omega), (y,\eta)\big) f(y,\eta)  d\mu_X(y,\eta) \\
&= \int_{X_n \times \Omega_n} \chi_{G_n}(x,y) f(y,\eta)  d\mu_{X_n}(y)  d\mu_n(\eta) \\
&= \int_{X_n} \chi_{G_n}(x,y) \left( \int_{\Omega_n} f(y,\eta)  d\mu_n(\eta) \right) d\mu_{X_n}(y) \\
&= \sum_{y \in X_n} (A_{G_n})_{x,y} \left( \int_{\Omega_n} f(y,\eta)  d\mu_n(\eta) \right) \mu_{X_n}(y) \\
&= \frac{1}{|X_n|} \sum_{y \in X_n} (A_{G_n})_{x,y} \int_{\Omega_n} f(y,\eta)  d\mu_n(\eta),
\end{align*}
while
\begin{align*}
(\jmath_{n,\infty}(A_{G_n}) f)(x,\omega)
& = \sum_{y,z \in X_n} (A_{G_n})_{y,z} (\jmath_{n,\infty}(\epsilon_{y,z}) f)(x,\omega)\\
&= \sum_{y \in X_n} (A_{G_n})_{x,y} \left( \int_{\Omega_n} f(y,\eta) \, d\mu_n(\eta) \right).
\end{align*}
Hence, (i) follows
as claimed. 
\end{proof}

\begin{lemma}\label{L:finite_infinite_intertwining}
Let $\mathfrak{G} \subseteq\Omega_X\times\Omega_X$ and $\mathfrak{H}\subseteq\Omega_A\times\Omega_A$ be Cantor graphs with finite-level graphs $G_n,H_n$, $n\in \bb{N}$, and let $D$ be a Cantor quantum permutation with 
associated sequence $(D_n)_{n\in \bb{N}}$. Then, for each $n\in\mathbb N$, the following are equivalent:
\begin{enumerate}
\item $D (T_{\chi_{\mathfrak G_n}}\otimes 1 )
    =
    (T_{\chi_{\mathfrak H_n}}\otimes 1 )D$;
\item $D_n\,(A_{G_n}\otimes 1)
    =
    (A_{H_n}\otimes 1)\,D_n$;
\item $D_n\,(\tilde{A}_{G_n}\otimes 1)
    =
    (\tilde{A}_{H_n}\otimes 1)\,D_n$.
\end{enumerate}
\end{lemma}

\begin{proof}
The equivalence of (ii) and (iii) is trivial. 
By Lemma~\ref{ident}, we have
\( 
T_{\chi_{\mathfrak G_n}}=\frac{1}{|X_n|}\jmath_{n,\infty}(A_{G_n})\),
\( T_{\chi_{\mathfrak H_n}}=\frac{1}{|A_n|}\jmath_{n,\infty}(A_{H_n})\) and 
\( \mathcal P_n=\jmath_{n,\infty}(I_n).
\)
Using Lemma~\ref{L:cantor_perm}, we have
\[
D(\mathcal P_n\otimes 1) = (\mathcal P_n\otimes 1)D(\mathcal P_n\otimes 1)
=
(\jmath_{n,\infty}\otimes\id)(D_n).
\]
Hence,
\[
D (T_{\chi_{\mathfrak G_n}}\otimes 1 )
    =|X_{n}|^{-1}D(\cl P_n \otimes 1)(\jmath_{n,\infty}\otimes\id)(A_{G_n}\otimes 1) =|X_{n}|^{-1}(\jmath_{n,\infty}\otimes\id)(D_n(A_{G_n}\otimes 1) )
\]
and
\[
(T_{\chi_{\mathfrak H_n}}\otimes 1 )D =|X_{n}|^{-1}(\jmath_{n,\infty}\otimes\id)(A_{H_n}\otimes 1)(\cl P_n \otimes 1) D=|X_{n}|^{-1}   (\jmath_{n,\infty}\otimes\id)((A_{H_n}\otimes 1)D_n ),
\]
and the claim follows from the injectivity of $\jmath_{n,\infty}\otimes\id$.
\end{proof}

Let $G$ be a simple finite graph with vertex set $V(G)$. 
For vertices $i,j\in V(G)$, we write $i\sim_G j$ (or $i\sim j$ if $G$ is clear from the context) if $\{i,j\}$ is an edge of $G$. 
The adjacency relation ${\rm rel}_G : V(G)\times V(G)\to \{0, 1, -1\}$ is 
defines as follows:
$$
\text{rel}_G(i,j) = 
\begin{cases}
0 & \text{if } i = j\\
1 & \text{if } i\sim_G j\\
-1 & \text{otherwise.}
\end{cases}
$$

\begin{remark}
\rm 
Suppose that $G$ and $H$ are finite graphs on the vertex sets $V(G)$ and $V(H)$, respectively, with $|V(G)| = |V(H)|$. 
The graph isomorphism game was originally introduced in \cite{atserias-etal}
as the game with input and output sets $V := V(G)\sqcup V(H)$. 
Denoting by $g_i$ (resp. $h_i$) the elements of $V(G)$ (resp. $V(H)$), 
$i = 1,2,3,4$, 
its rule function $\lambda = \lambda_{G,H}$ was defined to have the property that 
$$\lambda(g_1,g_2,g_3,g_4) = 
\lambda(h_1,h_2,h_3,h_4) = 
\lambda(g_1,h_1,g_2,h_2) = 
\lambda(h_1,g_1,h_2,g_2) = 0
$$
while the values $\lambda(g_1,g_2,h_1,h_2)$, 
$\lambda(h_1,h_2,g_1,g_2)$, $\lambda(g_1,h_1,h_1,g_2)$ and
$\lambda(h_1,g_1,g_1,h_2)$ are equal to $1$ precisely when 
${\rm rel}_G(g_1,g_2) = {\rm rel}_H(h_1,h_2)$. 
We will temporarily 
refer to this game as the \emph{doubled graph isomorphism game}.
Our definition of the Cantor graph isomorphism game is, on the other hand, based on the \emph{simplified graph isomorphism game}, having input (resp. output) 
set $V(G)\times V(G)$ (resp. $V(H)\times V(H)$) and a rule 
function $\tilde{\lambda} = \tilde{\lambda}_{G,H}$ defined by letting 
$\tilde{\lambda}(g_1,g_2,h_1,h_2) = 1$ precisely when 
${\rm rel}_G(g_1,g_2) = {\rm rel}_H(h_1,h_2)$.
Every perfect strategy for the doubled 
graph isomorphism game gives rise, by canonical restriction, to a perfect 
strategy for the simplified graph isomorphism game. 
Conversely, assume that $p$ is a perfect strategy for the 
simplified graph isomorphism game for the pair $(G,H)$ of quantum 
commuting type. 
In particular, $p$ is a bisynchronous correlation and thus, by 
\cite[Theorem 2.2]{paulsen-rahaman}, $p(a,b|x,y) = \tau(v_{x,a}v_{y,b})$, 
where $\tau$ is a tracial state on the quantum permutation group 
$C(S_V^+)$, where we have assumed $V = V(G) = V(H)$ and 
$\{v_{x,a} : x,a\in V\}$ are its canonical generators. 
Without loss of generality (after passing to a unital *-representation of $C(S_V^+)$), we may assume that $\tau$ is faithful. The winning conditions
of the simplified graph isomorphism game now imply the existence 
of a quantum permutation $(P_{x,a})_{x,a\in V}$ acting on a Hilbert space $H$, such that $P_{x,a}P_{y,b} = 0$ if ${\rm rel}_G(x,y)\neq {\rm rel}_H(a,b)$. 
By \cite[Theorem 2.2]{lmr}, the doubled graph isomorphism game 
for the pair $(G,H)$ has a perfect quantum commuting strategy.
Thus, there exists a canonical correspondence between the perfect 
quantum commuting strategies of the two games; similar arguments 
imply that the same remains true for perfect quantum, approximately quantum or local strategies. 
Since our interest lies in 
the tracial descriptions of quantum isomorphisms of types different from
general no-signalling ones, we have decided to restrict our attention to 
a Cantor version of the simplified graph isomorphism game. 
\end{remark}

To analyse the existence of quantum strategies for this game, we must investigate the associated algebras and ensure they respect the inductive structure of the Cantor space; this is done in the next subsection.

Let $ n\in \bb N$. Define $ \widetilde{\rm Iso}(G_n,H_n)$ as the tuples $(x,y,a,b)\in X_n\times X_n\times A_n\times A_n$ such that
\[\text{rel}_{G_k}(\pi_{n,k}x,\pi_{n,k}y)=\text{rel}_{H_k}(\pi_{n,k}a,\pi_{n,k}b), \quad \text{for all} \quad 1\leq k\leq n,
\]
where $ \pi_{n,k}: X_n \to X_k$ denote the canonical projections; similarly for $ A_n$. The games $\widetilde{\rm Iso}(G_n,H_n)$ define a bisynchronous and nested family of rule functions denoted by $( \widetilde \lambda_n)_{n \in \bb N}$. Define
\[
{\rm Iso}(\mathfrak{G},\mathfrak{H})=\bigcap_{n\in\mathbb N}(\mathfrak{p}_n\times\mathfrak{p}_n)^{-1}\bigl(\widetilde{\rm Iso}(G_n,H_n)\bigr).
\]

In particular, setting 
$${\rm Iso}(G_k,H_k) = \{(x,y,a,b) \in X_k\times X_k \times A_k\times A_k : 
\text{rel}_{G_k}(x,y) = \text{rel}_{H_k}(a,b)\},$$
we have
\( 
\widetilde{\rm Iso}(G_n,H_n)=\bigcap_{k=1}^{n}\bigl(\pi_{n,k}^{\times4}\bigr)^{-1}\bigl({\rm Iso}(G_k,H_k)\bigr),
\)
and thus,
\[
{\rm Iso}(\mathfrak G,\mathfrak H)
=\bigcap_{n\in\bb N}(\mathfrak p_n\times\mathfrak p_n)^{-1}\bigl(\widetilde{\rm Iso}(G_n,H_n)\bigr)
=\bigcap_{n\in\bb N}(\mathfrak p_n\times\mathfrak p_n)^{-1}\bigl({\rm Iso}(G_n,H_n)\bigr).
\]

Recall that, $X_n=\prod_{k< n}[d_k]$ and $A_n=\prod_{k< n}[c_k]$, and for $k< n$ we write
\( 
X_{k,n}:=\prod_{k\leq j< n}[d_j]\) and 
\( A_{k,n}:=\prod_{k\leq j<n}[c_j],
\)
so that $X_n=X_k\times X_{k,n}$ and $A_n=A_k\times A_{k,n}$. 
In the following, if $S$ is a finite set, 
we write $J_S$ for the $S\times S$ matrix all of whose entries are 
equal to $1$.

\begin{lemma}\label{l:coherence_is_intertwining}
Let $\cl C$ be a unital $*$-algebra, let $U=(e_{x,a})_{x\in X_n,\,a\in A_n}\in M_{X_n,A_n}(\cl C)$. Fix $0\le k < n$. Then the following are equivalent:
\begin{enumerate}
\item[(i)] $(I_{X_k}\otimes J_{X_{k,n}}\otimes 1_{\cl C})U=U(I_{A_k}\otimes J_{A_{k,n}}\otimes 1_{\cl C})$; 
\item[(ii)] $U$ satisfies the row and column coherence relations:
\[
\sum_{|a'|=n-k}e_{x_0x',\,a_0a'}
=
\sum_{|a'|=n-k}e_{x_0x'',\,a_0a'},
\qquad \sum_{|x'|=n-k}e_{x_0x',\,a_0a'}
=
\sum_{|x'|=n-k}e_{x_0x',\,a_0a''},
\]
for all $x_0\in X_k,\ a_0\in A_k,\ x',x''\in X_{k,n}$, $a',a''\in A_{k,n}$
\end{enumerate}
\end{lemma}

\begin{proof}
We recall that $A_k = X_k$ for all $k\in \bb{N}$ and hence $A_{k,n} = X_{k,n}$ for all $k < n$. 
For simplicity, write \( 
C_k:=I_{X_k}\otimes J_{X_{k,n}}\) and \(
R_k:=I_{A_k}\otimes J_{A_{k,n}},
\)
and note that $(C_k)_{x,y}=1$ if $\pi_{n,k}(x)=\pi_{n,k}(y)$ and
$(C_k)_{x,y}=0$ otherwise, and similarly for $R_k$.

Fix $x=x_0x'\in X_n$ and $a=a_0a'\in A_n$ with $x_0\in X_k$, $a_0\in A_k$, and
compute the two sides of (i) at the entry $(x,a)$:
\[
(U(R_k\otimes 1_{\cl C})_{x,a}
=\sum_{b\in A_n}e_{x,b}\,(R_k)_{b,a}
=\sum_{\substack{b\in A_n\\ \pi_{n,k}(b)=a_0}}e_{x_0x',\,b}
=\sum_{|c'|=n-k}e_{x_0x',\,a_0c'}=:T_{x_0,a_0;x'}.
\]
For the right-hand side, 
\[
((C_k\otimes 1_{\cl C})U)_{x,a}
=\sum_{y\in X_n}(C_k)_{x,y}\,e_{y,a}
=\sum_{\substack{y\in X_n\\ \pi_{n,k}(y)=x_0}}e_{y,\,a_0a'}
=\sum_{|z'|=n-k}e_{x_0z',\,a_0a'}=:S_{x_0,a_0;a'}.
\]

 (i)$\Rightarrow$(ii) By the computations above, $T_{x_0,a_0;x'}= S_{x_0,a_0;a'}$ for every $ x'$ and $ a'$, hence the claim follows.
 
 For (ii)$\Rightarrow$(i) we argue as in Remark~\ref{rmk:welldef}, using $|X_{k,n}|=|A_{k,n}|$, to obtain that the column and row coherence sums are equal and then it follows from the computation above.
\end{proof}

We record the following standard fact for the convenience of the reader.
Given a unital *-algebra $\cl C$, we continue to use the term \emph{magic unitary} for a square matrix over $\cl C$ whose rows and columns are 
PVM's.

\begin{lemma}\label{l:intertwiner_orthogonality}
Let $X_0$, $A_0$ be finite sets with $|X_0|=|A_0|$. Let $U=(e_{x,a})_{x\in X_0,\,a\in A_0}$ be a magic unitary with entries in a unital
$*$-algebra $\cl C$, 
and let $M\in M_{X_0}(\bb C)$, $N\in M_{A_0}(\bb C)$. Then 
$(M\otimes 1_{\cl C})U = U(N\otimes 1_{\cl C})$ if and
only if $(M_{x,y}-N_{a,b})\,e_{x,a}e_{y,b}=0$ for all $x,y\in X_0$ and $a,b\in A_0$. In
particular, if $M$ and $N$ are $\{0,1\}$-matrices, then 
$(M\otimes 1_{\cl C})U = U(N\otimes 1_{\cl C})$ if and only if
$e_{x,a}e_{y,b}=0$ whenever $M_{x,y}\neq N_{a,b}$.
\end{lemma}

\begin{proof}
Fix $x\in X_0$ and $b\in A_0$, and consider
\begin{align*}
   & D_{x,b}:=((M\otimes 1_{\cl C})U-U(N\otimes 1_{\cl C}))_{x,b}\\
   &=\sum_{y\in X_0}M_{x,y}\,e_{y,b}-\sum_{a\in A_0}e_{x,a}\,N_{a,b}\\
    &=\sum_{y\in X_0}\sum_{a\in A_0}M_{x,y}\,e_{x,a}e_{y,b} -\sum_{a\in A_0}\sum_{y\in X_0}N_{a,b}\,e_{x,a}e_{y,b}\\
    & =\sum_{y\in X_0}\sum_{a\in A_0}\bigl(M_{x,y}-N_{a,b}\bigr)\,e_{x,a}e_{y,b}.
\end{align*}

On the other hand, in the first sum defining
$D_{x,b}$,  multiplying on
the left by $e_{x,a}$ and on the right by $e_{y,b}$ and using the orthogonality relations, we obtain
\begin{align*}
    e_{x,a}\,D_{x,b}\,e_{y,b}&=
    e_{x,a}\Bigl(\sum_{z\in X_0}M_{x,z}\,e_{z,b}\Bigr)e_{y,b} - e_{x,a}\Bigl(\sum_{c\in A_0}e_{x,c}N_{c,b}\Bigr)e_{y,b}\\
    &=M_{x,y}\,e_{x,a}e_{y,b}-N_{a,b}\,e_{x,a}e_{y,b}\\
    &=\bigl(M_{x,y}-N_{a,b}\bigr)\,e_{x,a}e_{y,b}.
\end{align*}

Suppose first that $(M_{x,y}-N_{a,b})e_{x,a}e_{y,b}=0$ for all $x,y,a,b$. Then
every term in the first formula vanishes, so
$D_{x,b}=0$ for all $x\in X_0,\,b\in A_0$; that is, $(M\otimes 1_{\cl C})U=U(N\otimes 1_{\cl C})$.

Conversely, suppose $(M\otimes 1_{\cl C})U=U(N\otimes 1_{\cl C})$, so $D_{x,b}=0$ for all $x,b$. Fix arbitrary
$x,y\in X_0$ and $a,b\in A_0$. Applying the compression formula to this 
gives
\[
0=e_{x,a}\,D_{x,b}\,e_{y,b}
=\bigl(M_{x,y}-N_{a,b}\bigr)\,e_{x,a}e_{y,b},
\]
which is the desired relation. This proves the equivalence.

If $M,N$ are $\{0,1\}$-matrices, then for each quadruple $(x,y,a,b)$ either
$M_{x,y}=N_{a,b}$, in which case $M_{x,y}-N_{a,b}=0$ and the identity
$(M_{x,y}-N_{a,b})e_{x,a}e_{y,b}=0$ holds trivially, or $M_{x,y}\neq N_{a,b}$,
in which case $M_{x,y}-N_{a,b}\in\{1,-1\}$ 
and the identity $(M_{x,y}-N_{a,b})e_{x,a}e_{y,b}=0$ is equivalent to
$e_{x,a}e_{y,b}=0$. Hence, in this case, the family of identities from the
lemma is equivalent to
\[
e_{x,a}e_{y,b}=0 \quad\text{whenever } M_{x,y}\neq N_{a,b},
\]
as claimed.
\end{proof}

Now we study the presentation for the bisynchronous game algebra of  $\widetilde{\rm Iso}(G_n,H_n)$.
\begin{theorem}\label{t:presentation_tilde}
Let $n\in\bb N$. The algebra $\cl A^{\rm bs}(\widetilde{\rm Iso}(G_n,H_n))$ is the unital
$*$-algebra generated by the entries of a magic unitary
$U=(e_{x,a})_{x\in X_n,\,a\in A_n}$ subject to the relations
\begin{enumerate}
\item $(I_{X_k}\otimes J_{X_{k,n}}\otimes 1)\,U=U\,(I_{A_k}\otimes J_{A_{k,n}}\otimes 1)$ for $1\le k < n$;
\item $(A_{G_k}\otimes J_{X_{k,n}}\otimes 1)\,U=U\,(A_{H_k}\otimes J_{A_{k,n}}\otimes 1)$ for $1\le k < n$;
\item $(A_{G_n}\otimes 1)\,U=U\,(A_{H_n}\otimes 1)$.
\end{enumerate}
\end{theorem}

\begin{proof}
For $1\le k\le n-1$ put $C_k:=I_{X_k}\otimes J_{X_{k,n}}$, $R_k:=I_{A_k}\otimes J_{A_{k,n}}$,
$\bar A_{G_k}:=A_{G_k}\otimes J_{X_{k,n}}$ and
$\bar A_{H_k}:=A_{H_k}\otimes J_{A_{k,n}}$. Since
$G_k$ is reflexive,
$(C_k)_{x,y}=\delta_{\pi_{n,k}(x),\,\pi_{n,k}(y)}$, and 
$$(\bar A_{G_k})_{x,y}=(A_{G_k})_{\pi_{n,k}(x),\,\pi_{n,k}(y)},$$
we have that $(C_k)_{x,y}=1$ if and only if $\mathrm{rel}_{G_k}(\pi_{n,k}(x),\pi_{n,k}(y)) = 0$, and
$(\bar A_{G_k})_{x,y}=1$ if and only if
$\mathrm{rel}_{G_k}(\pi_{n,k}(x),\pi_{n,k}(y))\in\{0,1\}$; likewise for $R_k$,
$\bar A_{H_k}$ and $\mathrm{rel}_{H_k}$. 
Therefore
\begin{equation}\label{eq:rule_mismatch}
\widetilde\lambda_n(x,y,a,b)=0 \iff \exists\,k\in[n]:
(C_k)_{x,y}\neq(R_k)_{a,b} \text{ or }
(\bar A_{G_k})_{x,y}\neq(\bar A_{H_k})_{a,b}.
\end{equation}

Recall that $\cl A^{\rm bs}(\widetilde{\rm Iso}(G_n,H_n))$ is generated by projections $e_{x,a}$
with $\sum_{a'}e_{x,a'}=\sum_{x'}e_{x',a} = 1$ and $e_{x,a}e_{y,b}=0$ whenever $\widetilde\lambda_n(x,y,a,b)=0$.
The case $k=n$ of (\ref{eq:rule_mismatch}) 
shows that these relations contain those of
$\cl A^{\rm bs}({\rm Iso}(G_n,H_n))$; hence, by 
Lemma \ref{l:intertwiner_orthogonality} (see also
\cite[Proposition~2.3]{bigalois}), we obtain (iii). 
Consequently both
$\cl A^{\rm bs}(\widetilde{\rm Iso}(G_n,H_n))$ and the algebra in the statement are universal
unital $*$-algebras generated by the entries of a magic unitary, subject respectively to
the rule relations and to (i)--(ii). By Lemma~\ref{l:intertwiner_orthogonality}, for a
magic unitary, relation (i) at level $k < n$ is equivalent to $e_{x,a}e_{y,b}=0$ whenever
$(C_k)_{x,y}\neq(R_k)_{a,b}$, and relation (ii) at level $k < n$ is equivalent to
$e_{x,a}e_{y,b}=0$ whenever
$(\bar A_{G_k})_{x,y}\neq(\bar A_{H_k})_{a,b}$. By \eqref{eq:rule_mismatch},
the union of these relations over $1\le k< n$ is precisely the set of rule relations.
Hence, we infer the existence of mutually inverse unital *-homomorphisms.
\end{proof}

\begin{remark}\label{rem_generators}
\rm 
It follows from Theorem \ref{t:presentation_tilde}, its proof and Lemma \ref{l:coherence_is_intertwining}, that the algebra $\cl A^{\rm bs}(\widetilde{\rm Iso}(G_n,H_n))$ is generated by $e_{x,a}$, $x\in X_k$, $a\in X_k$, $k\leq n$, such that $(e_{x,a})_{x\in X_n, a\in A_n}$ is a magic unitary,  and satisfy the relations $e_{x,a}=\sum_{|a'|=n-k}e_{xx',aa'}
  =\sum_{|y'|=n-k}e_{xy',ab'}$ for all words $x\in X_k$, $a\in A_k$, $x'\in X_{n,k}$ and $b'\in A_{n,k}$,  and $e_{x,a}e_{y,b}=0$ whenever $\mathrm{rel}_{G_k}(x,y)\ne\mathrm{rel}_{H_k}(a,b)$, for all $x,y\in X_k$, $a,b\in A_k$.  
    
\end{remark}


\subsection{Quantum automorphisms and coherent game algebras}
\label{ss_cohgamealg}

In what follows, we apply the bi-Galois machinery first introduced in 
the theory of non-local games in \cite{bigalois} and later used in \cite{bch}. 
We recall some basic notions that we will use subsequently, and 
refer the reader to \cite{timmermann} and \cite{bichon99} for further details on 
Hopf algebras and Galois extensions.
If $(H,\Delta,\varepsilon,S)$ is a Hopf *-algebra over $\mathbb C$, a 
\emph{Hopf *-ideal} $J\subseteq H$ is a two–sided *-ideal such that
\[
\Delta(J)\ \subseteq\ J \otimes H+ H\otimes J, \ \ 
\varepsilon(J)=0  \ \mbox{ and } \ 
S(J)\subseteq J.
\]
In this case the quotient algebra \(\overline H:=H/J\) is a Hopf *-algebra with structure maps
\[
\overline\Delta\bigl( q(h)\bigr):=(q\otimes q)\Delta(h),\qquad
\overline\varepsilon\bigl( q(h)\bigr):=\varepsilon(h),\qquad
\overline S\bigl( q(h)\bigr):=q\bigl(S(h)\bigr),
\]
where \(q:H\twoheadrightarrow \overline H\) is the quotient map.

Let $\cl A$ be a Hopf *-algebra. A unital *-algebra $\cl Z$ is  a \emph{left $\cl A$-*-comodule algebra} if there exists a unital *-homomorphism $\delta:\cl Z\to \cl A\otimes \cl Z$ with
  \[
  (\mathrm{id}\otimes \delta)\circ \delta=(\Delta\otimes \mathrm{id})\circ \delta,
  \qquad
  (\varepsilon\otimes \mathrm{id})\circ \delta=\mathrm{id}_{\cl Z};
  \]
 and a \emph{right $\cl A$-*-comodule algebra} if there exists a unital *-homomorphism $\gamma:\cl Z\to \cl Z\otimes \cl A$ with
  \[
  (\gamma\otimes \mathrm{id})\circ \gamma=(\mathrm{id}\otimes \Delta)\circ \gamma,
  \qquad
  (\mathrm{id}\otimes \varepsilon)\circ \gamma=\mathrm{id}_{\cl Z}.
  \]

\smallskip

Let $\cl A$ and $\cl B$ be Hopf *-algebras. A unital *-algebra $\cl Z$ is an \emph{$\cl A$-$\cl B$ bi-Galois extension} if:
\begin{enumerate}
  \item[(i)] $\cl Z$ is a left $\cl A$-*-comodule algebra via $\delta:\cl Z\to \cl A\otimes \cl Z$;
  \item[(ii)] $\cl Z$ is a right $\cl B$-*-comodule algebra via $\gamma:\cl Z\to \cl Z\otimes \cl B$;
  \item[(iii)] the coactions commute: $(\delta\otimes \mathrm{id})\circ \gamma=(\mathrm{id}\otimes \gamma)\circ \delta$;
  \item[(iv)] the canonical maps
  \begin{align*}
  \kappa_\delta:\cl Z\otimes \cl Z\to \cl A\otimes \cl Z,\quad z\otimes z'\mapsto \delta(z)(1\otimes z')\\
  \kappa_\gamma:\cl Z\otimes \cl Z\to \cl Z\otimes \cl B,\quad z\otimes z'\mapsto (z\otimes 1)\gamma(z'),
  \end{align*}
  are bijections.
\end{enumerate}

In the sequel we will be interested in bi-Galois extensions that can be represented in some $ {\rm C}^*$-algebra equipped with a trace. The existence of such representations will involve a criterion that guarantees the existence of invariant states on bi-Galois extensions.
Let $\cl Z$ be an $\A$-$\cl B$ bi-Galois extension. For a linear functional $\tau:\cl Z\to \bb C$ we say: $\tau$ is a \emph{state} if $\tau(z^{*}z)\ge 0$ for all $z$ and $\tau(1)=1$;
 $\tau$ is a \emph{trace} if it is a state and $\tau(z_1 z_2)=\tau(z_2 z_1)$ for all $z_1,z_2$; $\tau$ is \emph{left invariant} if $(\mathrm{id}\otimes \tau)\circ \delta(z)=\tau(z)\,1_{\cl A}$, for all $z$; $\tau$ is \emph{right invariant} if $(\tau\otimes \mathrm{id})\circ \gamma=\tau(z)\,1_{\cl{B}}$, for all $z$; $\tau$ is \emph{bi-invariant} if it is both left and right invariant; $\tau$ is \emph{faithful} if $\tau(z^{*}z)=0\Rightarrow z=0$.

We recall that $(G_n)_{n\in \bb{N}}$ and $(H_n)_{n\in \bb{N}}$ are 
the sequences of graphs, associated with the Cantor graphs $\mathfrak{G}$ and $\mathfrak{H}$, respectively; here, 
$G_n$ and $H_n$ have vertex set $X_n$, and $A_n$, $n\in \bb{N}$, respectively.
Let $\mathcal O({\rm Aut^{+}}(G_n))$ denote the Hopf *-algebra of the quantum automorphism group of $G_n$;  recall that $\mathcal O({\rm Aut^{+}}(G_n))$ is the unital *-algebra generated by projections
\( 
\{v_{x,y} :x,y\in X_n\}
\)
such that $V=(v_{x,y})$ is a magic unitary and
\( 
(A_{G_n}\otimes 1)\,V=V\,(A_{G_n}\otimes 1).
\)
The Hopf *-algebra structure is given on generators by $\Delta(v_{x,y})=\sum_{z\in X_n} v_{x,z}\otimes v_{z,y}$, $\varepsilon(v_{x,y})=\delta_{x,y}$ and $S(v_{x,y})=v_{y,x}$.

Fix $0\le k<n$. For $x,y\in X_k$ and words $y',y'',x',x''$ of length $n-k$, recall the
row-coherence relations 
\begin{equation}\label{eq:coh-row-qaut}
\sum_{|y'|=n-k} v_{\,xx',\,y y'}
\;=\;
\sum_{|y'|=n-k} v_{\,xx'',\,y y'}.
\end{equation}
and the column-coherence
relations
\begin{equation}\label{eq:coh-col-qaut}
\sum_{|x'|=n-k} v_{\,xx',\,y y'}
\;=\;
\sum_{|x'|=n-k} v_{\,xx',\,y y''}.
\end{equation}
Let $J_n^{(\mathrm{coh})}$ be the two-sided *-ideal in $\mathcal O(\mathrm{Aut}^{+}(G_n))$
generated  by all instances of \eqref{eq:coh-row-qaut} and
\eqref{eq:coh-col-qaut}.

\begin{proposition}\label{prop:Hopf-ideal}
We have that 
$J_n^{(\mathrm{coh})}$ is a Hopf *-ideal of $\mathcal O({\rm Aut^{+}}(G_n))$. 
Letting $\mathcal O_{\mathrm{coh}}(G_n) = \mathcal O({\rm Aut^{+}}(G_n))/J_n^{(\mathrm{coh})}$ and $\pi : \mathcal O({\rm Aut^{+}}(G_n)) \to \mathcal O_{\mathrm{coh}}(G_n)$
be the quotient map, we have that $\mathcal O_{\mathrm{coh}}(G_n)$ 
inherits a Hopf *-algebra structure with structure maps $\overline\Delta$, $\overline\varepsilon$, $\overline S$ uniquely determined by
\[
\overline\Delta\circ \pi=(\pi\otimes \pi)\circ\Delta,\qquad
\overline\varepsilon\circ \pi=\varepsilon,\qquad
\overline S\circ \pi=\pi\circ S.
\]
Moreover, $ \cl O_{\rm coh}(G_n)$ is of Kac type. We call 
$\mathcal O_{\mathrm{coh}}(G_n)$
the \emph{coherent quantum automorphism Hopf *-algebra}.
\end{proposition}

\begin{proof}
Let
\[
R = R^{(n,k)}(x,y;x',x'') := \sum_{|y'|=n-k} v_{\,xx',\,y y'}
-
\sum_{|y'|=n-k} v_{\,xx'',\,y y'},
\]
as a canonical generator of $J_n^{(\mathrm{coh})}$. 
The counit satisfies
\[
\varepsilon\!\left(\sum_{|y'|=n-k} v_{xx',yy'}\right)=\sum_{|y'|=n-k}\delta_{xx',yy'}=\delta_{x,y},
\]
which is independent of $x'$; hence $\varepsilon(R)=0$.

For the comultiplication, compute
\begin{eqnarray*}
\Delta(R)
& = & 
\sum_{|y'|=n-k}\sum_{w\in X_n} \big(v_{xx',w}-v_{xx'',w}\big)\otimes v_{w,yy'}\\
& = & \sum_{|z| = k} \sum_{|z'|=n-k} \big(v_{xx',zz'}-v_{xx'',zz'}\big)\otimes\sum_{|y'|=n-k} v_{zz',yy'}.
\end{eqnarray*}
Now if we apply $ q_n\otimes q_n$ on $ \Delta(R)$ where $ q_n$ is the quotient map corresponding to $J_{n}^{\rm (coh)}$, the second leg becomes independent of $ z'$ and thus, the first leg vanishes as it belongs to $ J_{n}^{\rm (coh)}$. We conclude that 
$\Delta(R) \in \ker (q_n\otimes q_n)$. 
By symmetry, 
$$ \Delta(J_n^{(\mathrm{coh})}) \subseteq \ker (q_n\otimes q_n) = J_n^{(\mathrm{coh})}\otimes \mathcal O({\rm Aut^{+}}(G_n))\ +\ \mathcal O({\rm Aut^{+}}(G_n))\otimes J_n^{(\mathrm{coh})}. $$
Finally, $S(v_{xx',yy'})=v_{yy',xx'}$ sends row-coherence generators to column-coherence generators and vice versa, so $S(J_n^{(\mathrm{coh})})\subseteq J_n^{(\mathrm{coh})}$. 
Therefore $J_n^{(\mathrm{coh})}$ is a Hopf *-ideal, and the quotient carries the induced Hopf *-algebra structure.
Since $\overline{S}^2 = {\rm id}$, \cite[Proposition 1.7.9]{nesh} implies that $ \cl O_{\rm coh}(G_n)$ is of Kac type. 
\end{proof}

Let $G_n$ and $H_n$ be graphs with vertex sets $X_n$ and $A_n$, respectively. The level-$n$ game *-algebra of the graph isomorphism game is the *-algebra\marginpar{}
\( 
{\cl A}^{\rm bs}({\rm Iso}(G_n,H_n))
\)
generated by projections $\{u_{x,a}:x\in X_n,\ a\in A_n\}$ such that $U=(u_{x,a})\in M_{X_n,A_n}({\cl A}^{\rm bs}({\rm Iso}(G_n,H_n))$ is a magic unitary and
\[
(A_{G_n}\otimes 1)U=U (A_{H_n} \otimes 1),
\]
with $A_{G_n}$, $A_{H_n}$ the adjacency matrices of $G_n$, $H_n$.

Fix $0\le k<n$, $x\in X_k$, $a\in A_k$, and two words $x',x''$ of length $n-k$.
We impose the \emph{row-coherence} relations
\begin{equation}\label{eq:coh-row-GI}
\sum_{|a'|=n-k} u_{\,xx',\,a a'}
\;=\;
\sum_{|a'|=n-k} u_{\,xx'',\,a a'},
\end{equation}
and, by swapping the roles of $x$ and $a$, the \emph{column-coherence} relations
\begin{equation}\label{eq:coh-col-GI}
\sum_{|x'|=n-k} u_{\,xx',\,a a'}
\;=\;
\sum_{|x'|=n-k} u_{\,xx',\,a a''}.
\end{equation}
Let $I_n^{({\rm coh})}$ be the two-sided *-ideal generated through the relations 
\eqref{eq:coh-row-GI} and \eqref{eq:coh-col-GI}.

\begin{lemma} \label{l:quotient_reduction_tilde}
For each $n\in \mathbb N$ we have 
$$\cl A^{\rm bs}({\rm Iso}(G_n,H_n))\big/ I_n^{({\rm coh})}\cong \cl A^{\rm bs}(\widetilde{{\rm Iso}}(G_n,H_n)).$$
\end{lemma}

\begin{proof}
We only show the existence of the first of the two isomorphisms. 
For $x,y\in X_n$, let $u_{x,y}$ and $\widetilde u_{x,y}$ denote the generators of
\( 
  \cl A^{\rm bs}({\rm Iso}(G_n,H_n))\) and \( 
  \cl A^{\rm bs}(\widetilde{{\rm Iso}}(G_n,H_n)),
\)
respectively. Set
\( 
  [u_{x,y}]:=q(u_{x,y}),
\)
where
$$
  q:\cl A^{\rm bs}({\rm Iso}(G_n,H_n))\longrightarrow
  \cl A^{\rm bs}({\rm Iso}(G_n,H_n))/I_n^{(\mathrm{coh})}
$$
is the quotient map.
Define
\( 
  \pi:\cl A^{\rm bs}(\widetilde{{\rm Iso}}(G_n,H_n))\longrightarrow
  \cl A^{\rm bs}({\rm Iso}(G_n,H_n))/I_n^{(\mathrm{coh})}
\)
on generators by setting
\( 
  \pi(\widetilde u_{x,y})=[u_{x,y}].
\)
To show that $\pi$ is well defined, by Remark \ref{rem_generators} it remains to verify the relations involving the induced generators at levels $k<n$. For $x,a\in X_k$, define
\begin{equation}\label{eq_squarebr}
  [\widehat u_{x,a}]
  :=\sum_{|a'|=n-k}[u_{xx',aa'}]
  =\sum_{|y'|=n-k}[u_{xy',ab'}],   
\end{equation}
where the first sum is independent of the choice of $x'$ and the second is independent of the choice of $b'$. We claim that
\[
  [\widehat u_{x,a}][\widehat u_{y,b}]=0
  \qquad\text{whenever}\qquad
  {\rm rel}_{G_k}(x,y)\neq {\rm rel}_{H_k}(a,b),
\]
for all $x,y,a,b\in X_k$.

Suppose, towards a contradiction, that
\( 
  [\widehat u_{x,a}][\widehat u_{y,b}]\neq 0.
\)
Using the first representations of $[\widehat u_{x,a}]$ and $[\widehat u_{y,b}]$ in the equalities (\ref{eq_squarebr}), 
we see that for every $x',x''$, $|x'|=|x''|=n-k$ there exist  words $a',a''$ of length $n-k$  such that
\( 
  [u_{xx',aa'}][u_{yx'',ba''}]\neq 0.
\)
By the defining relations of $\cl A^{\rm bs}({\rm Iso}(G_n,H_n))$, this implies that 
\( 
  {\rm rel}_{G_n}(xx',yx'')={\rm rel}_{H_n}(aa',ba'').
\)

Similarly, using the second representations of the induced generators
in (\ref{eq_squarebr}), we obtain the converse statement: for every $a',a''$ there exist $x',x''$ such that
\( 
  {\rm rel}_{G_n}(xx',yx'')={\rm rel}_{H_n}(aa',ba'').
\)

We now distinguish the possible mismatches between the relations at level $k$.
Assume first that $x\not\sim_{G_k}y$ whereas $a\sim_{H_k}b$, where we use $\sim_H$ for the adjacency relation in a graph $H$.  If $x=y$, choose $x'=x''$. Then
\[
  {\rm rel}_{G_n}(xx',xx')={\rm rel}_{H_n}(aa',ba'')
\]
is impossible for every choice of $a',a''$, since $a\neq b$.

If $x\neq y$ and $x\not\sim_{G_k}y$, then there are no $x',x''$ for which
\[
  xx'\sim_{G_n}yx''.
\]
On the other hand, since $a\sim_{H_k}b$, there exist $a',a''$ of length $n-k$ such that
\[
  aa'\sim_{H_n}ba''.
\]
For these fixed $a',a''$, the equality
\( 
  {\rm rel}_{G_n}(xx',yx'')={\rm rel}_{H_n}(aa',ba'')
\)
cannot hold for any $x',x''$ of length $n-k$, contradicting the conclusion above.

The case
\[
  x\sim_{G_k}y
  \qquad\text{and}\qquad
  a\not\sim_{H_k}b
\]
is treated symmetrically. Hence
\( 
  [\widehat u_{x,a}][\widehat u_{y,b}]=0
\)
whenever ${\rm rel}_{G_k}(x,y)\neq{\rm rel}_{H_k}(a,b)$, and therefore $\pi$ is well defined.
The cases $x=y$, $a\ne b$, $a\not\sim_{H_k} b$ and $x\ne y$, $x\not\sim_{G_k} y$, $a=b$ can be treated similarly. 

Next, define a $*$-homomorphism
\( 
  \rho:\cl A^{\rm bs}({\rm Iso}(G_n,H_n))\longrightarrow
  \cl A^{\rm bs}(\widetilde{{\rm Iso}}(G_n,H_n))
\)
by setting
\( 
  \rho(u_{x,a})=\widetilde u_{x,a}.
\)
By Theorem \ref{t:presentation_tilde}
\[
  I_n^{(\mathrm{coh})}\subseteq\ker\rho.
\]
Therefore, $\rho$ factors through the quotient, yielding a $*$-homomorphism
\[
  \overline\rho:
  \cl A^{\rm bs}({\rm Iso}(G_n,H_n))/I_n^{(\mathrm{coh})}
  \longrightarrow
  \cl A^{\rm bs}(\widetilde{\rm Iso}(G_n,H_n)).
\]
Finally, the definitions on generators give $\overline\rho\circ\pi=\operatorname{id}$ and $\pi\circ\overline\rho=\operatorname{id}$.
Thus $\pi$ is a $*$-isomorphism, with inverse $\overline\rho$.

    \end{proof}

\begin{corollary}\label{c_cor79}
    For a Cantor quantum permutation $D$, the associated sequence $(D_m)_{m\in\bb N}$, and $n \in \bb N$, the condition $D_n(A_{G_k}\otimes J_{X_{k,n}} \otimes 1)=(A_{H_k}\otimes J_{X_{k,n}} \otimes 1)D_n$, for all $ 1\leq k \leq n$ is equivalent to $D_n(A_{G_n} \otimes 1)=(A_{H_n} \otimes 1)D_n$.
\end{corollary}
\begin{proof}
One direction is clear using $ k=n$. For the other direction, note that from the structure of Cantor quantum permutation, $D$ satisfies the row and column coherence relations. Thus, using the assumption,
Lemma~\ref{l:quotient_reduction_tilde} and Theorem~\ref{t:presentation_tilde} we obtain the converse.
\end{proof}

\smallskip

Recall \cite[Theorem 4.5]{bigalois} that if ${\cl A}^{\rm bs}({\rm Iso}(G_n,H_n))$ is non-zero then it is a
$\cl O({\rm Aut^{+}}(G_n))$--$\cl O({\rm Aut^{+}}(H_n))$ bi-Galois extension via the coactions
$$\delta_L(u_{x,a}) = \sum_{y\in X_n} v^G_{x,y}\otimes u_{y,a},\ \ \ 
\delta_R(u_{x,a}) = \sum_{b\in A_n} u_{x,b}\otimes v^H_{b,a}.$$

 Let $J_G:=J_{G,n}^{(\mathrm{coh})}\subseteq \cl O(\mathrm{Aut}^{+}(G_n))$,
$J_H:=J_{H,n}^{(\mathrm{coh})}\subseteq \cl O(\mathrm{Aut}^{+}(H_n))$ and  $I:=I_n^{(\mathrm{coh})}\subseteq {\cl A}^{\rm bs}({\rm Iso}(G_n,H_n))$ be the coherence ideals
from earlier with quotient maps
\[
q_G:\ \cl O(\mathrm{Aut}^{+}(G_n))\twoheadrightarrow \cl O_{\mathrm{coh}}(G_n),\quad
q_H:\ \cl O(\mathrm{Aut}^{+}(H_n))\twoheadrightarrow \cl O_{\mathrm{coh}}(H_n),
\]
and
\[
q:\ {\cl A}^{\rm bs}({\rm Iso}(G_n,H_n))\twoheadrightarrow {\cl A}^{\rm bs}(\widetilde{{\rm Iso}}(G_n,H_n)).
\]

\begin{theorem}\label{thm:COH-bi-Galois-quotients-kernels-commute}
If $\cl A^{\rm bs}(\widetilde{{\rm Iso}}(G_n,H_n))\neq 0$, then it is an $\cl O_{\mathrm{coh}}(G_n)$–$\cl O_{\mathrm{coh}}(H_n)$ bi-Galois extension with coactions
\[
\overline\delta_L\circ q \;=\; (q_G\otimes q)\circ \delta_L:\ 
\cl A^{\rm bs}({\rm Iso}(G_n,H_n))\longrightarrow \cl O_{\mathrm{coh}}(G_n)\otimes \cl A^{\rm bs}(\widetilde{{\rm Iso}}(G_n,H_n)),
\]
\[
\overline\delta_R\circ q \;=\; (q\otimes q_H)\circ \delta_R:\ 
\cl A^{\rm bs}({\rm Iso}(G_n,H_n))\longrightarrow \cl A^{\rm bs}(\widetilde{{\rm Iso}}(G_n,H_n))\otimes \cl O_{\mathrm{coh}}(H_n),
\]
where $\delta_L,\delta_R$ are the commuting coactions of  $\cl A^{\rm bs}({\rm Iso}(G_n,H_n))$.
\end{theorem}

\begin{proof}
By Proposition~\ref{prop:Hopf-ideal}, the spaces $J_G$ and $J_H$ are Hopf *-ideals, hence
$\cl O_{\mathrm{coh}}(G_n)$ and $\cl O_{\mathrm{coh}}(H_n)$ are Hopf *-algebras.
We claim that 
\begin{equation}\label{eq:coh-kern-left-right}
\delta_L(I)\ \subseteq\ \ker(q_G\otimes q) \ \mbox{ and } \ 
\delta_R(I)\ \subseteq\ \ker(q\otimes q_H).
\end{equation}
We check the containments on the canonical generators of $I$. For a row–coherence generator
\[
R_{x,a;x',x''}\ :=\ \sum_{|a'|=n-k}u_{xx',aa'}-\sum_{|a'|=n-k}u_{xx'',aa'}\in I,
\]
we have 
\[
\delta_L(R_{x,a;x',x''})
= \sum_{|y| = k} \sum_{|y'| = n-k}
\hspace{-0.2cm} \Big( v^G_{xx',yy'}-  v^G_{xx'',yy'}\Big)\hspace{-0.05cm}\otimes\hspace{-0.05cm}
\Big(\hspace{-0.1cm}\sum_{|a'|=n-k}\hspace{-0.1cm} u_{yy',aa'}\Big).
\]
Applying $q_G\otimes q$, the right leg becomes independent of $y'$ and thus the left leg is a row–coherence difference in
$\cl O({\rm Aut}^+(G_n))$ and vanishes modulo $J_G$, while the right factor is a row sum
whose class lies in $\cl A^{\rm bs}(\widetilde{{\rm Iso}}(G_n,H_n))$. Hence $(q_G\otimes q)\circ \delta_L(R_{x,a;x',x''})=0$.
The column–coherence generators are treated identically; the right coaction gives the second inclusion in \eqref{eq:coh-kern-left-right}. Therefore the formulas in the statement give rise to well–defined *-homomorphisms
\begin{align*}
    \overline\delta_L:\ \cl A^{\rm bs}(\widetilde{{\rm Iso}}(G_n,H_n))\to \cl O_{\mathrm{coh}}(G_n)\otimes \cl A^{\rm bs}(\widetilde{{\rm Iso}}(G_n,H_n)),\\
    \overline\delta_R:\ \cl A^{\rm bs}(\widetilde{{\rm Iso}}(G_n,H_n))\to \cl A^{\rm bs}(\widetilde{{\rm Iso}}(G_n,H_n))\otimes \cl O_{\mathrm{coh}}(H_n).
\end{align*}
The coassociativity and counit conditions descend from those of $\delta_L$ and $\delta_R$.

On $\cl A^{\rm bs}({\rm Iso}(G_n,H_n))$ we have the identity
\((\delta_L\otimes \id)\circ\delta_R=(\id\otimes \delta_R)\circ\delta_L\).
Applying \(q_G\otimes q\otimes q_H\) and using \eqref{eq:coh-kern-left-right} yields
\[
(\overline\delta_L\otimes \id)\circ \overline\delta_R
\ =\
(\id\otimes \overline\delta_R)\circ \overline\delta_L
\quad\text{on }\cl A^{\rm bs}(\widetilde{{\rm Iso}}(G_n,H_n)).
\]

\smallskip\noindent
Let
\[
\kappa_L: \cl A^{\rm bs}({\rm Iso}(G_n,H_n))\otimes \cl A^{\rm bs}({\rm Iso}(G_n,H_n)) \longrightarrow \cl O({\rm Aut}^+(G_n))\otimes \cl A^{\rm bs}({\rm Iso}(G_n,H_n)),
\]
with $ \kappa_L(a\otimes b)=\delta_L(a)\,(1\otimes b)$
be the left canonical map, which is a linear isomorphism.
We construct an explicit inverse. Define a unital *-homomorphism
$$\alpha: \cl O({\rm Aut}^+(G_n)) \longrightarrow \cl A^{\rm bs}({\rm Iso}(G_n,H_n))\otimes \cl A^{\rm bs}({\rm Iso}(G_n,H_n))$$
by letting  
$\alpha(v^G_{x,y}) = \sum_{a\in A_n} u_{x,a} \otimes u_{y,a}$. 
Let 
\[
\eta_L:
\cl O({\rm Aut}^+(G_n))\otimes \cl A^{\rm bs}({\rm Iso}(G_n,H_n))\ \longrightarrow \cl A^{\rm bs}({\rm Iso}(G_n,H_n))\otimes \cl A^{\rm bs}({\rm Iso}(G_n,H_n)),
\]
defined by $\eta_L := (\id\otimes m)\circ(\alpha\otimes \id)$,
where $m$ is multiplication on $\cl A^{\rm bs}({\rm Iso}(G_n,H_n))$.
One checks that
\[
\eta_L\circ \kappa_L=\id,\qquad \kappa_L\circ \eta_L=\id,
\]
so $\eta_L=\kappa_L^{-1}$.

By similar arguments to the ones in the first paragraph of the proof, 
there exists a well-defined *-homomorphism
\[
\overline\alpha:\ \cl O_{\rm coh}(G_n)\to \cl A^{\rm bs}(\widetilde{{\rm Iso}}(G_n,H_n))\otimes \cl A^{\rm bs}(\widetilde{{\rm Iso}}(G_n,H_n))
\]
satisfying the relation 
\[
(q\otimes q)\circ \alpha\ =\ \overline\alpha\circ q_G.
\]
Define $\overline\eta_L := (\id\otimes \overline m)\circ(\overline\alpha\otimes \id)$, that is,
\[
\overline\eta_L :
\cl O_{\rm coh}(G_n)\otimes \cl A^{\rm bs}(\widetilde{{\rm Iso}}(G_n,H_n)) \longrightarrow\ \cl A^{\rm bs}(\widetilde{{\rm Iso}}(G_n,H_n))\otimes \cl A^{\rm bs}(\widetilde{{\rm Iso}}(G_n,H_n)),
\]
where $\overline m$ is the multiplication on $\cl A(\widetilde{{\rm Iso}}(G_n,H_n))$.

Let
\[
\overline\kappa_L:\ \cl A^{\rm bs}(\widetilde{{\rm Iso}}(G_n,H_n))\otimes \cl A^{\rm bs}(\widetilde{{\rm Iso}}(G_n,H_n))\ \longrightarrow\ \cl O_{\rm coh}(G_n)\otimes \cl A^{\rm bs}(\widetilde{{\rm Iso}}(G_n,H_n)),
\]
be such that  $\overline\kappa_L( a\otimes  b)=\overline\delta_L( a)\,(1\otimes  b)$.
By construction,
\[
(q_G\otimes q)\circ \kappa_L\ =\ \overline\kappa_L\circ (q\otimes q),\qquad
(q\otimes q)\circ \eta_L\ =\ \overline\eta_L\circ (q_G\otimes q).
\]
Hence
\[
\overline\eta_L\circ \overline\kappa_L
=\id \ \mbox{ and } \ 
\overline\kappa_L\circ \overline\eta_L
=\id,
\]
so $\overline\kappa_L$ is a bijection.

The case for  $\overline\kappa_R$ is identical.
The two descended coactions commute and both canonical maps are bijections. Hence $\cl A^{\rm bs}(\widetilde{{\rm Iso}}(G_n,H_n))$ is an $\cl O_{\rm coh}(G_n)$–$\cl O_{\rm coh}(H_n)$ bi-Galois extension.
\end{proof}

Let $F \in M_n$ be an invertible matrix. The \emph{universal unitary quantum group} $\mathcal{O}(U_F^+)$ \cite{vdw} is the Hopf *-algebra 
generated by the entries of an $n\times n$ matrix
$u = [u_{i,j}]$ such that $u$ and 
\( 
 F\, \bar u\,  F^{-1}
\)
are unitaries in $M_n(\mathcal{O}(U_F^+))$.  Let $n \in \mathbb{N}$ and let $F_1, F_2 \in M_n$ be invertible. Define 
\(
\mathcal{O}(U_{F_1}^+,\, U_{F_2}^+)
\)
to be the unital *-algebra generated by the coefficients $z_{i,j}$ of an $n \times n$ matrix 
\( 
z = [z_{i,j}]_{1 \le i,j \le n} \in M_n\big(\mathcal{O}(U_{F_1}^+,\, U_{F_2}^+)\big)
\)
subject to the relations that make both $ z$ and $ F_1 \overline{z} F_2^{-1}$ unitaries, where $\overline{z} = [z_{i,j}^*]$. If $\mathcal{O}(U_{F_1}^+,\, U_{F_2}^+) \neq 0$, then by \cite{bichon99} it is an 
$\mathcal{O}(U_{F_1}^+)$–$\mathcal{O}(U_{F_2}^+)$ bi-Galois extension with respect to the bicomodule structure given on generators by
\[
\alpha_{F_1,F_2} : 
\mathcal{O}(U_{F_1}^+,\, U_{F_2}^+) \to 
\mathcal{O}(U_{F_1}^+) \otimes \mathcal{O}(U_{F_1}^+,\, U_{F_2}^+), 
\qquad
\alpha_{F_1,F_2}(z_{i,j}) = \sum_{k=1}^n u_{i,k} \otimes z_{kj},
\]
\[
\beta_{F_1,F_2} : 
\mathcal{O}(U_{F_1}^+,\, U_{F_2}^+) \to 
\mathcal{O}(U_{F_1}^+,\, U_{F_2}^+) \otimes \mathcal{O}(U_{F_2}^+), 
\qquad
\beta_{F_1,F_2}(z_{i,j}) = \sum_{l=1}^n z_{il} \otimes v_{lj},
\]
where $u = [u_{i,j}]$ and $v = [v_{i,j}]$ denote the fundamental co-representations of $U_{F_1}^+$ and $U_{F_2}^+$, respectively.

\begin{proposition}\label{prop:coh-bichon-left}
Assume that $\cl A^{\rm bs}(\widetilde{{\rm Iso}}(G_n,H_n))\neq 0$. Then the coherent game algebra
$\cl A^{\rm bs}(\widetilde{{\rm Iso}}(G_n,H_n))$ admits a left $\cl O_{\mathrm{coh}}(G_n)$-invariant state
\[
\omega_L: \cl A^{\rm bs}(\widetilde{{\rm Iso}}(G_n,H_n)) \longrightarrow \Bbb C.
\]
Moreover, $\omega_L$ is tracial since $\cl O_{\mathrm{coh}}(G_n)$ is of Kac type.
\end{proposition}

\begin{proof}
We show that  there exist matrices $F_G, F_H$, a surjective *-homomorphism
\(
\pi_G:\ \cl O(U^+_{F_G}) \to \cl O_{\mathrm{coh}}(G_n),
\)
and a surjective *-homomorphism
\( 
\sigma:\ \cl O(U^+_{F_G},U^+_{F_H})\ \twoheadrightarrow\ 
\cl A^{\rm bs}(\widetilde{{\rm Iso}}(G_n,H_n)),
\)
such that the compatibility relation
\begin{equation}\label{eq:coh-bichon-compat}
\overline\delta_L\circ \sigma\ =\ (\pi_G\otimes \sigma)\,\alpha_{F_G,F_H}
\end{equation}
holds. The result will follow by Bichon’s criterion \cite[Theorem 6.2.6]{bichon99}.

By \cite[Theorem 4.7]{bigalois} and its proof, there exist matrices $F_G,F_H$ and a surjective *-homomorphisms
\[
\pi_G^0:\ \cl O(U^+_{F_G})\twoheadrightarrow \cl O(\mathrm{Aut}^{+}(G_n)),
\qquad
\sigma^0:\ \cl O(U^+_{F_G},U^+_{F_H})\twoheadrightarrow 
\cl A^{\rm bs}({\rm Iso}(G_n,H_n)),
\]
and the left coaction satisfies the intertwining identity
\( 
\delta_L\circ \sigma^0 = (\pi_G^0\otimes \sigma^0)\circ\alpha_{F_G,F_H}.
\)
By Proposition~\ref{prop:Hopf-ideal},  the maps factor through the coherent quotients:
\(
\pi_G = q_G\circ \pi_G^0\) and \( \sigma = q\circ \sigma^0 \),
and the compatibility identity descends to \eqref{eq:coh-bichon-compat}. Since $\sigma$ is surjective and the condition \eqref{eq:coh-bichon-compat} is exactly the hypothesis of Bichon’s criterion applied to the left coaction, we conclude that
$\cl A^{\rm bs}(\widetilde{{\rm Iso}}(G_n,H_n))$ admits a left $\cl O_{\mathrm{coh}}(G_n)$-invariant state $\omega_L$.
Traciality follows from the Kac assumption (see  \cite[Theorem 3.15]{bigalois}).
\end{proof}


\subsection{Main characterisation theorem}\label{ss_maincharteo}

Let $ \mathfrak{G} \subseteq \Omega_X \times \Omega_X$ and $ \mathfrak{H} \subseteq \Omega_A \times \Omega_A$ be two Cantor graphs.
Let also $ {\rm Iso}(\mathfrak{G},\mathfrak{H})$ be the Cantor graph isomorphism game and $ {\cl A}^{\rm bs}({\rm Iso}(\mathfrak{G},\mathfrak{H}))$ its associated bisynchronous game *-algebra. 

The following result is probably known. We include a proof for the convenience of the reader. For a C*-algebra $C$ denote by $ T(C)$ the tracial state  space.
\begin{lemma}\label{l_trace_inductive_limit}
Let $(C_n,\rho_n)_{n\in\bb N}$ be an inductive sequence of separable
unital C$^*$-algebras. If $C_n$ admits a tracial state for each $ n \in \bb N$, then the inductive limit
\( 
C=\varinjlim(C_n,\rho_n)
\)
admits a tracial state.
\end{lemma}
\begin{proof}
Let $\tau_m\in T(C_m)$, $m\in \bb{N}$, and fix $n\in \bb{N}$. For $m\ge n$, set
\[
\tau_m^{(n)}:=\tau_m\circ \rho_{n,m}\in T(C_n).
\]
Since $C_n$ is separable, $T(C_n)$ is weak-$*$ compact and metrizable for each $ n \in \bb N$.  Choose a subsequence $(m_\ell^{})_\ell$ such that
$\tau_{m_\ell}^{(1)}$ converges to some $\sigma_1\in T(C_1)$. Passing to  a subsequence of $(m_\ell^{})_\ell$, denoted the same way, we may assume that $\tau_{m_\ell}^{(2)}$ converges to some
$\sigma_2\in T(C_2)$. Continuing inductively and using a diagonal subsequence, we obtain
a sequence, still denoted $(m_\ell^{})_\ell$, such that for every $n$,
\[
\tau_{m_\ell}^{(n)}\rightarrow \sigma_n\in T(C_n)
\]
in the weak-* topology. For $m_\ell\ge n+1$ we have
$\tau_{m_\ell}^{(n)}
=
\tau_{m_\ell}^{(n+1)}\circ \rho_n$. 
Passing to the limit gives
\[
\sigma_n=\sigma_{n+1}\circ \rho_n,\qquad n\in\mathbb N.
\]
Thus $(\sigma_n)_n$ is compatible and hence induces a tracial state on
\( 
C=\varinjlim(C_n,\rho_n).
\)
\end{proof}

\begin{theorem}\label{thm:main_equivalence}
    The following are equivalent:
    \begin{enumerate}
        \item[(i)]  $\cl A^{\rm bs} ({\rm Iso}(\mathfrak{G},\mathfrak{H})) \neq 0$;
        \item[(ii)] the Cantor game $ {\rm Iso}(\mathfrak{G},\mathfrak{H})$ has a perfect graded quantum commuting strategy;
        \item[(iii)] there exist a tracial von Neumann algebra $\cl N$ and a Cantor quantum permutation $ D\in \cl B(L^2(\Omega_X))\bar\otimes {\cl N}$  such that 
        \begin{equation}\label{commutation}
        (T_{\chi_{\mathfrak{G}_n}}\otimes 1)D = D (T_{\chi_{\mathfrak{H}_n}}\otimes 1), \ \ \ n\in \bb{N};
        \end{equation}
        \item[(iv)] there exist a Hilbert space $H$ and a Cantor quantum permutation 
        $D\in\cl B(L^2(\Omega_X))$ $\bar\otimes$ $\cl B(H)$ such that
        (\ref{commutation}) holds.
       \end{enumerate}
\end{theorem}

\begin{proof}
    (i)$\Rightarrow$(ii)
  Assume  $\cl A^{\rm bs}({\rm Iso}(\mathfrak{G},\mathfrak{H})) \neq 0$. Then, by the definition of  $\cl A^{\rm bs}({\rm Iso}(\mathfrak{G},\mathfrak{H})$,  each coherent algebra 
  $\cl B_{n}=\cl A^{\rm bs}(\widetilde{{\rm Iso}}(G_n,H_n))$
is non-zero. By Proposition \ref{prop:coh-bichon-left} and \cite[Theorem 3.15]{bigalois}, for each $n$, the universal C*-algebra $C_n:=C^*(\cl B_n) $ is non-trivial and admits a tracial state $\tau_n$. We may lift the connecting maps $ \pi_n^{\rm bs} : \cl B_n \to \cl B_{n+1} $, $ \pi_n^{\rm bs}(e_{x,a})= \sum_{|a'| = 1}e_{xx',aa'}$, to unital *-homomorphisms $  \rho_n : C_n \to C_{n+1} $. Consider the $C^*$-algebra inductive limit  $(C,\rho_{n,\infty})=\varinjlim(C_n,\rho_n)$ and recall that 
$ \cl A^{\rm bs}({\rm Iso}(\mathfrak{G},\mathfrak{H}))=\varinjlim(\cl B_n,\pi_n)$. Then there is a unital *-homomorphism $ \pi : \cl A^{\rm bs}({\rm Iso}(\mathfrak{G},\mathfrak{H}))\to C $ such that $ \pi \circ \pi^{\rm s}_{n,\infty}=  \rho_{n,\infty} \circ j_{n}$ where $ j_n : \cl B_n \to C_n$ the canonical map to the $C^*$-completion. 

Finally, using Lemma~\ref{l_trace_inductive_limit} the C*-algebra $C$ possesses a tracial state and the proof is complete by Theorem \ref{th_bs_game_alg_reps} since we found a *-homomorphism from the graded game *-algebra into a $C^*$-algebra equipped with a faithful tracial state (by passing to the GNS representation we may assume that the trace is faithful).

(ii)$\Rightarrow$(iii)
It follows from Theorem \ref{th_bs_game_alg_reps} that there exists a unital *-homo\-mor\-phism $ \pi : \cl A^{\rm bs}({\rm Iso}(\mathfrak{G},\mathfrak{H})) \to \cl A$ where $ \cl A$ is a unital C*-algebra equipped with a faithful tracial state $\tau$. Taking the GNS representation $\pi_\tau$ and replacing $\cl A\simeq\pi_\tau(\cl A)$  by $\pi_\tau(\cl A)''$ we view $\cl A$ as a von Neumann algebra. 
Let $D\in\cl B(L^2(\Omega_X))\bar\otimes\cl A$ be the quantum permutation on $\Omega_X$ associated to $\pi$
 and set $D_n=\sum_{z,x\in X_n} E_{z,x}\otimes\pi(p_{z,x}) $.   We have $(A_{G_n}\otimes 1)D_n=D_n(A_{H_n}\otimes 1)$. Using Lemma \ref{L:finite_infinite_intertwining}, 
the latter relation is equivalent to 
$(T_{\chi_{\mathfrak{G}_n}} \otimes 1)D=D(T_{\chi_{\mathfrak{H}_n}}\otimes 1)$. 
    
(iii)$\Rightarrow$(iv) is trivial.

 (iv)$\Rightarrow$(i)
Assume there exists a Cantor quantum permutation $D$ acting on $H$ such that $(T_{\chi_{\mathfrak{G}_n}}\otimes I_H)D = D (T_{\chi_{\mathfrak{H}_n}}\otimes I_H)$. 
Since $D$ is a Cantor quantum permutation, each $D_n \in M_{X_n} \otimes \mathcal{B}(H)$ is a quantum permutation matrix on the finite set $X_n$. Let $D_n = (u_{x,a})_{x,a \in X_n}$ where $u_{x,a}$ are projections in $\mathcal{B}(H)$.
The commutation relation $(T_{\chi_{\mathfrak{G}_n}}\otimes 1)D = D (T_{\chi_{\mathfrak{H}_n}}\otimes 1)$ implies via Lemma \ref{L:finite_infinite_intertwining} that for every $n$, the finite approximations satisfy
$$(A_{G_n} \otimes 1) D_n = D_n (A_{H_n} \otimes 1);$$
by Corollary \ref{c_cor79}, there is a canonical unital *-representation of $\cl B_n = \cl A^{\rm bs}(\widetilde{{\rm Iso}}(G_n,H_n))$ into 
$\cl B(H)$. Using Lemma \ref{L:compat}, we obtain a unital *-representation 
of $\cl B$ on $H$. 
\end{proof}

The following is an extension of \cite[Theorem 4.3]{paulsen-rahaman} to Cantor graphs. 
We note that if $\Phi : \cl B(H)\to \cl B(H)$ is a 
completely positive trace-preserving map, 
$\Phi$ leaves the trace class $\cl S_1(H)$ invariant, 
and hence the restriction $\Phi|_{\mathcal{S}_1} : \cl S_1(H)\to \cl S_1(H)$ has a dual 
$(\Phi|_{\mathcal{S}_1})^* : \cl B(H)\to \cl B(H)$ defined via the 
relation 
$\langle(\Phi|_{\mathcal{S}_1})^*(S),\rho\rangle = 
\langle S,\Phi(\rho)\rangle$.

\begin{corollary}\label{cor:iso_fact}
 Let $\mathfrak{G}$, $\mathfrak{H}$ be Cantor graphs on $\Omega_X$. The following are equivalent:
 \begin{enumerate}
 \item the graph isomorphism game ${\rm Iso}(\mathfrak{G}, \mathfrak{H})$ has a perfect graded quantum commuting strategy;
\item there is a factorisable map $\Phi:\mathcal{B}(L^2(\Omega_X))\to \mathcal{B}(L^2(\Omega_X))$ via quantum permutation on $\Omega_X$ such that $\Phi(T_{\chi_{\mathfrak{G}_n}})=T_{\chi_{\mathfrak{H}_n}}$ and $(\Phi|_{\mathcal{S}_1})^*(T_{\chi_{\mathfrak{H}_n}})=T_{\chi_{\mathfrak{G}_n}}$, $n\in \bb{N}$. 
 \end{enumerate}
\end{corollary}

\begin{proof}
(i)$\Rightarrow$(ii)    Assume there exists a perfect graded quantum commuting strategy for the isomorphism game ${\rm Iso}(\mathfrak{G}, \mathfrak{H})$. Then, by Theorem \ref{thm:main_equivalence}, there are a tracial von Neumann algebra $(\mathcal{N},\tau)$ and  a Cantor quantum permutation $D\in \mathcal{B}(L^2(\Omega_X))\bar\otimes\mathcal{N}$ such that $(T_{\chi_{\mathfrak{G}_n}}\otimes I_H)D = D (T_{\chi_{\mathfrak{H}_n}}\otimes I_H)$, $n\in \bb{N}$. Let $\Phi: \mathcal{B}(L^2(\Omega_X))\to \mathcal{B}(L^2(\Omega_X))$, $\Phi(X)=(\text{id}\otimes\tau)(D^*(X\otimes 1_H)D)$. Using $D^*(T_{\chi_{\mathfrak{G}_n}}\otimes I_H)D = (T_{\chi_{\mathfrak{H}_n}}\otimes I_H)$ we have
$\Phi(T_{\chi_{\mathfrak{G}_n}})=T_{\chi_{\mathfrak{H}_n}}$. As $\Phi$ is factorisable, its restriction to $\mathcal{S}_1$ has the image in $\mathcal{S}_1$. In particular, for $S\in \mathcal{S}_1$ we have  
    \begin{eqnarray*}
    &&\langle (\Phi|_{\mathcal{S}_1})^*(T_{\chi_{\mathfrak{H}_n}}), S\rangle
    =\langle T_{\chi_{\mathfrak{H}_n}},\Phi (S)\rangle
    =\langle T_{\chi_{\mathfrak{H}_n}}, (\text{id}\otimes\tau)(D^*(S\otimes 1)D)\rangle\\
    &&=(\text{Tr}\otimes\tau)\big((T_{\chi_{\mathfrak{H}_n}}\otimes 1)D^*(S\otimes 1)D\big)
    =(\text{Tr}\otimes\tau)\big(D(T_{\chi_{\mathfrak{G}_n}}\otimes 1)D^*(S\otimes 1)\big)\\
    &&=(\text{Tr}\otimes\tau)\big((T_{\chi_{\mathfrak{G}_n}}\otimes 1)(S\otimes 1)\big)
    =\langle T_{\chi_{\mathfrak{G}_n}}, S\rangle,
    \end{eqnarray*}
    where we used traciality of $\text{Tr}\otimes\tau$. Hence
    \( 
    (\Phi|_{\mathcal{S}_1})^*(T_{\chi_{\mathfrak{H}_n}})=T_{\chi_{\mathfrak{G}_n}}.
    \)
    
(ii)$\Rightarrow$(i)   
Suppose that $\Phi$ is factorisable via a Cantor quantum permutation $D\in \mathcal B(L^2(\Omega_X))\bar\otimes\mathcal N$ and satisfies
\( 
\Phi(T_{\chi_{\mathfrak G_n}})=T_{\chi_{\mathfrak H_n}}\) and
\( (\Phi|_{\mathcal S_1})^*(T_{\chi_{\mathfrak H_n}})=T_{\chi_{\mathfrak G_n}}\), \(n\in\mathbb N.
\)
Let $D_n=(p_{x,a})_{x,a\in X_n}\in M_{X_n}\otimes\mathcal N$ be the finite-level component of $D$ and set $\Phi_n$ to be the corresponding factorisable maps.
Using Lemma \ref{ident} and the relations $\Phi\circ\jmath_{n,\infty}=\jmath_{n,\infty}\circ\Phi_n$, we obtain
\[
\frac{1}{|X_n|}\jmath_{n,\infty}(\Phi_n(A_{G_n}))=\frac{1}{|X_n|}\Phi(\jmath_{n,\infty}(A_{G_n}))=\Phi(T_{\chi_{\mathfrak G_n}})=T_{\chi_{\mathfrak H_n}}=\frac{1}{|X_n|}\jmath_{n,\infty}(A_{H_n}),
\]
which by the injectivity of $\jmath_{n,\infty}$ yields
\( 
A_{H_n}=(\id_n\otimes\tau)(D_n^*(A_{G_n}\otimes 1)D_n).
\)
Similarly, the adjoint condition gives
\( 
A_{G_n}=(\id_n\otimes\tau)(D_n(A_{H_n}\otimes 1)D_n^*).
\)
Arguing as in \cite[Theorem~4.3]{paulsen-rahaman}, it follows that
\( 
\tau(p_{x,a}p_{y,b})=0
\) whenever
\( {{\rm rel}}_{G_n}(x,y)\neq {{\rm rel}}_{H_n}(a,b).
\)
After passing to a quotient if necessary, we may assume that $\tau$ is faithful, and hence
\( 
p_{x,a}p_{y,b}=0\)
 whenever
\({{\rm rel}}_{G_n}(x,y)\neq {{\rm rel}}_{H_n}(a,b).
\)
Equivalently,
\( 
(A_{G_n}\otimes 1)D_n=D_n(A_{H_n}\otimes 1)\), \(n\in\mathbb N.
\)
By Lemma~\ref{L:finite_infinite_intertwining},
\( 
(T_{\chi_{\mathfrak G_n}}\otimes 1)D=D(T_{\chi_{\mathfrak H_n}}\otimes 1)\), \( n\in\mathbb N
\)
and the conclusion follows from Theorem~\ref{thm:main_equivalence}.
\end{proof}

Specifying the type of the ancilla in Theorem \ref{thm:main_equivalence} and Corollary \ref{cor:iso_fact} one can obtain similar characterisations for classes other than the qc class, see \cite{paulsen-rahaman} for the finite case. 


\subsection{Connection to locally finite graphs}\label{sec:loc_fin}

In this subsection, we relate the Cantor graph isomorphism game to quantum isomorphism of locally finite graphs as defined by Rollier and Vaes in \cite{rollier-vaes}. 
The notion of quantum isomorphism of two finite graphs is typically defined through perfect strategies for the graph isomorphism game. By \cite{lmr}, the existence of a perfect quantum commuting strategy is equivalent to the existence of a magic unitary that intertwines the adjacency matrices of the corresponding graphs. The latter was taken in \cite{rollier-vaes} as the definition of quantum isomorphism of infinite graphs. Here we show that the 
two approaches are equivalent. 

For a (not necessarily finite) graph $\Pi$ we write $V(\Pi)$ for the set of vertices and $E(\Pi)\subset V(\Pi)\times V(\Pi)$ for the set of edges, where we assume $(i,j)\in E(\Pi)$ if and only if $(j,i)\in E(\Pi)$, and that $(i,i)\not\in E(\Pi)$, 
that is, that the graph $\Pi$ is undirected and has no loops. We write $i\sim_{\Pi} j$ (or $i\sim j$ when $\Pi$ is clear from the context) to signify $(i,j)\in E(\Pi)$. 
For a vertex $i$ of $\Pi$, its degree is the number (allowed to be infinity) 
$\text{deg}(i) = |\{j\in V(\Pi) : (i,j)\in E(\Pi)\}|$. 
We call 
$\Pi$ \emph{locally finite} if $\text{deg}(i)<\infty$ for all $i\in V(\Pi)$. 

Let $I$, $J$ be finite or infinite sets. A matrix $(v_{i,j})_{(i,j)\in I\times J}$ of bounded operators on a Hilbert space $H$ is called a {\it magic unitary} if
\begin{equation}\label{magic}
v_{i,j}=v_{i,j}^*=v_{i,j}^2 \text{ and } \sum_{k\in I}v_{k,j}=1=\sum_{k\in J}v_{i,k} \text{ strongly, for all } i\in I, j\in J.
\end{equation}
It can be easily verified that a magic unitary $(v_{i,j})_{(i,j)\in I\times J}$ defines a 
unitary operator \( 
V:\ell^2(J)\otimes H\to \ell^2(I)\otimes H
\)
with matrix coefficients the projections $v_{i,j}$. 

In \cite{rollier-vaes}, two locally finite graphs $\Pi$ and $\Pi'$ were called quantum isomorphic if there exists a non-zero Hilbert space $H$ and a magic unitary $V=(v_{i,j})_{(i,j)\in V(\Pi)\times V(\Pi')}$ on $H$ such that
\begin{equation}\label{iso_rel}
\sum_{k\in V(\Pi'): k\sim j} v_{i,k}=\sum_{k\in V(\Pi): k\sim i}v_{k,j} \text{ for all }i\in V(\Pi) \text{ and }j\in V(\Pi').
\end{equation}
We note that the relations are equivalent to $(A_\Pi\otimes 1_H)V=V(A_{\Pi'}\otimes 1_H)$, where $A_\Pi$ is the adjacency matrix for $\Pi$.  Multiplying (\ref{iso_rel}) on the left by $v_{i,t}$ and on the right by $v_{s,j}$ and using that $v_{l,k'}v_{l',k'}=0=v_{k,m}v_{k,m'}$ if $l\ne l'$ and $m\ne m'$, one finds that 
\begin{equation}\label{adj_rel}
v_{i,t}v_{s,j}=0 \text{ whenever }i\sim s \text{ and  }t\not\sim j, \text{ or }i\not\sim s \text{ and }t\sim j.
\end{equation} Conversely, if (\ref{adj_rel}) holds and $i,j\in V(\Pi)$ are vertices with $i\sim j$ then 
\begin{eqnarray*}
(V(A_{\Pi'}\otimes 1_H)V^*)_{i,j}
& = &
\sum_{k\sim l} v_{i,k}v_{j,l}=\sum_{k\sim l}v_{i,k}v_{j,l}+\sum _{k\not\sim l}v_{i,k}v_{j,l}\\
& = &
\left(\sum_{k\in J} v_{i,k}\right)\left(\sum_{l\in J} v_{j,l}\right) = 1_H = (A_{\Pi}\otimes 1_H)_{i,j},
\end{eqnarray*}
while if $i \nsim j$, we have 
$$(V(A_{\Pi'}\otimes 1_H)V^*)_{i,j}=0=(A_{\Pi}\otimes 1_H)_{i,j},$$
thus giving (\ref{iso_rel}). 

To a Cantor graph $\mathfrak{G}\subset\Omega_X\times\Omega_X$, 
we associate a locally finite graph $\Pi_{\mathfrak{G}}$ with vertex set $V(\Pi_{\mathfrak{G}}) = \cup_{n=1}^{\infty} X_n$ as a disjoint union, 
and set $E(\Pi_{\mathfrak{G}})$ of edges given by 
$$E(\Pi_{\mathfrak{G}})=\{(x,y)\in V(\Pi_{\mathfrak{G}})\times V(\Pi_{\mathfrak{G}}): x\sim_{\mathfrak{G}} y \},$$
where $x\sim_{\mathfrak{G}}y$ if and only if either $x\sim_{G_n}y$, when $x, y\in X_n$ and $x\neq y$, $n\in\mathbb N$, or $x\in X_n$, $y\in X_{n+1}$ and $y=xx'$, $n\in\mathbb N$, or $x\in X_{n+1}$, $y\in X_n$ and $x=yy'$.
The graph $\Pi_{\mathfrak{G}}$ has a level structure where vertices at the same level are the initial words with the same length and each edge of $\Pi_{\mathfrak{G}}$ either has both of its endpoints within the same level, or its two endpoints are in consecutive levels. For such graphs it is natural to consider isomorphisms that preserve the levels, that is, the levels of the two graphs are isomorphic. 

\begin{definition}\label{def: iso_level}
\rm 
Let $\mathfrak{G}$ and $\mathfrak{H}$ be two Cantor graphs with the corresponding locally finite graphs $\Pi_{\mathfrak{G}}$ and $\Pi_{\mathfrak{H}}$. We say that $\Pi_{\mathfrak{G}}$ and $\Pi_{\mathfrak{H}}$ are 
\emph{quantum isomorphic as level graphs} if they are quantum isomorphic with a magic unitary $V=(v_{x,a})$ such that $v_{x,a}=0$, when $|x|\ne |a|$. 
\end{definition}

Observe that if $\Pi_{\mathfrak{G}}$ and $\Pi_{\mathfrak{H}}$ are quantum isomorphic as level graphs then the corresponding levels $(\Pi_{\mathfrak{G}})_n$ and $(\Pi_{\mathfrak{H}})_n$ are quantum isomorphic, with the magic unitary $(v_{x,a})_{x,a\in X_n}$ implementing the isomorphism. 
The next theorem shows that quantum isomorphism of $\Pi_\mathfrak G$ graphs considered as level graphs can be characterised via perfect graded strategies for Cantor graph isomorphism games.  

\begin{theorem}\label{level_iso}
There is a perfect graded quantum commuting strategy for the isomorphism game of two Cantor graphs $\mathfrak{G}$ and $\mathfrak{H}$ if and only if $\Pi_{\mathfrak{G}}$ and $\Pi_{\mathfrak{H}}$ are quantum isomorphic as level graphs.  
\end{theorem}

\begin{proof}
    If there  is a perfect graded quantum commuting strategy for the isomorphism game of $\mathfrak{G}$ and $\mathfrak{H}$,
then by Theorem \ref{thm:main_equivalence} there is a Cantor quantum permutation $D\in \mathcal{B}(L^2(\Omega)\otimes H)$ such that 
    $(T_{\chi_{\mathfrak{G}_n}}\otimes 1_H)D=D(T_{\chi_{\mathfrak{H}_n}}\otimes 1_H)$, $n\in \bb{N}$. 
By Lemma \ref{L:finite_infinite_intertwining}, there are projections $p_{x,a}\in\mathcal{B}(H)$, $x,a\in X_n$ such that $D_n=(p_{x,a})_{x,a\in X_n}$ is a quantum permutation matrix, satisfying 
    \begin{equation}\label{commutation2}
    (A_{G_n}\otimes 1_H)D_n=D_n(A_{H_n}\otimes 1), \ \ \ n\in\mathbb N
    \end{equation}
    and $\sum_{x'\in[d_n]} p_{xx',ab}=\sum_{a'\in[d_n]}p_{xy,aa'}=p_{x,a}$ for all $x,a\in X_n$ and all $y,b\in[d_n]$. From (\ref{commutation2}) we get  $p_{x,a}p_{y,b}=0$ if $x,y,a,b\in X_n$ and $\text{rel}_{G_n}(x,y)\ne\text{rel}_{H_n}(a,b)$. 
    
    As $p_{x,a}p_{xx',b}=\sum_{a'\in [d_n]} p_{xx',aa'}p_{xx',b}$, we obtain that 
    $p_{x,a}p_{xx',b}=0$ if $b\in X_{n+1}$ and $b\ne aa'$ for any $a'\in [d_n]$, and similarly $p_{x,a}p_{y, aa'}=0$ if $y\in X_{n+1}$ and $y\ne xx'$, for any $x'\in[ d_n]$; here $x$, $a\in X_n$, $x'\in [d_n]$. 
Letting now $V=(v_{x,a})_{x\in V(\Pi_{\mathfrak{G}}), a\in V(\Pi_{\mathfrak{H}})}$, where $v_{x,a}=p_{x,a}$ if $x$, $a$ are initial words of the same length and $v_{x,a}=0$, otherwise,
we see that relations (\ref{iso_rel}) are satisfied; 
we thus obtain that $V:\ell^2(V(\Pi_{\mathfrak{H}}))\otimes H\to \ell^2(V(\Pi_{\mathfrak{G}}))\otimes H$ is unitary giving a quantum level-isomorphism of the locally finite graphs $\Pi_{\mathfrak{G}}$ and $\Pi_{\mathfrak{H}}$.  
    
Conversely, if $V=(v_{x,a})_{x\in V(\Pi_{\mathfrak{G}}), a\in V(\Pi_{\mathfrak{H}})}$ is unitary with $v_{x,a}=0$ if $|x|\ne |a|$ and $V$ implements the quantum isomorphism of $\Pi_{\mathfrak{G}}$ and $\Pi_{\mathfrak{H}}$, then as $\sum_{cc'\in X_{n+1}}v_{xx',cc'}=1$ for $xx'\in X_{n+1}$, we obtain
$$v_{x,b}=\sum_{cc'\in X_{n+1}}v_{x,b} v_{xx',cc'}, \ \ \ x, b\in X_n, x'\in[d_n].$$
Now, if $c\neq b$, then $x\sim xx'$ in $\Pi_{\mathfrak G}$, whereas
$b\not\sim cc'$ in $\Pi_{\mathfrak H}$. Thus, by the forbidden-adjacency
relations,
\( v_{x,b}v_{xx',cc'}=0 
\), when $ c \neq b$.
It follows that
\[
v_{x,b} = \sum_{c'\in[d_n]} v_{x,b}v_{xx',bc'}.
\]
On the other hand, if $y\neq x$, then $y\not\sim xx'$ while
$b\sim bc'$. Again by the forbidden-adjacency relations,
\( 
v_{y,b}v_{xx',bc'}=0\) when  $y\neq x$ .
Using the relation $\sum_{y\in X_n}v_{y,b}=1$, we obtain
\[
v_{xx',bc'} = \left(\sum_{y\in X_n}v_{y,b}\right)v_{xx',bc'} =  v_{x,b}v_{xx',bc'}.
\]
Consequently,
$v_{x,b} =\sum_{c'\in[d_n]} v_{xx',bc'}$.

Similarly, $v_{x,b}=\sum_{x'\in[d_n]} v_{xx',bb'}$ for all $b'\in[d_n]$. Letting $D_n=(v_{x,a})_{x,a\in X_n}$, we get $(A_{G_n}\otimes 1_H)D_n=D_n(A_{H_n}\otimes 1_H)$ for each $n\in\mathbb N$. 
On the other hand, by Remark \ref{r_qX}, the family $\{v_{x,a} : |x| = |a|\}$ gives rise to a Cantor quantum permutation $D$ which  
further satisfies the relations $(T_{\chi_{\mathfrak{G}_n}}\otimes 1_H)D=D(T_{\chi_{\mathfrak{H}_n}}\otimes 1_H)$, $n\in \bb{N}$. 
By Theorem \ref{thm:main_equivalence}, $\mathfrak{G}$ and $\mathfrak{H}$ are isomorphic via a graded bisynchronous correlation of quantum commuting type.
\end{proof}

Recall the homeomorphism  $\tilde f$ of $\Omega_X$  for $f\in \mathrm{GB}(X)$ from Proposition \ref{p_cohhome}.

\begin{proposition}\label{p_gradedlocallyloc}
The following conditions are equivalent: 
\begin{enumerate}
    \item 
there exists a perfect local graded strategy for the isomorphism game on two Cantor graphs $\mathfrak G$ and $\mathfrak H$; 
\item the graphs $\Pi_{\mathfrak G}$ and $\Pi_{\mathfrak H}$ are isomorphic as level graphs; 
\item $\mathfrak G$ and $\mathfrak H$ are homeomorphic via a graded homeomorphism implementing the isomorphism, i.e. $(\tilde f\times \tilde f)(\mathfrak G)=\mathfrak H$, for some $f\in \mathrm{GB}(X)$.
\end{enumerate}
\end{proposition}
\begin{proof}
(i)$\Leftrightarrow$(ii) 
    By Theorem \ref{th_bs_game_alg_reps}, the isomorphism game has a perfect graded local strategy if and only if there exists a one-dimensional representation of the game algebra 
    $\cl A^{\rm bs}(\text{Iso}(\mathfrak G,\mathfrak H))$; the latter is given by $\chi(p_{x,a})=\delta_{f_n(x),a}$, $n\in\mathbb N$, $x,a\in X_n$,  where $f=(f_n)_{n\in\mathbb N}\in {\rm GB}(X)$ (see Proposition \ref{p_cohhome}), and $$\delta_{f_n(x),a}\delta_{f_n(y),b}=0 \text{ if }\text{rel}_{G_n}(x,y)\ne\text{rel}_{H_n}(a,b), n\in \mathbb N, x,y, a,b\in X_n,$$ showing that $f_n:X_n\to X_n$ gives the isomorphism of the graphs $G_n$ and $H_n$. 
    As for $n\in\mathbb N$, $x,a\in X_n$ and  $xx',b\in X_{n+1}$, $p_{x,a}p_{xx',b}=\sum_{a'\in [d_n]} p_{xx',aa'}p_{xx',b}$, we obtain that 
    $p_{x,a}p_{xx',b}=0$ if $b\ne aa'$ for any $a'\in [d_n]$, and similarly $p_{x,a}p_{y, aa'}=0$ if $y\in X_{n+1}$ and $y\ne xx'$, for any $x'\in[ d_n]$. This implies $\delta_{f_n(x),a}\delta_{f_n(xx'),b}=0$ if $b\ne aa'$, and $\delta_{f_n(x),a}\delta_{f_{n+1}(y),aa'}=0$, if $y\ne xx'$;  hence for $x\in X_n$, $y\in X_{n+1}$,  $x\sim y$ if and only if 
    $f(x)=f_n(x)\sim f_{n+1}(y)=f(y)$, showing that $f$ is the isomorphism of the level graphs  $\Pi_{\mathfrak G}$ and $\Pi_{\mathfrak H}$. 
    
Conversely, any level isomorphism  $f=(f_n)_{n\in\mathbb N}$, $f_n:X_n\to X_n$, $n\in \mathbb N$, of $\Pi_{\mathfrak G}\simeq \Pi_{\mathfrak H}$ determines a bijective graded map. Indeed, for $n\in\mathbb N$, $x\in X_n$ and $x'\in[d_n]$, we have $f_{n+1}(xx')=f(xx')\sim f(x)=f_n(x)$ giving $f_{n+1}(xx')=f_n(x)a'$ for some $a'\in [d_n]$.  By Proposition \ref{p_cohhome}, $f$ determines a one-dimensional representation 
of $\cl B_X$. Since $f_n$ is a graph isomorphism between $G_n$ and $H_n$, 
$f$ descends to a character of 
${\cl A}^{\rm bs}(\text{Iso}(\mathfrak G,\mathfrak H))$;
according to Theorem \ref{th_bs_game_alg_reps} it gives rise to a perfect local strategy for $\text{Iso}(\mathfrak G,\mathfrak H)$. 

(ii)$\Leftrightarrow$(iii) If $f=(f_n)_{n\in\mathbb N}$ is a level isomorphism of the graphs $\Pi_{\mathfrak G}$ and $\Pi_{\mathfrak H}$ then $f\in\text{GB}(X)$ and by Proposition \ref{p_cohhome} $\tilde f:\Omega_X\to\Omega_X$ is a homeomorphism. 
Moreover, observing that $\mathfrak{p}_n=\iota_{n,\infty}^*\times\iota_{n,\infty}^*$,  we obtain
$$\mathfrak{p}_n((\tilde f\times\tilde f)(\mathfrak G))=(f_n\times f_n)(\mathfrak{p_n}(\mathfrak G))=(f_n\times f_n)(G_n)=H_n=\mathfrak p_n(\mathfrak H), n\in\mathbb N,$$ 
giving $(\tilde f\times\tilde f)(\mathfrak G)=\mathfrak H.$
\end{proof}

\begin{proposition}\label{notiso_but_qc}
    There exist Cantor graphs $\mathfrak G$ and $\mathfrak H$ which are isomorphic
    via a graded bisynchronous correlation of quantum commuting type
but not via a graded bisynchronous correlation of local type. 
\end{proposition}

\begin{proof} 
By \cite{atserias-etal}, there exist finite graphs $G$ and $H$ which are
quantum isomorphic but not isomorphic. Let
\( p=(p_{x,a})_{x\in V(G),\,a\in V(H)}
\)
be a magic unitary implementing a quantum isomorphism between $G$ and $H$. 
Choose a sequence of finite sets $[d_i]$ of cardinality at least $2$, set
\(  X_n=V(G)\times [d_1]\times\cdots\times[d_{n-1}]\), and \( A_n=V(H)\times [d_1]\times\cdots\times[d_{n-1}]\), and consider the associated Cantor sets $ \Omega_X$ and $\Omega_A$ (we have set $X_1 = V(G)$ and $A_1 = V(H)$). 
Define 
$$\mathfrak{G}(G)=\{(x_i)_{i\in\mathbb N},(y_i)_{i\in \mathbb N})\in\Omega_X\times\Omega_X: x_1\sim_G y_1 \mbox{ or } x_1 = y_1; x_i,y_i\in[d_{i-1}], i\geq 2\}$$ and $\mathfrak{G}(H)$ defined similarly.

For each level $n$, write a vertex of $X_n$ as $xw$, where
$x\in V(G)$ and $w\in [d_1]\times\cdots\times[d_{n-1}]$, and similarly write
a vertex of $A_n$ as $az$. Define
\(  v_{xw,az} =  \delta_{w,z} p_{x,a}.
\)
If $|xw|\neq |az|$, set $v_{xw,az}=0$. Then
\(  V=(v_{\beta,\alpha})_{\beta\in V(\Pi_{\mathfrak G(G)}),\,\alpha\in V(\Pi_{\mathfrak G(H)})}
\)
is a level-preserving magic unitary.

Indeed, at each fixed level $n$, the matrix $(v_{xw,az})$ is the tensor
product of the magic unitary $(p_{x,a})$ with the identity permutation matrix
on the tail coordinates. Since
adjacency inside level $n$ depends only on the first coordinate, the quantum
isomorphism relations for $G$ and $H$ imply the horizontal quantum
isomorphism relations for the level graphs. The vertical prefix edges are also
preserved, because the tail coordinates are matched identically. Thus $\Pi_{\mathfrak G(G)}$ and $\Pi_{\mathfrak G(H)}$ are quantum isomorphic
as level graphs. Hence, by Theorem \ref{level_iso}, the Cantor graphs
$\mathfrak G(G)$ and $\mathfrak G(H)$ are graded qc-isomorphic.

On the other hand, $\Pi_{\mathfrak G(G)}$ and $ \Pi_{\mathfrak G(H)}$ are not isomorphic as level graphs. Indeed, any
level-preserving graph isomorphism
\(  \Pi_{\mathfrak G(G)}\cong \Pi_{\mathfrak G(H)}
\)
restricts at level $1$ to a graph isomorphism $G\cong H$, contradicting the
choice of $G$ and $H$. The proof is completed in view of Proposition \ref{p_gradedlocallyloc}. 
\end{proof}

Let $G=(V(G),E(G)$, $H=(V(H),E(H))$ be simple finite graphs and $|V(G)|=|V(H)|$. It is known (see \cite{bigalois}) that 
the graph isomorphism game between $G$ and $H$ has a perfect strategy 
of type ${\rm t}$ if and only if there exists a unital *-homomorphism 
$$\alpha: \cl D_{V(H)}\hspace{-0.1cm}\to \hspace{-0.1cm} \cl D_{V(G)}\hspace{-0.05cm}\otimes \hspace{-0.05cm} \cl N \text{ such that }\alpha\hspace{-0.05cm}\circ \hspace{-0.05cm} A_H \hspace{-0.05cm} = \hspace{-0.05cm} (A_G\hspace{-0.05cm}\otimes \hspace{-0.05cm}\id)\hspace{-0.05cm}\circ \hspace{-0.05cm}\alpha, 
(\Tr\hspace{-0.05cm}\otimes \hspace{-0.05cm}\id)\hspace{-0.05cm}\circ \hspace{-0.05cm}\alpha \hspace{-0.05cm} = \hspace{-0.05cm} \Tr(\cdot) 1,$$ 
where 
$\cl N=\cl B(\cl H)$ for some Hilbert space $\cl H$
(resp.~$\cl N= \cl R^\omega$, $\cl N= M_n$ and $\cl N=\mathbb C$) when $\rm t=qc$ (resp.~$\rm t=qa$, $\rm t=q$ and $\rm t=loc$). Then $$\alpha (\delta_v)=\sum_{w\in V(G)} \delta_w\otimes p_{w,v}, v\in V(H),$$
and $P=(p_{w,v})_{w,v}$ is a magic unitary which satisfies $(A_G\otimes I)P=P(A_H\otimes I)$. In the qc-case, one can take the $C^*$-algebra 
$\cl N=C^*({\cl A}^{\rm bs}(\text{Iso}(G,H)))$. If $G=H$, the map $\alpha$ is a coaction of the quantum group $C^*({\cl A}^{\rm bs}(\text{Iso}(G,G)))$ acting on $V(G)$. 

\begin{corollary}
Let $\mathfrak{G}$ and $\mathfrak{H}$ be Cantor graphs on $\Omega_X$. The following are equivalent:
\begin{enumerate}
\item there is a perfect graded quantum commuting strategy for the isomorphism game of  $\mathfrak{G}$ and $\mathfrak{H}$;
\item there exists a system of unital *-homomorphisms
\begin{align*}
    \alpha =(\alpha_n: \cl D_{X_n}\to \cl D_{X_n}\otimes \cl B(H), \alpha_n\circ A_{H_n}=(A_{G_n}\otimes I)\alpha_n, (\Tr\otimes\id)\alpha_n=\Tr(\cdot)1 )
\end{align*}
such that for any $m<n$ the diagram 
\[
\begin{tikzcd}
\cl D_{X_m} \arrow[r,"\iota_{m,n}"] \arrow[d,"\alpha_m"'] 
& \cl D_{X_n} \arrow[d,"\alpha_n"] \\
\cl D_{X_m}\otimes \cl B(H) \arrow[r,"{  \iota_{m,n}} \otimes \mathrm{id}"'] 
& \cl D_{X_n}\otimes\cl B(H)
\end{tikzcd}
\] commutes, where $\iota_{m,n}: \cl D_{X_m}\to \cl D_{X_n}$, is the injective homomorphism given by $\iota_{m,n}(\delta_x)=\sum_{|x'|=n-m} \delta_{xx'}$, $x\in X_m$.
\end{enumerate}
 \end{corollary}
 \begin{proof}
 Let $V=(v_{x,a})_{x\in V(\Pi_{\mathfrak{G}}), a\in V(\Pi_{\mathfrak{H}}})$ be the magic unitary that implements the quantum level-graph isomorphism of $\Pi_{\mathfrak{G}}$ and  $\Pi_{\mathfrak{H}}$
 and let $\alpha_n(\delta_a)=\sum_{x\in X_n}\delta_x\otimes v_{x,a}$, $a\in X_n$. 
 As each $v_{x,a}$ is a projection and $\sum_{a\in X_n}v_{x,a}=I$,  for any $x\in X_n$, $\alpha_n$ is  a unital *-homomorphism. The condition $\sum_{x\in X_n} v_{x,a}=I$, $a\in X_n$, gives the trace preservation.  Moreover, the family $\alpha_n$ satisfies the conditions of (ii):
 \begin{align}\label{alpha_rel}
    && \alpha_n(\iota_{m,n}(\delta_a))
    = \alpha_n\left(\sum_{|a'|=n-m}\delta_{aa'}\right)
    = \sum_{|a'|=n-m}\sum_{xx'\in X_n} \sum_{|x'|=n-m}\delta_{xx'}\otimes v_{xx',aa'}, \\
     &&(\iota_{m,n}\otimes{\rm id})(\alpha_m(\delta_a))=\sum_{x\in X_m}i_{m,n}(\delta_x)\otimes v_{x,a}=\sum_{x\in X_m}\sum_{|x'|=n-m}\delta_{xx'}\otimes v_{x,a};\nonumber
 \end{align}
 as $\sum_{|a'|=n-m}v_{xx',aa'}=v_{x,a}$  for any $x'$, we obtain the 
  equality $\alpha_n\circ \iota_{m,n}=(\iota_{m,n}\otimes{\rm id})\circ\alpha_m$. With $D_n=(v_{x,a})_{x,a\in X_n}$, the equality $(A_{G_n}\otimes I)D_n=D_n(A_{H_n}\otimes I)$ is equivalent to
 $\alpha_n\circ A_{H_n}=(A_{G_n}\otimes \id)\circ\alpha_n$. 

 To see the converse we note first that any $\alpha_n$ from (ii) is given by $\alpha_n(\delta_a)=\sum_{x\in X_n}\delta_x\otimes v_{x,a}$, where $D_n:=(v_{x,a})_{x,a\in X_n}$ is a magic unitary. From (\ref{alpha_rel}), if $m<n$, we get $\sum_{|a'|=n-m}v_{xx',aa'}=v_{x,a}$ for all $a\in X_m$ and $xx'\in X_n$ with $|x'|=n-m$. 
 In particular, as $v_{xx',aa'}v_{yy',aa'}=0$ if $xx'\ne yy'$, we obtain $v_{x,a}v_{yy',aa'}=0$ if $x\ne y$. Furthermore,
 $v_{xx',aa'}\leq v_{x,a}$ and hence 
 $$v_{x,a}=v_{x,a}\left(\sum_{yy'\in X_n} v_{yy',aa'}\right)=\sum_{|y'|=n-m}v_{xy',aa'}.$$
 Letting now  $V=(v_{x,a})_{x\in V(\Pi_{\mathfrak{G}}), a\in V(\Pi_{\mathfrak{H}})})$ we get that $V$ implements the quantum level-graph isomorphism  of $\Pi_{\mathfrak{G}}$ and $\Pi_{\mathfrak{H}}$ and hence by Theorem \ref{level_iso} gives a qc-isomorphism of Cantor graphs $\mathfrak G$ and $\mathfrak H$. 
 \end{proof}

\end{document}